\documentclass[usenatbib,fleqn]{mnras}

\makeatletter
\newlength{\abovecaptionskip}%
\makeatother

\usepackage{threeparttable}

\usepackage{amsmath,amssymb}
\usepackage{mathrsfs}
\usepackage{graphicx}
\usepackage{epstopdf}
\graphicspath{{./figures/}}
\usepackage{url}
\usepackage{aas_macros}
\usepackage{astro}
\usepackage{deluxetable}

\newcommand{\Mdot}{\dot{M}}
\newcommand{\eddr}{\dot{M}/\dot{M}_{\rm Edd}}

\newcommand\lsim{\mathrel{\rlap{\lower4pt\hbox{\hskip1pt$\sim$}}
    \raise1pt\hbox{$<$}}}
\newcommand\gsim{\mathrel{\rlap{\lower4pt\hbox{\hskip1pt$\sim$}}
    \raise1pt\hbox{$>$}}}
\newcommand{\rs}{r_{\rm s}}
\newcommand{\rb}{r_{\rm b}}

\newcommand{\vw}{\tilde{v}_{w}}

\newcommand{\dxdy}[2]{\frac{d #1}{d #2} }
\newcommand{\ddr}[1]{\dxdy{#1}{r}}

\newcommand{\dpdr}{\dxdy{p}{r}}

\newcommand{\dsdr}{\dxdy{s}{r}}

\newcommand{\ke}{\frac{v^2}{2}}
\newcommand{\kew}{\frac{\tilde{v}_{w}^2}{2}}

\newcommand{\gammaf}{\frac{\gamma}{\gamma-1}}
\newcommand{\gammafi}{\frac{\gamma-1}{\gamma}}
\newcommand{\cs}{\frac{p}{\rho}}
\newcommand{\Q}{q (\ke+\kew-\gammaf \cs)}

\newcommand{\kb}{k_{\rm b}}
\renewcommand{\mp}{m_{\rm p}}
\newcommand{\pc}{\rm pc}

\newcommand{\Menc}{M_{\rm enc}}
\newcommand{\rhostar}{\rho_*}
\newcommand{\Mstar}{M_{\star}}

\newcommand{\Mbh}[1][]{M_{\bullet#1}}
\newcommand{\Mbheight}{M_{\bullet,8}}

\newcommand{\rinf}{r_{\rm inf}}
\newcommand{\rIa}{r_{\rm Ia}}

\newcommand{\RateIa}{R_{\rm Ia}}

\newcommand{\vwO}{v_{w}}

\newcommand{\pyear}{{\rm yr}^{-1}}

\renewcommand{\th}{t_{\rm h}}
\newcommand{\tcool}{t_{\rm cool}}
\newcommand{\tff}{t_{\rm ff}}

\newcommand{\densSlope}{\nu}

\author[Generozov, Stone, \& Metzger]{Aleksey Generozov\thanks{E-mail: ag@astro.columbia.edu}, Nicholas C. Stone, Brian~D.~Metzger\\
  Columbia Astrophysics Laboratory, Columbia University, 550 West
  120th Street, New York, NY 10027} \title[Circumnuclear Media of
Quiescent SMBHs]{Circumnuclear Media of Quiescent Supermassive Black
  Holes}

\begin{document}
\maketitle

\begin{abstract}
  We calculate steady-state, one-dimensional hydrodynamic profiles of
  hot gas in slowly accreting (``quiescent") galactic nuclei for a
  range of central black hole masses $M_{\bullet}$, parametrized gas
  heating rates, and observationally-motivated stellar density
  profiles.  Mass is supplied to the circumnuclear medium by stellar
  winds, while energy is injected primarily by stellar winds,
  supernovae, and black hole feedback.  Analytic estimates are derived
  for the stagnation radius (where the radial velocity of the gas
  passes through zero) and the large scale gas inflow rate, $\dot{M}$,
  as a function of $M_{\bullet}$ and the gas heating efficiency, the
  latter being related to the star-formation history.  We assess the
  conditions under which radiative instabilities develop in the
  hydrostatic region near the stagnation radius, both in the case of a
  single burst of star formation and for the average star formation
  history predicted by cosmological simulations.  By combining a
  sample of measured nuclear X-ray luminosities, $L_{X}$, of nearby
  quiescent galactic nuclei with our results for
  $\dot{M}(M_{\bullet})$ we address whether the nuclei are consistent
  with accreting in a steady-state, thermally-stable manner for
  radiative efficiencies predicted for radiatively inefficiency
  accretion flows.  We find thermally-stable accretion cannot explain
  the short average growth times of low mass black holes in the local
  Universe, which must instead result from gas being fed in from large
  radii, due either to gas inflows or thermal instabilities acting on
  larger, galactic scales.  Our results have implications for attempts
  to constrain the occupation fraction of SMBHs in low mass galaxies
  using the mean $L_{\rm X}-M_{\bullet}$ correlation, as well as the
  predicted diversity of the circumnuclear densities encountered by
  relativistic outflows from tidal disruption events.
\end{abstract}

\section{Introduction}
\label{sec:introduction}

Supermassive black holes (SMBHs) lurk in the centres of most, if not
all nearby galaxies (see reviews by,
e.g. \citealt{KormendyRichstone:1995a};
\citealt{FerrareseFord:2005a}). However, only a few percent of these
manifest themselves as luminous active galactic nuclei (AGN).  Nearly
quiescent SMBHs, such as those hosting low luminosity AGN, constitute
a silent majority (e.g.~\citealt{Ho:2009a}).

Understanding why most SMBHs appear to be inactive requires
characterizing their gaseous environments.  Gas near the SMBH sphere
of influence, hereafter denoted the `circumnuclear medium' (CNM),
controls the mass accretion rate, $\dot{M}_{\bullet}$.  The accretion
rate in turn determines the SMBH luminosity and the feedback of its
energy and momentum output on larger scales.  Dense gas in the nucleus
may lead to runaway cooling, resulting in bursty episodes of star
formation and AGN activity (e.g.~\citealt{Ciotti&Ostriker07}).

Knowledge of how $\dot{M}_{\bullet}$ depends on the SMBH mass, $\Mbh$,
and other properties of the nucleus informs key questions related to
the co-evolution of SMBHs and their host galaxies with cosmic time
(e.g.~\citealt{Kormendy&Ho13}; \citealt{HeckmanBest:2014a}).  In the
low redshift Universe, SMBH growth is dominated by low mass black
holes, $M_{\bullet} \lesssim 10^{8}M_{\odot}$
(e.g.~\citealt{Heckman+04}), a fact often attributed to the trend of
`cosmic down-sizing' resulting from hierarchical structure growth
(e.g, \citealt{Gallo+08}).  However, the physical processes by which
typical low mass black holes accrete could in principle be distinct
from those operating at higher SMBH masses, or those in AGN.  Of key
importance is whether SMBHs grow primarily by the accretion of gas fed
in directly from galactic or extragalactic scales, or whether
significant growth can result also from local stellar mass loss in the
nuclear region.

A better understanding of what mechanisms regulate accretion onto
quiescent SMBHs would shed new light on a variety of observations,
such as the occupation fraction of SMBHs in low mass galaxies.
\citet{miller+2015} use the average relationship between the nuclear
X-ray luminosities, $L_{X}$, of a sample of early type galaxies and
their associated SMBH masses to tentatively infer that the SMBH
occupation fraction becomes less than unity for galaxies with stellar
masses $M_{\star} \lesssim 10^{10}M_{\odot}$ ($M_{\bullet} \lesssim
10^{7}M_{\odot}$).  This method relies on extrapolating a power-law
fit of the $L_X-\Mbh$ distribution to low values of $L_{\rm X}$ below
the instrument detection threshold, an assumption which is
questionable if different physical processes control the accretion
rates onto the lowest mass SMBHs.

The gas density in galactic nuclei also influences the emission from
stellar tidal disruption events (TDEs), such as the high energy
transient {\it Swift} J1644+57 (\citealt{Bloom+11}, \citealt{Burrows+11}; \citealt{Levan+11}; \citealt{Zauderer+11}).
This event was powered by an impulsive relativistic jet, which
produced synchrotron radio emission as the jet material decelerated
from shock interaction with the CNM of the previously quiescent SMBH
(\citealt{Giannios&Metzger11}; \citealt{Zauderer+11}).  Modeling of J1644+57 showed that the CNM density was much lower than
that measured surrounding Sgr A$^{\star}$ on a similar radial scale
(\citealt{BergerZauderer+:2012a}; \citealt{Metzger+2012}).  However, a
TDE jet which encounters a denser CNM would be decelerated more rapidly,
producing different radio emission than in J1644+57.
Variations in the properties of the CNM could help explain why most
TDEs appear to be radio quiet (e.g.~\citealt{Bower+13};
\citealt{VanVelzen+13}).

Gas comprising the CNM of quiescent (non-AGN) galaxies can in
principle originate from several sources: (1) wind mass loss from
predominantly evolved stars; (2) stellar binary collisions; and (3)
unbound debris from a recent TDE.  Stellar wind mass loss is probably
the dominant source insofar as collisions are relevant only in
extremely dense stellar environments for very young stellar
populations (\citealt{Rubin&Loeb11}), while \citet{MacLeod+13} find
that TDEs contribute subdominantly to the time-averaged accretion rate
of quiescent SMBHs.

The gas inflow rate on large scales is much easier to constrain both
observationally and theoretically than the black hole (horizon scale)
accretion rate.  \citet{Ho:2009a} determines the inflow rates in a
sample of early-type galaxies by using X-ray observations to determine
the Bondi accretion rate, and also by using estimated mass loss rates of
evolved stars. Both methods lead him to conclude that the
available gas reservoir is more than sufficient to power the observed
low-luminosity AGN, assuming the standard $\sim$ 10 per cent radiative
efficiency for thin disc accretion.  Several lines of evidence now
suggest that low-luminosity AGN result from accretion proceeding in a
radiatively inefficient mode (\citealt{Yuan&Narayan14}), due either to
the advection of gravitationally-released energy across the SMBH
horizon (e.g.~\citealt{Narayan&Yi95}) or due to disc outflows, which
reduce the efficiency with which the inflowing gas ultimately reaches
the SMBH (e.g.~\citealt{Blandford&Begelman99}; \citealt{Li+13}).

Another approach to determine the inflow rates, which we adopt
in this paper, is to directly calculate the density, velocity and
temperature profiles of the CNM using a physically motivated
hydrodynamic model.  Mass is injected into the nuclear environment via
stellar winds, while energy is input from several sources including
stellar winds, supernovae (SNe), and AGN feedback
(\citealt{Quataert:2004a,De-ColleGuillochon+:2012a,ShcherbakovWong+:2014a}).
Unlike previous works, which focused primarily on modeling individual
galaxies, here we model the CNM properties across a
representative range of galaxy properties, including different SMBH
masses, stellar density profiles, and star formation histories (SFHs).

Previous studies, employing multi-dimensional numerical hydrodynamics and
including variety of (parametrized) physical effects, have focused on
massive elliptical galaxies (e.g.~\citealt{Ciotti&Ostriker07};
\citealt{Ciotti+10}).  These works show the periodic
development of cooling instabilities on galactic scales, which
temporarily increase the gas inflow rate towards the nucleus until
feedback becomes strong enough to shut off the flow and halt
SMBH growth.  

In this paper we focus on time-independent models, in which the
nuclear gas receives sufficient heating to render radiative cooling
negligible.  This approach allows us to systematically explore the
relevant parameter space and to derive analytic expressions that prove
useful in determining under what conditions cooling instabilities
manifest in the nuclear region across the expected range of galaxy
properties, or whether other (non-AGN) forms of feedback can produce a
prolonged state of steady, thermally stable accretion.  Even if
cooling instabilities develop on galactic scales over longer $\sim$
Gyr time-scales, we aim to explore whether a quasi steady-state may
exist between these inflow events on smaller radial scales comparable
to the sphere of influence.

In the presence of strong heating, one-dimensional steady-state flow
is characterized by an inflow-outflow structure, with a critical
radius known as the``stagnation radius" $\rs$ where the radial
velocity passes through zero.  Mass loss from stars interior to the
stagnation radius is accreted, while that outside $\rs$ is unbound in
an outflow from the nucleus.  The stagnation radius, rather than the
Bondi radius, thus controls the inflow rate (although we will
show that $\rs$ usually resides near the nominal Bondi radius).  When
heating is sufficiently weak, however, the stagnation radius may move
to much larger radii or not exist at all, significantly increasing the
inflow rate the SMBH, i.e. a ``cooling
flow".  However, the hydrostatic nature of gas near the stagnation
radius also renders the CNM at this location particularly susceptible
to {\it local} thermal instabilities, the outcome of which could well
be distinct from the development of a cooling flow.

This paper is organized as follows.  In $\S\ref{sec:model}$ we
describe our model, including the sample of galaxy properties used in
our analysis ($\S\ref{sec:gal_model}$) and our numerical procedure for
calculating the steady-state hydrodynamic profile of the CNM
($\S\ref{sec:hydro}$).  In $\S\ref{sec:results}$ we describe our
analytic results, which are justified via the numerical solutions we
present in $\S\ref{sec:numerical}$.  We move from a general and
parametrized treatment of heating to a physically motivated one in \S
\ref{sec:heating}, where we consider a range of physical processes
that can inject energy into the CNM.  In $\S\ref{sec:discussion}$ we
discuss the implications of our results for topics which include the
nuclear X-ray luminosities of quiescent black holes, jetted TDEs, and
the growth of SMBHs in the local Universe.  In
$\S\ref{sec:conclusions}$ we summarize our conclusions.  Table
\ref{table:definitions} provides the definitions of commonly used
variables.  Appendix \ref{app:rs} provides useful analytic results
for the stagnation radius, while Appendix \ref{app:windheat} provides
the details of our method for calculating stellar wind heating and
mass input.

\begin{table*}
\begin{threeparttable}
\begin{minipage}{18cm}
\caption{Definitions of commonly used variables}
\begin{tabular}{ll}
  \hline
  {Variable} & {Definition} \\
  \hline
  $M_{\bullet}$ & Black hole mass \\
  $\tilde{v}_{w}$ & Total heating parameter, including minimum heating
  rate from stellar and black hole velocity dispersion \\
  $v_{w}$ & Total heating parameter, excluding minimum heating rate from
  velocity dispersion \\
   $\sigma_0$ & Stellar velocity dispersion (assumed to be radially
   constant).\\
   $t_{\rm dyn}$ & $r/\sigma_0$: Dynamical time at large radii (where stellar
   potential dominates)\\
   $t_{\rm ff}$ & $(r^3/(G \Mbh))^{1/2}$: Free fall time (where the
   black hole potential dominates)\\
  $\zeta$ & Alternative heating parameter, $\zeta \equiv \sqrt{1 + (v_w/\sigma_0)^2}$ (eq.~[\ref{eq:zetacrit}])\\
  $r_{s}$ & Stagnation radius, where gas radial velocity goes to zero \\
  $r_{\rm inf}$ & Radius of sphere of influence (eq.~[\ref{eq:rsoi}]) \\
  $r_{\rm b}$ & Outer break radius of stellar density profile \\ 
  $r_{\rm Ia}$ & Radius interior to which SN Ia are infrequent compared to the dynamical time-scale (eq.~[\ref{eq:rIa}]) \\ 
  $\rho_{\star}(r)$ & 3D radial stellar density profile \\
  $\rho(r)$ & Gas density of CNM \\
  $M_{\star}(r)$ & Total enclosed stellar mass inside radius $r$ \\
  $M_{\rm enc}(r)$ & Total enclosed mass inside radius $r$ (SMBH + stars) \\
  $q(r)$ & Mass source term due to stellar winds, $q \propto \rho_{\star}$ (eq.~[\ref{eq:q}]) \\
  $\eta$ & Parameter setting normalization of mass input from stellar winds (eq.~[\ref{eq:q}]) \\
  $\tau_{\star}$ & Age of stellar population, in case of a single burst of star formation \\
  $\Gamma$ & Power-law slope of radial stellar surface brightness
  profile interior to the break radius \\
  $\delta$ & Power-law slope of the 3D stellar density profile inside of
  the break radius, $\delta \equiv \Gamma+1$.\\
  $\dot{M}$ & Large scale inflow rate (not necessarily equal to the SMBH accretion rate)   \\
  $\dot{M}_{\bullet}$ & SMBH accretion rate, $\dot{M}_{\bullet} = \alpha\dot{M}$, where $\alpha < 1$ accounts for outflows from the accretion disc on small scales.   \\
  $\dot{M}_{\rm Ia}$ & Maximum accretion rate as limited by SN Ia (eq.~[\ref{eq:eddr_Ia}]) \\
  $\dot{M}_{\rm C}$ & Equilibrium inflow rate set by Compton heating acting alone (eq.~[\ref{eq:MdotC}]) \\
  $\dot{M}_{\rm TI}$ & Maximum accretion rate for thermally stable accretion (eq.~[\ref{eq:Mdotmax}]) \\
  $\dot{q}_{\rm heat}/|\dot{q}_{\rm rad}|$ & Ratio of external heating (stellar winds, SN Ia, MSPs) to radiative cooling (eq.~[\ref{eq:cooling2}]) \\
  $t_{\rm h}$ & Hubble time-scale \\
  $\nu$ & Gas density power-law slope at the stagnation radius
  (eq.~[\ref{eq:densSlope}]) \\
  \hline
\label{table:definitions}  
\end{tabular}
\end{minipage}
\end{threeparttable}

\end{table*}

\section{Model}
\label{sec:model}
\subsection{Galaxy models}
\label{sec:gal_model}
\citet{LauerFaber+:2007a} use {\it Hubble Space Telescope} WFPC2 imaging to
measure the radial surface brightness profiles for hundreds of nearby
early type galaxies. The measured profile is well fit by a ``Nuker" law parameterization:
\begin{equation}
  I(\xi)=I_b 2^{(\beta-\Gamma)/\alpha} \xi^{-\Gamma} (1+\xi^\alpha)^{-(\beta-\Gamma)/\alpha}, \,\,\,\xi\equiv\frac{r}{\rb},
\end{equation}
i.e., a broken power law that transitions from an inner power law
slope, $\Gamma$, to an outer power law slope, $\beta$, at a break
radius, $\rb$.  If the stellar population is spherically symmetric,
then this corresponds to a 3D stellar density $\rhostar \propto
r^{-1-\Gamma}$ for $r \ll \rb$ and $\rhostar\propto r^{-1-\beta}$ for
$r \gg \rb$.  We can write the deprojected stellar density
approximately (formally, this is the $\alpha \rightarrow \infty$
limit) as
\begin{align}
\rho_\star = 
\begin{cases}
\rho_\star|_{\rinf} \left(r/\rinf\right)^{-1-\Gamma} & r \leq \rb\\
\rho_\star|_{\rb} \left(r/\rb\right)^{-1-\beta} & r > \rb,
\label{eq:rhostar}
\end{cases}
\end{align}
where $\rho_\star|_{\rinf}$ is the stellar density at the radius of
the black hole sphere of influence\footnote{This approximate
  expression agrees surprisingly well with the mean empirical scaling
  relation for $r_{\rm inf}(M_\bullet)$ calculated in appendix C of
  \citet{Stone&Metzger15}, although we note that this relation has
  significant scatter. Also, we note that the  scaling is somewhat
  different for cores and cusps. Fixing the power law slope,
  $\rinf\simeq 25 (8) \Mbheight^{0.6}$ for core (cusp)
  galaxies. We will use separate scaling relations for $\rinf$ for
  cores and cusps unless otherwise noted.},
\begin{equation}
  \rinf \simeq G \Mbh/\sigma_{\bullet}^2 \approx 14 M_{\bullet,8}^{0.6}\,{\rm pc},
\label{eq:rsoi}
\end{equation}
where $M_{\bullet,8} \equiv M_{\bullet}/10^{8}M_{\odot}$ and the
second equality in (\ref{eq:rsoi}) employs the $\Mbh-\sigma$
relationship of \citet{McConnellMa+:2011a},
 \begin{align}
M_{\bullet} \simeq 2\times 10^{8}\left(\frac{\sigma_{\bullet}}{200\,{\rm
      km\,s^{-1}}}\right)^{5.1}M_{\odot}.
\label{eq:Msigma}
\end{align}
This may be of questionable validity for low mass black holes (e.g.,
\citealt{Greene+2010, Kormendy+2013}). Also, several of the black hole
masses used in \citet{McConnellMa+:2011a} were underestimated
\citep{Kormendy+2013}. However, our results are not overly sensitive
to the exact form of the $\Mbh-\sigma$ relationship that we use.

A galaxy model is fully specified by four parameters: $\Mbh$,
$\Gamma$, $\rb$, and $\beta$.  We compute models for three different
black hole masses, $\Mbh = 10^6$, $10^7$, $10^8 \Msun$.  The
distribution of $\Gamma$ in the \citet{LauerFaber+:2007a} sample is
bimodal, with a concentration of ``core'' galaxies with $\Gamma <
0.3$ and a concentration of ``cusp'' galaxies with $\Gamma > 0.5$.
We bracket these possibilities by considering models with $\Gamma=0.8$
and $\Gamma=0.1$.  

We fix $\beta = 2$ but find that our results are not overly sensitive
to the properties of the gas flow on radial scales $\gtrsim \rb$.  The
presence of the break radius $r_{b}$ is, however, necessary to obtain
a converged steady state for some regions of our parameter
space\footnote{In addition, for $\beta\leq 2$ the stellar density must
  steepen at still larger radii to avoid the mass enclosed diverging
  to infinity. However, we do not find it necessary to include this
  outer break.}.  We consider solutions calculated for up to four
values of $\rb$: 50 pc, 100 pc, 200 pc, and 400 pc, motivated by the
range of break radii from the \citet{LauerFaber+:2007a}
sample.\footnote{For the core galaxies the break radius follows the
  scaling relationship $\rb\sim 90 \, (\Mbheight)^{0.5}$ pc, with
  scatter of approximately one dex. Most of the cusp galaxies have $\rb$
  between 100 and 1000 pc, and lack a clear trend with $\Mbh$. The
  mean $\rb$ for cusp galaxies in the \citet{LauerFaber+:2007a} sample
  is $\sim 240$ pc.} We neglect values of $\rb$ which would give
unphysically large bulges for a given $\Mbh$.

Finally, we note that \citealt{LauerFaber+:2005a} find that $\sim
60\%$ of cusp galaxies and $\sim29\%$ of core galaxies have (generally
unresolved) emission in excess of the inward extrapolation of the
Nuker law.  Indeed, some low mass galaxies with $\Mbh\lsim 10^{8}
\Msun$ possess nuclear star clusters (\citealt{GrahamSpitler:2009a}),
which are not accounted for by our simple parametrization of the
stellar density.  In such cases gas and energy injection could be
dominated by the cluster itself, i.e. concentrated within its own
pc-scale ``break radius" which is much smaller than the outer break in
the older stellar population on much larger scales.  Although our
analysis does not account for such an inner break, we note that for
high heating rates the stagnation radius and concomitant inflow
rate are not sensitive to the break radius.

\subsection{Hydrodynamic Equations}
\label{sec:hydro}

Following Quataert (2004, see also
\citealt{HolzerAxford:1970a,De-ColleGuillochon+:2012a,ShcherbakovWong+:2014a}),
we calculate the density $\rho$, temperature $T$, and radial velocity
$v$ of the CNM for each galaxy model by solving the equations of
one-dimensional, time-dependent hydrodynamics,

\begin{align}
  &\frac{\partial \rho}{\partial t}+\frac{1}{r^2}\frac{\partial}{\partial r}\left(\rho r^2 v\right)=q \label{eq:drhodt}\\
  &\rho \left(\frac{\partial v}{\partial t} + v\frac{\partial
      v}{\partial r}\right) =-\frac{\partial p}{\partial r}- \rho\frac{GM_{\rm enc}}{r^{2}} -q v \label{eq:dvdt}\\
  &\rho T\left(\frac{\partial s}{\partial t} + v\frac{\partial
      s}{\partial r}\right)=q\left[\ke+\kew-\gammaf \cs \right] ,
\label{eq:dsdt}
\end{align}

where $p$ and $s$ are the pressure and specific entropy, respectively,
and $M_{\rm enc} = M_{\star}(r) + \Mbh$ is the enclosed mass (we
neglect dark matter contributions).  We adopt an ideal gas equation of
state with $p = \rho kT/\mu m_p$ with $\mu = 0.62$ and $\gamma = 5/3$.  The source term in equation (\ref{eq:drhodt}),
\begin{align}
  q=\frac{\eta \rhostar}{\th},
\label{eq:q}
\end{align}
represents mass input from stellar winds, which we parametrize in
terms of the fraction $\eta$ of the stellar density $\rhostar$ being
recycled into gas on the Hubble time $\th = 1.4 \times 10^{10}$ yr.
To good approximation $\eta\simeq 0.02 (\tau_{\star}/\th)^{-1.3}$ at
time $\tau_{\star}$ following an impulsive starburst (e.g.,
\citealt{Ciotti+91})\footnote{\citet{Ciotti+91} give the mass return
  rate from evolved stars as a function of B-band luminosity instead
  of volumetrically, but our expressions are equivalent.}, although
$\eta$ is significantly higher for continuous SFHs
(bottom panel of Fig.~\ref{fig:vwSources}).

Source terms $\propto q$ also appear in the momentum and entropy
equations (eqs.~[\ref{eq:dvdt}] and [\ref{eq:dsdt}]) because the
isotropic injection of mass represents, in the SMBH rest frame, a
source of momentum and energy relative to the mean flow.  Physically,
these result from the mismatch between the properties of virialized
gas injected by stellar winds and the mean background flow.  The term
$\propto p/\rho = c_{s}^{2}$ is important because it acts to stabilize
the flow against runaway cooling ($\S\ref{sec:instability}$).

The term $\propto \vw^2 = \sigma(r)^2+v_{w}^2$ in the entropy equation
accounts for external heating sources (e.g.,
\citealt{ShcherbakovWong+:2014a}), where

\begin{equation}
\sigma \approx \sqrt{\frac{3 G \Mbh}{(\Gamma+2)
    r}+\sigma_0^2},
\label{eq:sigmarel}
\end{equation}
is the stellar velocity dispersion. This accounts for the minimal
amount of shock heating from stellar winds due to the random motion of
stars in the SMBH potential. We take $\sigma_0$ to be constant and use
$\sigma_0^{2} \approx 3 \sigma_{\bullet}^2$, where
$\sigma_{\bullet}\simeq 170 \Mbheight^{0.2}$ km/s is the velocity
dispersion from the \citet{McConnellMa+:2011a} $\Mbh-\sigma$
relation.  The second term, $v_{w}^{2}$, parametrizes additional sources of
energy input, including faster winds from young stars, millisecond
pulsars (MSPs), supernovae, AGN feedback, etc ($\S\ref{sec:heating}$).  We
assume that $v_w$ is constant with radius, i.e. that the volumetric
heating rate is proportional to the local stellar density. 

Our model does not take into account complications such as more
complicated geometries or the discrete nature of real stars
\citep{Cuadra+2006, Cuadra+2008}. In the case of Sgr A* these effects
reduce the time-averaged inflow rate to $\sim10^{-6} \Msun \pyear$ --
an order of magnitude less than a 1D spherical model
\citep{Cuadra+2006}.  Motions of individual stars can produce ~order
of magnitude spikes in the accretion rate \citep{Cuadra+2008}.

To isolate the physics of interest, our baseline calculations
neglects three potentially important effects: heat conduction,
radiative cooling, and rotation.  Heat conduction results in
an an additional heating term in equation (\ref{eq:dsdt}),
\begin{equation}
\dot{q}_{\rm cond} = \nabla\cdot(\kappa \nabla T),
\label{eq:qdotcond}
 \end{equation}
 where $\kappa =\kappa_{\rm spitz}/(1+\psi)$
 (\citealt{DaltonBalbus:1993a}) is the conductivity and $\kappa_{\rm
   spitz} = \kappa_0 T^{5/2}$ is the classical \citet{Spitzer62} value
 ($\kappa_0\simeq 2\times 10^{-6}$ in cgs units).  The flux limiter
 $\psi = \kappa_{\rm spitz} \nabla T/(5\phi \rho c_s^3)$ saturates the
 conductive flux if the mean free path for electron coulomb scattering
 exceeds the temperature length scale, where $c_s \equiv (kT/\mu
 m_p)^{1/2}$ is the isothermal sound speed and $\phi \lesssim 1$ is an
 uncertain dimensionless constant (we adopt $\phi = 0.1$).  Even a
 weak magnetic field that is oriented perpendicular to the flow could
 suppress the conductivity by reducing the electron mean free path.
 However, for radially-decreasing temperature profiles of interest,
 the flow is susceptible to the magneto-thermal instability
 (\citealt{Balbus01}), the non-linear evolution of which results in a
 radially-directed field geometry (\citealt{Parrish&Stone07}).  In
 $\S\ref{sec:conductivity}$ we show that neglecting conductivity
 results in at most order-unity errors in the key properties of the
 solutions.

 Radiative cooling contributes an additional term to equation
 (\ref{eq:dsdt}), of the form
\begin{equation}
\dot{q}_{\rm rad} = -\Lambda(T)n^{2},
\label{eq:qdot_rad}
\end{equation}
where $n \equiv \rho/\mu m_p$ and $\Lambda(T)$ is the cooling
function.  We neglect radiative cooling in our baseline calculations,
despite the fact that this is not justified when the wind heating
$\vwO$ is low or if the mass return rate $\eta$ is high.  Once
radiative cooling becomes comparable to other sources of
heating and cooling, its presence can lead to thermal instability
(e.g.~ \citealt{Gaspari+2012}, \citealt{McCourt+12},
\citealt{Li&Bryan14a}) that cannot be accounted for by our 1D
time-independent model.  Our goal is to use solutions which neglect
radiation to determine over what range of conditions cooling
instabilities will develop ($\S\ref{sec:instability}$).

Equations (\ref{eq:drhodt})-(\ref{eq:dsdt}) are solved using a sixth
order finite difference scheme with a third order Runge-Kutta scheme
for time integration and artificial viscosity terms in the velocity
and entropy equations for numerical stability
(\citealt{Brandenburg:2003a})\footnote{code is available at
  https://github.com/alekseygenerozov/hydro}.  We assume different
choices of $v_{w} = 300, 600, 1200$ km s$^{-1}$ spanning a physically
plausible range of thermally stable heating rates.  Although we are
interested in the steady-state inflow/outflow solution (assuming one
exists), we solve the time-dependent equations to avoid numerical
issues that arise near the critical sonic points.

Our solutions can be scaled to any value of the mass input parameter,
$\eta$, since the mass and energy source terms scale linearly with
$\rho$ or $\rho_{\star}$; however, the precise value of $\eta$ must be
specified when cooling or thermal conduction are included.  We check
the accuracy of our numerical solutions by confirming that mass is
conserved across the grid, in addition to the integral constraint on
the energy (Bernoulli integral).

\section{Analytic Results}
\label{sec:results}

We first describe analytic estimates of physical quantities, such as
the stagnation radius $\rs$ and the mass inflow rate, the detailed
derivation of which are given in Appendix \ref{app:rs}.  

\subsection{Flow Properties Near the Stagnation Radius}

Continuity of the entropy derivative at the stagnation radius where $v = 0$ requires that the temperature at this location be given by (eq.~[\ref{eq:first_law}])
\begin{align}
T|_{r_{s}}& = \frac{\gamma-1}{\gamma}\frac{\mu m_p
  \tilde{v}_{w}^{2}}{2k} \underset{v_w \gg \sigma_0}\approx \frac{\gamma-1}{\gamma}\frac{\mu
  m_p v_w^2}{2k} \frac{13+8\Gamma}{13+8\Gamma-6\nu} \nonumber\\ 
 &\approx
 \begin{cases}
  5.0\times 10^6\ \, v_{500}^2 \,\,{\rm K} & \text{core} \\
  5.5\times 10^6\ \, v_{500}^2 \,\,{\rm K} & \text{cusp},
 \end{cases}
\label{eq:Tanalytic}
\end{align}
where $v_{500} \equiv v_{w}/(500$ km s$^{-1}$)  and $\densSlope \equiv -d{\rm ln}\rho/d{\rm ln\,r}|_{r_{\rm s}}$ is the density power-law slope at $r = r_{\rm s}$.
Empirically, we find from our numerical solutions that
\begin{align}
\densSlope \simeq \frac{1}{6} \left(4 \Gamma+3\right)
\label{eq:densSlope}
\end{align}

Hydrostatic equilibrium likewise determines the value of the stagnation radius (Appendix \ref{app:rs}, eq.~\ref{eq:rs2main})
\begin{align}
  \rs&=\frac{G \Mbh}{\densSlope \vw^2}\left[\frac{13+ 8\Gamma}{4+2\Gamma}\right]\nonumber\\
  &=\frac{G \Mbh}{\densSlope (v_w^{2}+\sigma_0^2)}\left[4
    \frac{\Mstar|_{\rs}}{\Mbh} +\frac{13+
      8\Gamma}{4+2\Gamma}- \frac{3\nu}{2+\Gamma}\right]
\label{eq:stag_analytic}
\end{align}
For high heating rates $v_w \gg \sigma_0$, the stagnation radius resides well inside the SMBH sphere of influence.  In this case $\Mstar|_{\rs}/\Mbh \ll1$, such that equation~\eqref{eq:stag_analytic} simplifies to
\begin{align}
  &\rs \underset{v_w \gtrsim \sigma_0}\approx
  \left(\frac{13+8\Gamma}{4+2\Gamma}-
    \frac{3\nu}{2+\Gamma}\right)\frac{G \Mbh}{\nu v_w^2}\nonumber\\
  &\approx \begin{cases} 8
    \, \pc \,\, \Mbheight v_{500}^{-2}\, \pyear& \text{core} \\
    4 \, \pc \,\, \Mbheight v_{500}^{-2} \, \pyear & \text{cusp},
  \end{cases}
  \label{eq:stag_simple}
\end{align}
where we have used equation~\eqref{eq:densSlope} to estimate
$\densSlope$ separately for
core ($\Gamma = 0.1$; $\densSlope\approx 1$) and cusp ($\Gamma = 0.8$; $\densSlope \approx 0.6$) galaxies.  This
expression is similar to that obtained by \citet{Volonteri+11} on more
heuristic grounds (their eq.~6). 

In the opposite limit of weak heating ($v_w\lsim \sigma_0$) the
stagnation radius moves to large radii, approaching the break radius
$\rb$ in the stellar density profile, implying that all of the
interstellar medium (ISM) inside of $\rb$ is inflowing.  In
particular, we find that $r_s$ approaches $\rb$ for heating below a
critical threshold

\begin{equation} 
\zeta\equiv \sqrt{1 + \left(\frac{v_w}{\sigma_0}\right)^2} <
\zeta_c \approx \sqrt{\left(\frac{\rb}{\rinf}\right)^{(1-\Gamma)}+1}
\label{eq:zetacrit}
\end{equation}

Condition (\ref{eq:zetacrit}) approximately corresponds to the
requirement that the heating rate exceed the local escape speed at the
break radius, $\rb$.  This result makes intuitive sense: gas is
supplied to the nucleus by stars which are gravitationally bound to
the black hole, so outflows are possible only if the specific heating
rate $\sim v_{w}^{2}$ significantly exceeds the specific gravitational
binding energy.

\subsection{Inflow Rate}

The large scale inflow rate towards the SMBH is given by the total
mass loss rate interior to the stagnation radius (eq.~[\ref{eq:q}]),
\begin{eqnarray}
  \dot{M} &=& 4\pi \int_{0}^{r_{s}}q r^{2}dr = \frac{\eta
    M_{\star}|_{r_{\rm s}}}{\th} = \frac{\eta \Mbh (\rs/\rinf)^{2-\Gamma}}{\th} \nonumber \\
  &\approx&
  \begin{cases}
    4.5 \times 10^{-5} M_{\bullet,8}^{1.76}
    v_{500}^{-3.8}  \eta_{0.02} \Msun \, \pyear& \text{core} \\
    3.2 \times 10^{-5} M_{\bullet,8}^{1.48} 
    v_{500}^{-2.4}  \eta_{0.02} \Msun \, \pyear  & \text{cusp}, 
  \end{cases}
  \label{eq:mdot_analytic}            
\end{eqnarray}
where we have assumed $v_{w} \gg \sigma_0$ by adopting equation
(\ref{eq:stag_simple}) for $r_s$.  The resulting Eddington ratio is
given by
\begin{eqnarray}
\frac{\dot{M}}{\dot{M}_{\rm edd}} &\approx&
  \begin{cases}
    2.0 \times 10^{-5} M_{\bullet,8}^{0.76}
    v_{500}^{-3.8}  \eta_{0.02}   & \text{core}, \\
    1.4 \times 10^{-5} \Mbheight^{0.48} 
    v_{500}^{-2.4}  \eta_{0.02}   & \text{cusp}, 
  \end{cases}
  \label{eq:eddr_analytic}
\end{eqnarray}
where $\dot{M}_{\rm edd} = 2.2M_{\bullet,8}M_{\odot}$ yr$^{-1}$ is the
Eddington accretion rate, assuming a radiative efficiency of ten per
cent.  Note the sensitive dependence of the inflow rate on the wind
heating rate.  Equation~\eqref{eq:eddr_analytic} is the radial mass
inflow rate on relatively large scales and does not account for outflows
from the SMBH accretion disc (e.g.~\citealt{Blandford&Begelman99};
\citealt{Li+13}), which may significantly reduce the fraction of
$\dot{M}$ that actually reaches the SMBH. Thus we distinguish between
the large scale inflow rate, $\dot{M}$, and the accretion rate onto
the black hole, $\dot{M}_{\bullet}$.

The gas density at the stagnation radius, $\rho|_{r_{\rm s}}$, is more
challenging to estimate accurately.  By using an alternative estimate of
$\dot{M}$ as the gaseous mass within the stagnation radius divided by
the free-fall time $t_{\rm ff}|_{r_{\rm s}} = (r_{\rm
  s}^{3}/GM_{\bullet})^{1/2}$,
\begin{align}
  &\dot{M}\sim\frac{(4 \pi/3) \rs^3 \rho|_{r_{\rm s}}}{\tff|_{r_{\rm s}}},
  \label{eq:mdot_gas}
\end{align}
 in conjunction with eqs. (\ref{eq:mdot_analytic}) and (\ref{eq:stag_simple}), we find that
\begin{align}
  \rho|_{r_{\rm s}}\approx
  \begin{cases}
    5.2 \times 10^{-26} \Mbheight^{-0.2} v_{500}^{-0.8}  \eta_{0.02} \,
    \, {\rm g \, cm^{-3}}& \text{core},\\
    1.0 \times 10^{-25}  \Mbheight^{-0.5} v_{500}^{0.6}  \eta_{0.02} \,\, {\rm g \,cm^{-3}} & \text{cusp}
  \end{cases}
  \label{eq:rhors}
\end{align}

It is useful to compare our expression for $\dot{M}$ (eq.~[\ref{eq:mdot_analytic}]) to the standard Bondi rate for accretion onto a point source from an external medium of specified density and temperature (\citealt{Bondi52}):
\begin{align}
  \dot{M}_{\rm B} =4\pi \lambda r_{\rm B}^2 \rho|_{r_{\rm B}}v_{\rm
    ff}|_{r_{\rm B}},
\label{eq:bondi}
\end{align}
where $r_{\rm B} \equiv GM/c_{\rm s,ad}^{2}$ is the Bondi radius,
$c_{\rm s,ad} = \sqrt{\gamma kT/\mu m_p}$ is the adiabatic sound
speed, $v_{\rm ff}|_{r_{\rm B}} = r_{\rm B}/t_{\rm ff}|_{r_{\rm B}} =
(GM_{\bullet}/r_{\rm B})^{1/2}$ and $\lambda$ is a parameter of order
unity.

Equation (\ref{eq:mdot_gas}) closely resembles the Bondi formula
(eq.~[\ref{eq:bondi}]) provided that $r_{\rm B}$ is replaced by $\rs$.
Indeed, for $r_{s} < r_{\rm inf}$ we have that
(eq.~[\ref{eq:stag_analytic}])
\begin{align}
  \rs\approx\frac{13+8\Gamma}{4+2\Gamma}\frac{G \Mbh}{\densSlope
    \vw^2} \approx \frac{13+8\Gamma}{(2+\Gamma)(3+4\Gamma)}r_{\rm B},
  \label{eq:rbondi}
\end{align}
where the second equality makes use of equation (\ref{eq:Tanalytic}).  

\subsection{Heat Conduction}
\label{sec:conductivity}

Our analytic derivations neglect the effects of heat conduction, an
assumption we now check.  The ratio of the magnitude of the conductive
heating rate (eq.~[\ref{eq:qdotcond}]) to the external heating rate at
the stagnation radius is given by
\begin{align}
  \left.\frac{\nabla\cdot(\kappa \nabla T)}{q v_{\rm
w}^{2}/2}\right|_{r_{\rm s}} &\sim \frac{2t_{\rm h}\kappa_0
T|_{r_{\rm s}}^{7/2}}{r_{\rm s}^{2}\eta \rho_{\star}|_{r_{\rm s}}
\tilde{v}_{w}^{2}}
 \left(1+\frac{\kappa_0 T^{7/2}|_{r_{\rm s}}}{5 \rs
    \phi \rho c_s^3}\right)^{-1}
\nonumber \\ &\sim {\rm min}
  \begin{cases}
  20 \eta_{0.02}^{-1}
M_{\bullet,8}^{-0.8} v_{500}^{6.8} &  \text{unsaturated (core)}\\
 30 \eta_{0.02}^{-1}
M_{\bullet,8}^{-0.5} v_{500}^{5.4} &  \text{unsaturated (cusp)}\\
  2 \phi & \text{saturated},
  \end{cases}
 \label{eq:conduction}
\end{align}
where the second equality makes use of equations (\ref{eq:Tanalytic}),
(\ref{eq:stag_simple}), and we have made the approximations $\nabla^{2} \sim
1/r_{\rm s}^{2}$, $\nabla \sim 1/r_{\rm s}$.  The
stellar density profile is approximated as (eq.~[\ref{eq:rhostar}])
\begin{eqnarray}
  \rho_{\star}|_{r_{\rm s}} &\simeq& \frac{M_{\bullet}(2-\Gamma)}{4\pi r_{\rm inf}^{3}}\left(\frac{r_{\rm s}}{r_{\rm inf}}\right)^{-1-\Gamma} \nonumber \\
 &\approx& \begin{cases}
    7.9 \times 10^{-19}M_{\bullet,8}^{-1.2}v_{500}^{2.2}\,{\rm g\,cm^{-3}}
    & \text{core} \\
    2.3 \times 10^{-18} M_{\bullet,8}^{-1.5}v_{500}^{3.6}
    \,{\rm g\,cm^{-3}}  & \text{cusp}, 
  \end{cases}
  \label{eq:rhostarrs}
\end{eqnarray}
where the stagnation radius is assume to reside well inside the Nuker
break radius.  

Equation (\ref{eq:conduction}) shows that, even when conduction is
saturated, our neglecting of heat conduction near the stagnation
radius results in at most an order unity correction for causal values
of the saturation parameter $\phi < 0.1$.  Our numerical experiments
which include conductive heating confirm this
($\S\ref{sec:numerical}$).  We do not consider the possibility that
the conduction of heat from the inner accretion flow can affect
the flow on much larger scales \citep{Johnson+2007}.

\subsection{Thermal Instability}
\label{sec:instability}

\begin{figure}
  \includegraphics[width=\columnwidth]{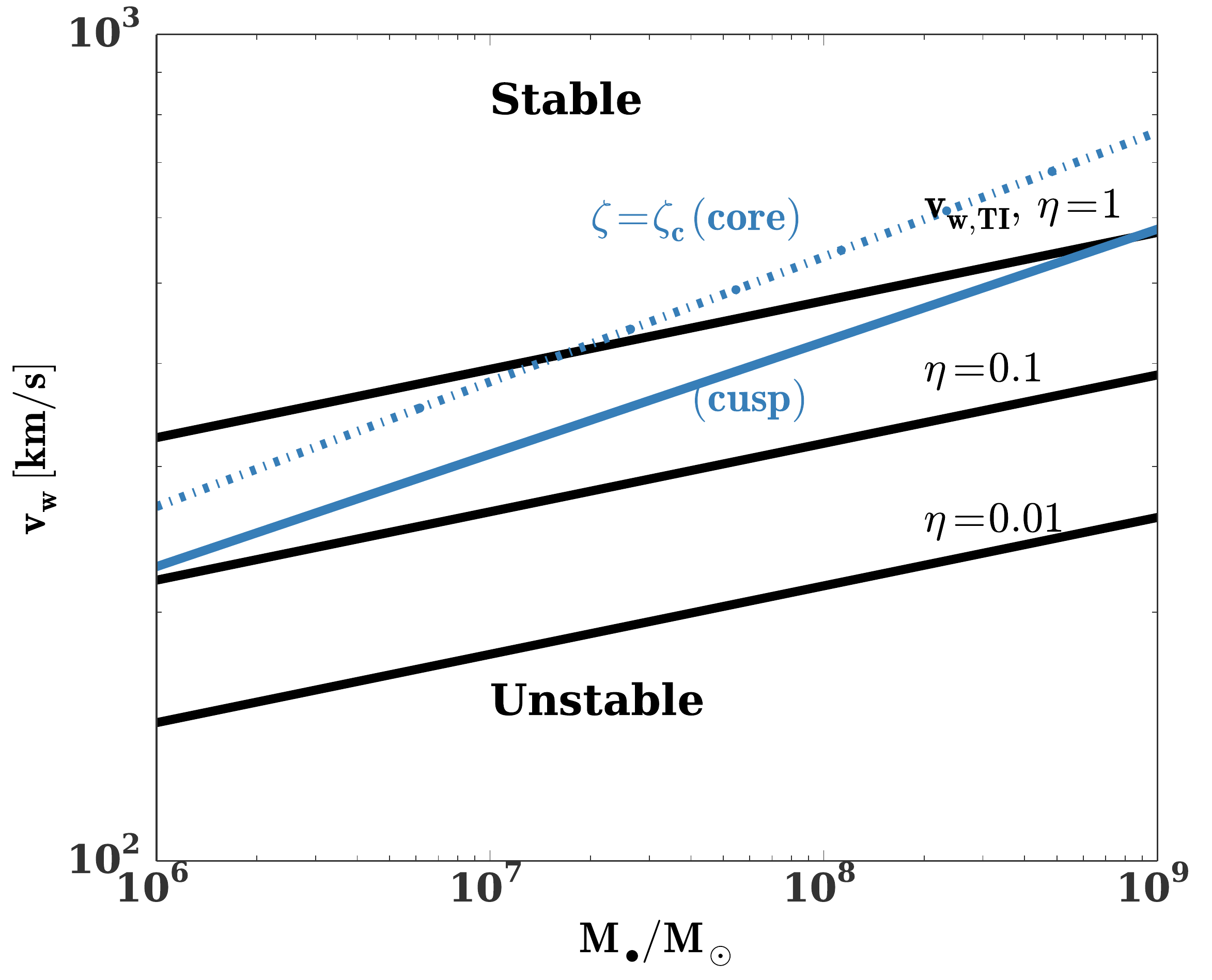}
  \caption{Minimum effective wind heating parameter required for
    thermal stability as a function of SMBH mass.  Black lines show
    $v_{\rm TI}$ (eq.~[\ref{eq:cooling3}]), the heating rate required
    for $(\dot{q}_{\rm heat}/|\dot{q}_{\rm rad}|)_{r_{s}} > 10$ in the
    high-heating limit when the stagnation radius lies interior to the
    influence radius, for different values of the mass loss parameter
    $\eta$ as marked.  Blue lines show the minimum heating parameter
    required to have $\zeta > \zeta_c = (r_{b}/r_{\rm
      inf})^{0.5(1-\Gamma)}$ (eq.~[\ref{eq:zetacrit}]), separately for
    cusp (solid) and core (dashed) galaxies. Based on the
    \citet{LauerFaber+:2007a} sample we take $\rinf=25 (8) \Mbh^{0.6}
    pc$ and $\rb=90 \Mbheight^{0.5}$ pc (240 pc) for cores
    (cusps). For $\zeta < \zeta_c$ the stagnation radius moves from
    inside the influence radius, out to the stellar break radius
    $\rb$. This renders the flow susceptible to thermal runaway, even
    if $v_{w} > v_{\rm TI}$.  }
\label{fig:TI}
\end{figure}

Radiative cooling usually has its greatest impact near or external to
the stagnation radius, where the gas resides in near hydrostatic
balance.  If radiative cooling becomes important, it can qualitatively
alter key features of the accretion flow. Initially hydrostatic gas is
thermally unstable if the cooling time is much less than the free-fall
time , potentially resulting in the formation of a multi-phase medium
(\citealt{Gaspari+2012,Gaspari+2013,Gaspari+2015,Li&Bryan14b}).

Even if the hot plasma of the CNM does not condense into cold clouds,
the loss of pressure can temporarily increase the inflow and accretion
rates by producing a large-scale cooling flow.  When coupled to
feed-back processes which result from such enhanced accretion, this
can lead to time-dependent limit cycle behavior
(e.g.~\citealt{Ciotti&Ostriker07}; \citealt{Ciotti+10};
\citealt{Yuan&Li11}, \citealt{Gan+14}), which is also inconsistent
with our assumption of a steady, single-phase flow.

Cooling instability can, however, be prevented if destabilizing
radiative cooling ($\dot{q}_{\rm rad} \propto T^{-2.7}$) is
overwhelmed by other sources of cooling, namely the {\it stabilizing}
term $\propto -q c_{s}^{2} \propto -T$ in the entropy equation
(eq.~[\ref{eq:dsdt}]).  Neglecting radiative cooling to first order,
this term is balanced at the stagnation radius by the external heating
term, $\dot{q}_{\rm heat} = q \tilde{v}_{w}^{2}/2$.  One can therefore
assess thermal stability by comparing the ratio of external heating to
radiative cooling $|\dot{q}_{\rm rad}|$ (eq.~[\ref{eq:qdot_rad}]),
\begin{align}
\left.\frac{\dot{q}_{\rm heat}}{|\dot{q}_{\rm rad}|}\right|_{\rm r_{\rm s}} \simeq
  \begin{cases}
   630 \eta_{0.02}^{-1} M_{\bullet,8}^{-0.76}v_{500}^{7.2}  &, \text{core}\\
   660 \eta_{0.02}^{-1} M_{\bullet,8}^{-0.48}v_{500}^{5.8}  &, \text{cusp},     
  \end{cases}
  \label{eq:cooling2}
\end{align}
where we have used equations (\ref{eq:rhors}) and (\ref{eq:rhostarrs})
for the gas and stellar densities at the stagnation radius,
respectively.  We have approximated the cooling function for $T <
2\times 10^{7}$ K as $\Lambda(T) = 1.1 \times 10^{-22} \left(T/10^6
  \text{K}\right)^{-0.7} $erg cm$^3 $s$^{-1}$, which assumes solar
metallicity gas (\citealt{Draine:2011a}; his Fig. 34.1).  

To within a constant of order unity, equation (\ref{eq:cooling2}) also
equals the ratio of the gas cooling time-scale $t_{\rm cool} \equiv (3n
kT/2 \mu)/|\dot{q}_{\rm rad}|$ to the free-fall time $t_{\rm ff}$ at the
stagnation radius.  This equivalence can be derived using the equality
\begin{align}
\rho|_{r_{\rm s}}=\frac{3 q\tff|_{r_{\rm s}}}{2-\Gamma}.
\label{eq:rhors2}
\end{align}
that results by combining equations
\eqref{eq:mdot_analytic},\eqref{eq:mdot_gas}, and
\eqref{eq:rhostarrs}.  \citet{Sharma+2012} argue cooling instability
develops in a initially hydrostatic atmosphere if $t_{\rm cool}
\lesssim 10 t_{\rm ff}$, so equation (\ref{eq:cooling2}) represents a
good proxy for instability in this case as well.

Based on our numerical results (\S \ref{sec:numerical}) and the work
of \citet{Sharma+2012} we define thermally stable flows according to the
criterion $\dot{q}_{\rm heat} > 10|\dot{q}_{\rm rad}|$ ($t_{\rm cool}
> 10t_{\rm ff}$) being satisfied near the stagnation radius.  This
condition translates into a critical minimum heating rate
\begin{align}
v_{w} > v_{\rm TI} \simeq
  \begin{cases}
   280 \eta_{0.02}^{0.14} M_{\bullet,8}^{0.11}\,{\rm km\,s^{-1}}  &, \text{core}\\
   240 \eta_{0.02}^{0.17} M_{\bullet,8}^{0.08}\,{\rm km\,s^{-1}}   &, \text{cusp}  \\
  \end{cases}
  \label{eq:cooling3}
\end{align}
Equations (\ref{eq:cooling2}) and (\ref{eq:cooling3}) are derived
using expressions for the stagnation radius and gas density in the
high heating limit of $\zeta > \zeta_c = (r_{b}/r_{\rm
  inf})^{0.5(1-\Gamma)}$ (eq.~[\ref{eq:zetacrit}]).  However, for
$\zeta < \zeta_c$, the stagnation radius diverges to the break radius
$\rb$.  The quasi-hydrostatic structure that
results in this case greatly increases the gas density, which in
practice renders the flow susceptible to thermal runaway, even if
$v_{w} > v_{\rm TI}$ according to equation (\ref{eq:cooling3}).  In
other words, the true condition for thermal instability  can be
written

\begin{equation}
v_w^2 > {\rm max}[v_{\rm TI}^2,\sigma_0 (r_{b}/r_{\rm
  inf})^{(1-\Gamma)}]
\label{eq:cooling4}
\end{equation}

Figure \ref{fig:TI} shows $v_{\rm TI}(M_{\bullet})$ for different mass
input parameters $\eta$, as well as the [$\eta$-independent] $\zeta >
\zeta_c$ criterion, shown separately for cusp and core galaxies.
Based on the \citet{LauerFaber+:2007a} sample we take $\rinf = 25(8)M
^{0.6}$ pc and $\rb = 90 \Mbheight^{0.5}$ (240 pc) and thus
$\rb/\rinf\simeq 4 \Mbheight^{-0.1} (30 \Mbheight^{-0.6})$ for cores
(cusps). We see that the for high $\eta$ and low $\Mbh$ the $v_w
\gtrsim v_{\rm TI}$ criterion is more stringent, while for low $\eta$
and high $\Mbh$, the $\zeta>\zeta_c$ criterion is more stringent.

The minimum heating rate for thermal stability corresponds to the {\it
  maximum} thermally-stable inflow rate. From
equations~\eqref{eq:cooling3} and~\eqref{eq:cooling4} this is

\begin{align}
  \frac{\dot{M}_{\rm TI}}{\dot{M}_{\rm edd}} \simeq \begin{cases} {\rm
      min} \begin{cases}
      2\times 10^{-4} \eta_{0.02}^{0.47}M_{\bullet,8}^{0.36} \\
      2\times 10^{-5} \eta_{0.02} \Mbheight^{0.19}
    \end{cases}  & \text{core},  \\
    {\rm min}
    \begin{cases}
     8\times 10^{-5} \eta_{0.02}^{0.59} M_{\bullet,8}^{0.28}   \\
      2\times10^{-5} \eta_{0.02}  \Mbheight^{0.15}
    \end{cases}  & \text{cusps}. \\
  \end{cases}
  \label{eq:Mdotmax}
\end{align}

Note that since the SMBH accretion rate cannot exceed the large scale
inflow rate, $\dot{M}_{\rm TI}$ also represents the maximum thermally
stable accretion rate.

What we describe above as ``thermal instability" may in practice
simply indicate an abrupt transition from a steady inflow-outflow
solution to a global cooling flow, as opposed a true thermal
instability.  In the former case the stagnation radius diverges to
large radii, increasing the density in the inner parts of the flow,
which increases cooling and creates a large inflow of cold gas towards
the nucleus.  A true thermal instability would likely result in a
portion of the hot ISM condensing into cold clouds, a situation which
may or may not be present in a cooling flow.  In this paper we do not
distinguish between these possibilities, although both are likely
present at some level.  Finally, note that if the CNM were to
``regulate" itself to a state of local marginal thermal instability
(as has previously been invoked on cluster scales; e.g.,
\citealt{Voit+15}), then equation (\ref{eq:Mdotmax}) might naturally
reflect the characteristic mass fall-out rate and concomitant star
formation rate.

\begin{figure}
  \includegraphics[width=\columnwidth]{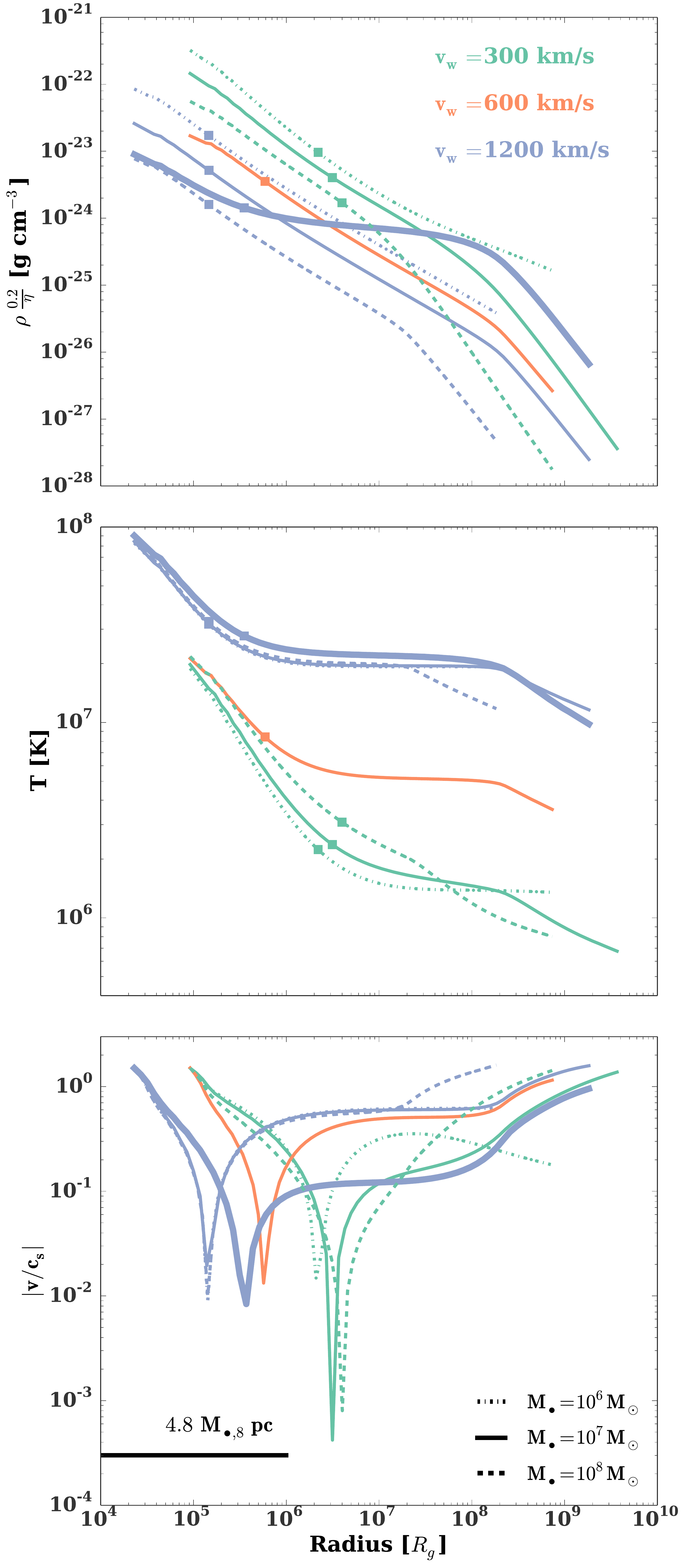}
  \caption{\label{fig:profiles}Radial profiles of the CNM density
    ({\it top}), temperature ({\it middle}), and velocity ({\it
      bottom}), calculated for a representative sample of galaxies.
    Colors denote values of the effective wind heating rate,
    $\vwO=1200$ km s$^{-1}$ ({\it blue}), 600 km s$^{-1}$ ({\it
      orange}), and 300 km s$^{-1}$ ({\it green}).  Line styles denote
    different black hole masses: $\Mbh=10^6 \Msun$ ({\it dot-dashed}),
    $10^7 \Msun$ ({\it solid}), and $10^8 \Msun$ ({\it dashed}). Thin
    and thick lines denote cusp galaxies ($\Gamma$=0.8) and core
    galaxies ($\Gamma$=0.1), respectively.  Squares mark the locations
    of the stagnation radius.  }
\end{figure}


  \subsection{Angular Momentum}
  \label{sec:rotation}

  Our spherically symmetric model neglects the effects of angular
  momentum on the gas evolution.  However, all galaxies possess some
  net rotation, resulting in centrifugal forces becoming important at
  some radius $r_{\rm circ} = l^{2}_{s}/(GM_{\bullet})$.  Here
  $l_{s} = \langle r V_{\phi}\rangle |_{r_s}$ is the stellar
  specific angular momentum near the stagnation radius, from which
  most of the accreted mass originates, where $V_{\phi}$ is the
  stellar azimuthal velocity.

  \citet{EmsellemCappellari+:2007a} use two-dimensional kinematic data
  to measure the ratio of ordered to random motion in a sample of
  early type galaxies, which they quantify at each galactic radius $R$
  by the parameter
  \begin{align}
    \lambda_R \equiv \frac{\langle r|V|\rangle}{\langle R\sqrt{V^2+\sigma^2}\rangle} \underset{R \ll r_{\rm inf}}\sim \frac{V_{\phi}}{\sigma},
  \end{align}
  where $\sigma$ is the velocity dispersion and the brackets indicate
  a luminosity-weighted average.  The circularization radius of the
  accretion flow can be written in terms of $\lambda_R$ as

\begin{equation}
\frac{r_{\rm circ}}{r_{s}} \approx \frac{r_{s}}{r_{\rm inf}}\lambda_{R}^{2} \lesssim \lambda_{R}^{2},
\label{eq:rcirc}
\end{equation}
where we have used the definition $r_{\rm inf} \equiv
GM_{\bullet}/\sigma_{\bullet}^{2}$ and in the second inequality have assumed
that $r_{\rm s} \lesssim r_{\rm inf}$, a condition which is satisfied
for the thermally-stable solutions of interest.

\citet{EmsellemCappellari+:2007a} (their Fig.~2) find that $\lambda_R$
is generally $< 0.1$ on radial scales $<$ 10 per cent of the galaxy
half-light radius and that $\lambda_R$ decreases with decreasing $R$
interior to this point.  From equation (\ref{eq:rcirc}) we thus
conclude that $r_{\rm circ} \lesssim 0.01r_{s}$. For the low inflow
rates considered the gas ($\eddr<0.01$) would be unable to cool on a
dynamical time and would likely drive equatorial and polar outflows
\citep{Li+13}. Our model cannot capture such two dimensional
structures, but would still be relevant for intermediate polar angles
where the gas is inflowing.
\section{Numerical Results}
\label{sec:numerical}

\begin{table*}
\begin{threeparttable}
\begin{minipage}{18cm}
  \caption{Summary of Numerical Solutions}
\begin{tabular}{lccccccccc}
  \hline
  {$M_{\bullet}$} & {$v_{w}$} & {$\rb^{(a)}$} &   $r_{\rm s}/r_{\rm
    inf}^{(b)}$ &  {$\left(\frac{\dot{M}}{\dot{M}_{\rm edd}}\right)_{r_{s}}^{(c)}
      (\eta=0.2)$} & {$\left(\frac{\dot{q}_{\rm heat}}{|\dot{q}_{\rm
            rad}|}\right)_{r_s}^{(d)} (\eta=0.2)$} & Unstable $\eta$'s  \\
    ($M_{\odot}$) & (km s$^{-1}$) & (pc) &- & - & - &  - & \\ 
    \hline
    Cusp Galaxies, $\Gamma = 0.8$ & & & & & & & & \\
    $    10^{ 6 }$ & 300 & -- & 0.12 & $ 5.1 \times 10^{ -5 }$ & 28 & 0.6\\
    ... & 600 & -- & $ 3.2 \times 10^{ -2 }$ & $ 1.0 \times 10^{ -5 }$ & $ 1.9 \times 10^{ 3 }$ \\
    ... & 1200 & -- & $ 7.9 \times 10^{ -3 }$ & $ 1.8 \times 10^{ -6 }$ & $ 7.7 \times 10^{ 4 }$ \\
   10$^{7}$ & 300 & 50 & 0.38 & $ 2.0 \times 10^{ -4 }$ & 4.5 & 0.2, 0.6\\
... & ... & 100 & $\mathbf{TI}$ & $\mathbf{TI}$ & $\mathbf{TI}$ \\
... & ... & 200 & $\mathbf{TI}$ & $\mathbf{TI}$ & $\mathbf{TI}$ \\
... & ... & 400 & $\mathbf{TI}$ & $\mathbf{TI}$ & $\mathbf{TI}$ \\
... & 600 & 25 & $ 8.1 \times 10^{ -2 }$ & $ 3.2 \times 10^{ -5 }$ & $ 6.2 \times 10^{ 2 }$ \\
... & ... & 100 & $ 8.1 \times 10^{ -2 }$ & $ 3.2 \times 10^{ -5 }$ & $ 6.2 \times 10^{ 2 }$ \\
... & 1200 & 100 & $ 2.0 \times 10^{ -2 }$ & $ 6.0 \times 10^{ -6 }$ & $ 2.6 \times 10^{ 4 }$ \\
$    10^{ 8 }$ & 300 & 25 & 0.69 & $ 4.2 \times 10^{ -4 }$ & 5.5 & 0.6\\
... & ... & 50 & 0.92 & $ 6.0 \times 10^{ -4 }$ & 2 & 0.6\\
 ... & ... & 100 & 1.4 & $ 9.5 \times 10^{ -4 }$ & 0.48 & 0.2, 0.6 \\
 ... & ... & 200 & 2.5 & $ 1.9 \times 10^{ -3 }$ & $5.3 \times  10^{
   -2 }$ & 0.2, 0.6  \\
... & 450 & 100 & 0.43 & $ 2.4 \times 10^{ -4 }$ & 19 \\
...$^{\dagger}$ & 450 & 100 & 1.2 & $ 8.1 \times 10^{ -4 }$ & 2.6 \\
... & 600 & 25 & 0.21 & $ 1.0 \times 10^{ -4 }$ & $ 2.1 \times 10^{ 2 }$ \\
... & ... & 100 & 0.22 & $ 1.1 \times 10^{ -4 }$ & $ 1.7 \times 10^{ 2
}$ \\
...$^{\dagger}$ & ... & 100 & 0.22 & $ 1.1 \times 10^{ -4 }$ & $ 1.6 \times 10^{ 2 }$ \\
...$^{\ddagger}$  & ... & 100 & 0.07 & $ 2.8 \times 10^{ -5 }$ & $ 4.7 \times 10^{ 2
}$ \\
... & ... & 200 & 0.23 & $ 1.1 \times 10^{ -4 }$ & $ 1.6 \times 10^{ 2 }$ \\
... & ... & 400 & 0.23 & $ 1.1 \times 10^{ -4 }$ & $ 1.5 \times 10^{ 2 }$ \\
... & 1200 & 100 & $ 5.0 \times 10^{ -2 }$ & $ 1.8 \times 10^{ -5 }$ & $ 8.3 \times 10^{ 3 }$ \\
\hline
Core Galaxies, $\Gamma = 0.1$  &  & & & & & & & & \\
$    10^{ 6 }$ & 600 & 25 & $\mathbf{TI}$ & $\mathbf{TI}$ & $\mathbf{TI}$  & \\
... & 1200 & -- & $ 1.5 \times 10^{ -2 }$ & $ 2.3 \times 10^{ -7 }$ & $ 1.7 \times 10^{ 5 }$ \\
$    10^{ 7 }$ & 600 & 25 & 0.19 & $ 2.9 \times 10^{ -5 }$ & 79 & 0.6 \\
... & ... & 50 & $\mathbf{TI}$ & $\mathbf{TI}$ & $\mathbf{TI}$ \\
... & 1200 & 25 & $ 3.9 \times 10^{ -2 }$ & $ 1.4 \times 10^{ -6 }$ & $ 2.9 \times 10^{ 4 }$ \\
... & ... & 50 & $ 4.1 \times 10^{ -2 }$ & $ 1.5 \times 10^{ -6 }$ & $ 2.3 \times 10^{ 4 }$ \\
$    10^{ 8 }$ & 600 & 25 & 0.27 & $ 5.7 \times 10^{ -5 }$ &
$ 1.0 \times 10^{ 2 }$ \\
... & ... & 50 & 0.46 & $ 1.5 \times 10^{ -4 }$ & 9.4 & 0.2, 0.6\\
... & 1200 & 25 & $ 8.5 \times 10^{ -2 }$ & $ 6.1 \times 10^{ -6 }$ & $ 8.2 \times 10^{ 3 }$ \\
... & ... & 50 & $ 9.2 \times 10^{ -2 }$ & $ 7.0 \times 10^{ -6 }$ & $ 6.0 \times 10^{ 3 }$ \\
... & ... & 100 & 0.1 & $ 8.8 \times 10^{ -6 }$ & $ 3.8 \times 10^{ 3 }$ \\
... & ... & 200 & 0.15 & $ 1.7 \times 10^{ -5 }$ & $ 8.0 \times 10^{ 2 }$ \\
\hline
\label{table:models}  
\end{tabular}
\begin{tablenotes}
\item $^{(a)}$ Break radius of stellar density profile. $^{(b)}
  \rinf=14 \Mbh[8]^{0.6} $ as fixed in our numerical runs.
  $^{(c)}$Inflow rate in Eddington units, normalized to a stellar mass
  input parameter $\eta = 0.2$.  $^{(d)}$Ratio of wind heating rate to
  radiative cooling rate at the stagnation radius. $\dagger$Calculated
  with radiative cooling included, assuming mass loss parameter $\eta
  = 0.2$.  $\ddagger$Calculated with radiative cooling and conductivity
  included, assuming mass loss parameter $\eta = 0.2$ and conductivity
  saturation parameter $\phi = 0.1$.  Solutions including radiative
  cooling were performed for cusp galaxies with $v_w=300 $ km s$^{-1}$
  and core galaxies with $v_w=600 $ km s$^{-1}$ for $\eta$ = 0.02,
  0.2, and 0.6.  Values of $\eta$ resulting in thermally unstable
  solutions are marked in the final column.  Solutions found to be
  thermally unstable for all $\eta \geq 0.02$ are denoted as {\bf TI}.
\end{tablenotes}
\end{minipage}
\end{threeparttable}

\end{table*}

Our numerical results, summarized in Table \ref{table:models}, allow
us to study a range of CNM properties and to assess the validity of
the analytic estimates from the previous section.  

Figure~\ref{fig:profiles} shows profiles of the density $\rho(r)$,
temperature $T(r)$, and radial velocity $|v(r)|/c_s$, for the cusp
($\Gamma=0.8$) solutions within our grid.  As expected, the gas
density increases towards the SMBH $\rho\propto r^{-\densSlope}$ with
$\densSlope\simeq1$, i.e. shallower than the $-3/2$ power law for
Bondi accretion. This power law behavior does not extend through all
radii, however, as the gas density profile has a break coincident with
the location of the break in the stellar light profile ($\rb=100$
pc). The temperature profile is relatively flat at large radii, but
increases as $\propto 1/r^{k}$ interior to the sphere of influence,
where $k\lesssim 1$, somewhat shallower than expected for virialized
gas within the black hole sphere of influence.  The inwardly directed
velocity increases towards the hole with a profile that is somewhat
steeper than the local free-fall velocity $v \propto v_{\rm ff}\propto
r^{-1/2}$.  The flow near the stagnation radius is subsonic, but
becomes supersonic at two critical points. The inner one at $r \simeq
0.1 r_{\rm s}$ is artificially imposed for numerical stability,
although we have verified that moving the inner boundary has a small
effect on the solution properties near the stagnation radius. The
outer one is located near the break radius, $r \simeq \rb$ and is
caused by the transition to a steeper stellar density profile
exterior to the outer Nuker break radius.

\begin{figure}
  \includegraphics[width=\columnwidth]{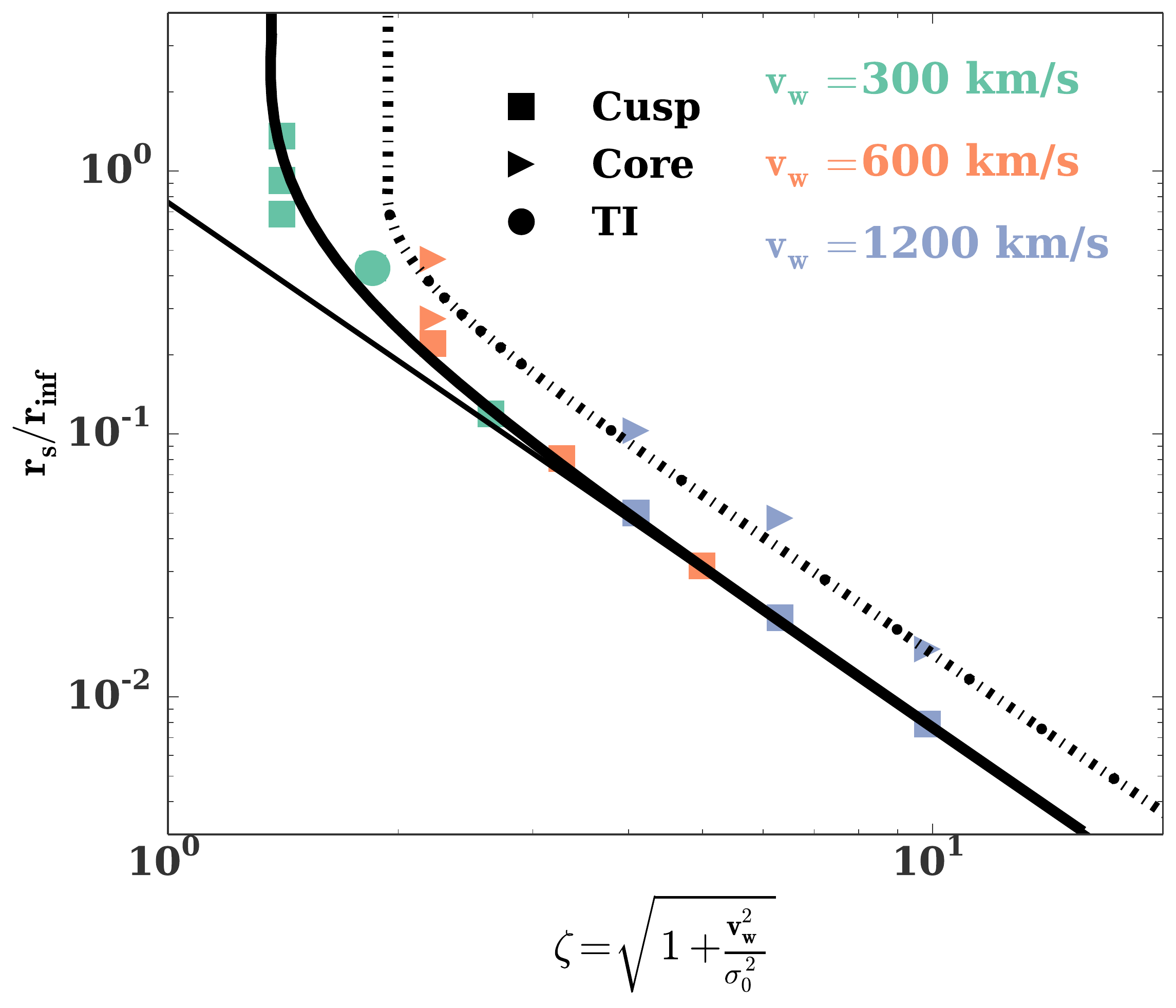}
  \caption{\label{fig:stag} Stagnation radius $r_{s}$ in units of the
    sphere of influence radius $r_{\rm inf}$ (eq.~[\ref{eq:rsoi}]) for
    galaxies in our sample as a function of the stellar wind heating
    parameter $\zeta \equiv \sqrt{1+(v_w/\sigma_0)^{2}}$.  Green,
    orange, and blue symbols correspond to different values of $v_{w}
    =$ 300, 600, and 1200 km s$^{-1}$, respectively.  Squares
    correspond to cusp galaxies ($\Gamma = 0.8$), while triangles
    correspond to cores ($\Gamma = 0.1$). Green circles correspond to
    cusp solutions which would be thermally unstable.  The black
    curves correspond to the analytic prediction from equation
    (\ref{eq:rsRinf}), with thick solid and dot-dashed curves
    calculated for parameters $(\Gamma=0.8, \densSlope\simeq 1)$ and
    $(\Gamma=0.1,\densSlope\simeq0.6)$, respectively. The thin black
    solid line corresponds to the simplified analytic result for $\rs$
    from equation~\eqref{eq:stag_simple} (recall that $\rinf \simeq G
    \Mbh/ \sigma_\bullet^2$).}
\end{figure}

Figure~\ref{fig:stag} shows our calculation of the stagnation radius
$r_{\rm s}/r_{\rm inf}$ as a function of the wind heating parameter
$\zeta = \sqrt{1+(v_w/\sigma_0)^{2}}$, with different colors showing
different values of $v_{w}$.  Cusp and core galaxies are marked with
square and triangles, respectively.  Shown for comparison are our
analytic results (eq.~[\ref{eq:rsRinf}])
with solid and dashed lines for cusp and core galaxies, respectively,
calculated assuming $\densSlope = 1$ and $\densSlope= 0.6$,
respectively.

Our analytic estimates accurately reproduce the numerical results in
the high heating limit $\zeta \gg 1$ ($v_w \gg \sigma_0$; $r_{\rm s}
\lesssim r_{\rm inf}$).  However, for low heating the stagnation
radius diverges above the analytic estimate, approaching the stellar
break radius $\rb \gg r_{\rm inf}$.  This divergence occurs
approximately when $\zeta < \zeta_c \propto (\rb/r_{\rm
  soi})^{0.5(\Gamma-1)}$ (eq.~[\ref{eq:zetacrit}]). Physically this
occurs because the heating rate is insufficient to unbind the gas from
the stellar potential.  Thus, for small values of the heating rate
(small $\zeta$) the location of the stagnation radius will vary
strongly with the break radius. This explains the behavior of the
three vertically aligned green squares. These are three cusp
($\Gamma=0.8$) galaxies with $\Mbh=10^8 \Msun$ and $v_w=300$ km
s$^{-1}$ but with different break radii ($\rb=$ 25, 50, and 100 pc
from top to bottom). This divergence of the stagnation radius to large
radii occurs at a higher value of $\zeta$ in core galaxies ($\Gamma =
0.1$), explaining the behavior of the two core galaxies shown in
Fig.~\ref{fig:stag} as vertically aligned orange triangles.

\begin{figure}
  \includegraphics[width=\columnwidth]{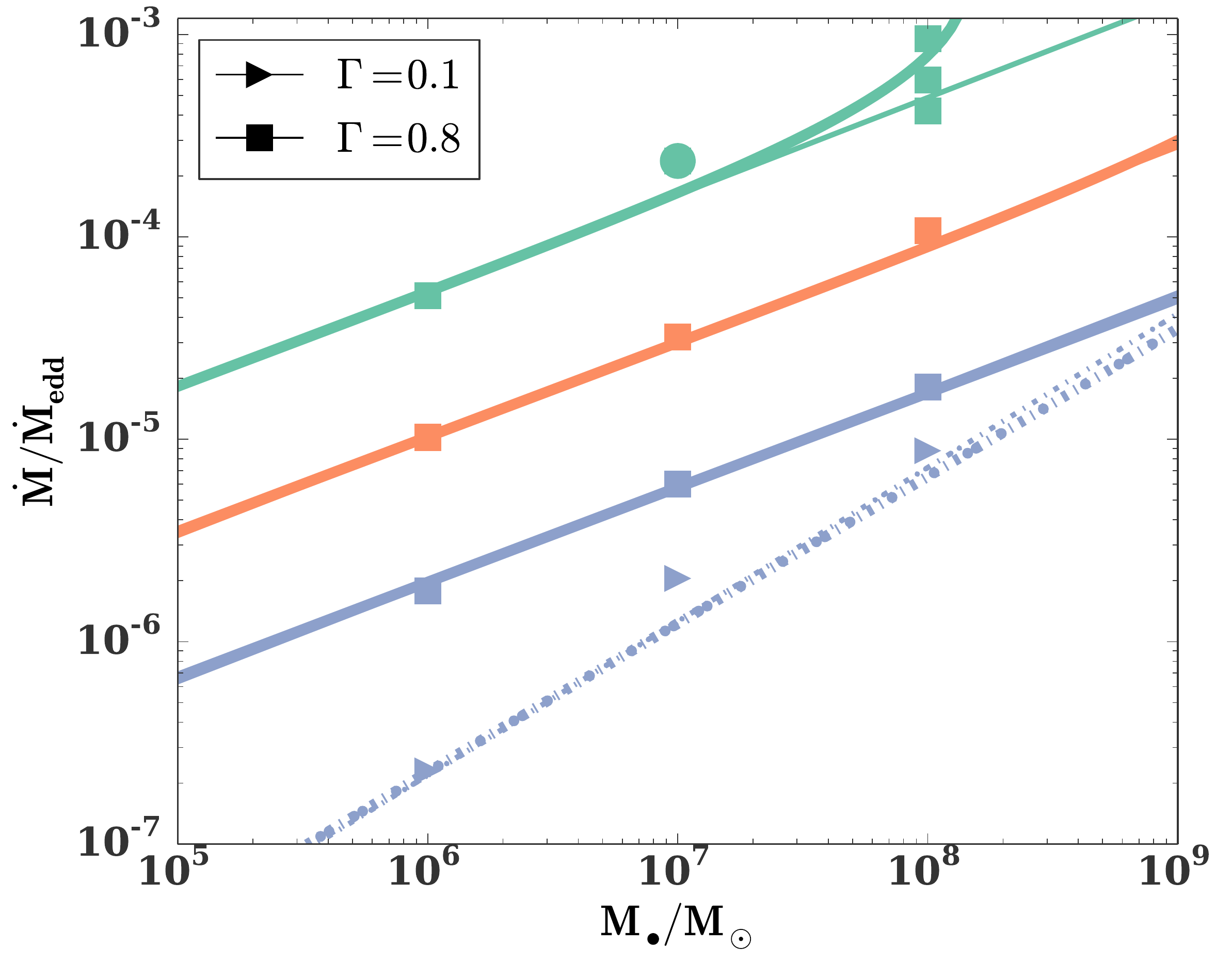}
  \caption{\label{fig:mdot_mass} Inflow rate $\dot{M}/\dot{M}_{\rm
      edd}$ versus SMBH mass for galaxies in our sample, calculated
    for different values of the wind heating parameter $\vwO =$ 300 km
    s$^{-1}$ ({\it green}), 600 km s$^{-1}$ ({\it orange}), and 1200
    km s$^{-1}$ ({\it blue}).  Squares correspond to cusp galaxies
    ($\Gamma=0.8$), while triangles correspond to cores
    ($\Gamma$=0.1). The green circle corresponds to a cusp solution
    which would be thermally unstable.  Thin solid and dot-dashed
    curves correspond to our simple analytic estimates of
    $\dot{M}/\dot{M}_{\rm edd}$ (eq.~[\ref{eq:eddr_analytic}]) for
    cusp and core galaxies, respectively.  Thick curves correspond to
    the more accurate implicit analytic expression given by equation
    \eqref{eq:rs2main}.}
\end{figure}

Figure~\ref{fig:mdot_mass} shows the inflow rate for a sample of
our numerical solutions for different values of
$\vwO =$ 300, 600 and 1200 km s$^{-1}$, and for both core
($\Gamma$=0.1) and cusp galaxies ($\Gamma$=0.8).  Shown for comparison
is our simple analytic estimate of $\dot{M}/\dot{M}_{\rm edd}$ from equation
(\ref{eq:eddr_analytic}).  For high wind velocities ($v_{w} \gg
\sigma_0$) the stagnation radius lies well inside the black hole sphere of
influence and our analytic estimate provides a good fit to the
numerical results.  However, for low wind velocities and/or high
$M_{\bullet}$ (large $\sigma_0$), the numerical accretion rate
considerably exceeds the simple analytic estimate as the stagnation radius
diverges to large radii (Fig.~\ref{fig:stag}).

\begin{figure}
  \includegraphics[width=\columnwidth]{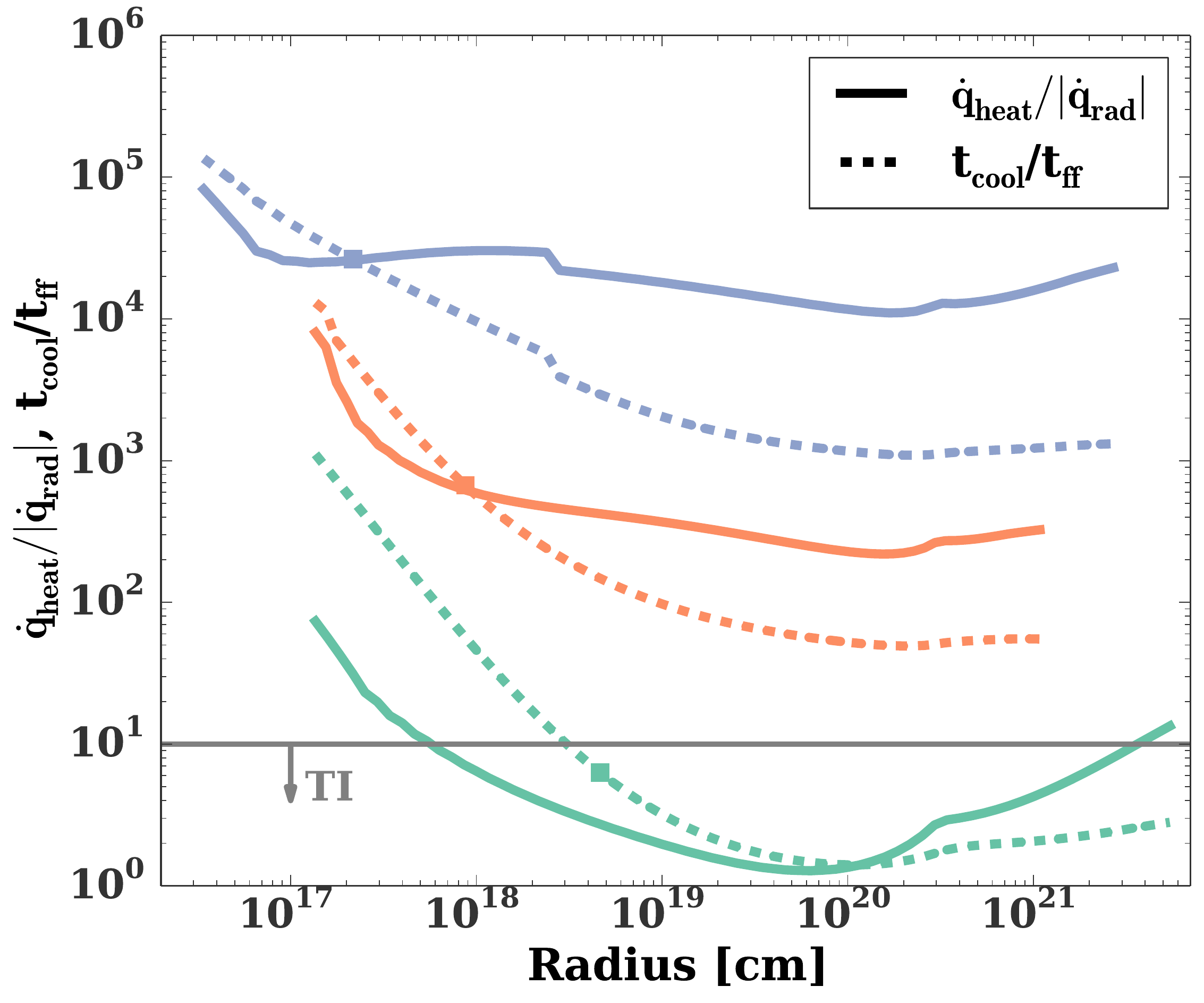}
  \caption{\label{fig:cooling} Ratio of the rates of heating to
    radiative cooling, $\dot{q}_{\rm heat}/|\dot{q}_{\rm rad}|$, as a
    function of radius (solid lines) for a $\Mbh=10^7 \Msun$ cusp
    galaxy ($\Gamma=0.8$).  Dashed lines show the ratio of the cooling
    time-scale to the free-fall time-scale $t_{\rm cool}/t_{\rm ff}$.
    For high heating rates both ratios are approximately equal at the
    stagnation radius (squares).  When $\dot{q}_{\rm
      heat}/\dot{q}_{\rm rad} \lesssim 10$ (or, equivalently, $t_{\rm
      cool}/t_{\rm ff} \lesssim 10$ near the stagnation radius), then
    the flow is susceptible to thermal instabilities.}
\end{figure}

Fig.~\ref{fig:cooling} shows the ratio of wind heating to radiative
cooling, $\dot{q}_{\rm heat}/|\dot{q}_{\rm rad}|$
(eq.~[\ref{eq:cooling2}] and surrounding discussion) as a function of
radius for $v_w=300, 600,$ and 1200 km/s, $\Mbh=10^{7} \Msun$, and
$\Gamma=0.8$.  Radiative cooling is calculated using the cooling
function of \citet{Draine:2011a} for solar metallicity.  For high
heating rates of $v_{w} = 600$ and $1200$ km s$^{-1}$ cooling is
unimportant across all radii, while for $v_{w} = 300$ km s$^{-1}$ we
see that $\dot{q}_{\rm heat}/|\dot{q}_{\rm rad}|$ and $\tcool/\tff$
can be less than unity, depending on the wind mass loss parameter
$\eta$.  Because at the stagnation radius $\dot{q}_{\rm
  heat}/|\dot{q}_{\rm rad}|$ is within a factor of two of its
minimum across the entire grid, the value of $(\dot{q}_{\rm
  heat}/|\dot{q}_{\rm rad}|)_{r_{s}}$ is a global diagnostic of
thermal instability for cusp galaxies. For core
galaxies, the minimum value for the ratio of wind heating to radiative
cooling may be orders of magnitude less than the value at the
stagnation radius. However, we find that the $\zeta/\zeta_c$
criterion (see eq.~\eqref{eq:zetacrit}) combined with the ratio of
wind heating to radiative cooling at $\rs$ still gives a reasonable
diagnostic of thermal stability for core galaxies.

Also note that for $v_{w}=600$ and $1200$ km s$^{-1}$ we have
$\dot{q}_{\rm heat}/|\dot{q}_{\rm rad}| \sim t_{\rm cool}/t_{\rm ff}$
near the stagnation radius, but these ratios diverge from each other
at low heating ($v_{w} \lesssim 300$ km s$^{-1}$).  This results
because equations (\ref{eq:mdot_gas}), (\ref{eq:rhors}) underestimate
the true gas density in the case of subsonic flow (weak heating).

Although most of our solutions neglect thermal conduction and
radiative cooling, these effects are explored explicitly in a subset of
simulations.  Dagger symbols in Table \ref{table:models} correspond to
solutions for which we turned on radiative cooling.  For cases which
are far from being thermal instability when cooling is neglected (e.g.,
$M_{\bullet} = 10^{8}M_{\odot}$, $v_{w} = 600$ km s$^{-1}$, $r_{b} =
100$ pc, $\eta = 0.2$, $\dot{q}_{\rm heat}/|\dot{q}_{\rm rad}| = 76$),
including radiative cooling has little effect on the key properties of
the solution, such as the stagnation radius, mass inflow rate, and
the cooling ratio $\dot{q}_{\rm heat}/|\dot{q}_{\rm rad}|$.  However,
for solutions that are marginally thermally stable, including
radiative losses acts to significantly decrease $\dot{q}_{\rm
  heat}/|\dot{q}_{\rm rad}|$.  We calculated solutions with radiative
cooling for all cusp galaxies with $v_w=300$ km s$^{-1}$ and for all
core galaxies with $v_w=600 $ km s$^{-1}$, each for three different
values of $\eta = 0.02, 0.2, 0.6$, spanning the physical range of
expected mass loss for continuous star formation
(Fig.~\ref{fig:vwSources}).  Solutions found to be unstable for all
values of $\eta$ are marked in boldface as ``TI" in Table
\ref{table:models}, with the unstable values of $\eta$ provided in the
final column.

Including thermal conduction, by contrast, results in only order unity
changes to our solutions for our fiducial value of the saturation
parameter $\phi = 0.1$, consistent with our analytic expectations
(eq.~[\ref{eq:conduction}]). However, we found that for runs which
included conductivity, the stagnation radius varied with the location
of the inner grid boundary: as the boundary is moved inwards, more heat
can be conducted outwards from deeper in the gravitational potential
well of the black hole. We expect conductivity would begin to have a
significant effect for an inner boundary $\lsim 0.01 \rs$ (the
conductive run in Table~\ref{table:models} has an inner grid boundary
at $\sim 0.03 \rs$). However, it is not clear physically if the
magnetic field could remain coherent over so many decades in radius in
the face of turbulence from stellar winds.

\section{Heating Sources}
\label{sec:heating}

Key properties of the flow, such as the mass inflow rate and the
likelihood of thermal instability, depend sensitively on the assumed
heating rate $\propto qv_{w}^{2}$.  In this section we estimate the
heating rate, $v_w$, taking into account contribution from stellar
winds, supernovae, millisecond pulsars and SMBH feedback. We make the
simplifying assumption that all sources of energy injection are
efficiently mixed into the bulk of CNM gas. In fact, slower stellar
wind material may cool and form high density structures as in
\citet{Cuadra+2005}. Additionally, some of the injected energy may
leak away from the bulk of the CNM material through low density holes,
as described by \citet{Harper-Clark+2009} for stellar feedback and by
\citet{Zubovas+2014} for AGN feedback.

We summarize the individual heating sources below, providing estimates
of the value of the effective wind velocity
\begin{align}
  v_{w} = \sqrt{\frac{2 \th \dot{e}}{\eta \rhostar}}
  \label{eq:vw_eff}
\end{align}
for each, where $\dot{e}$ is volumetric heating rate.  

\subsection{Stellar winds} 

Energy and mass input to the CNM by stellar winds is the sum of
contributions from main sequence and post-main sequence populations.
At early times following star formation, energy input is dominated by
the fast line-driven outflows from massive stars (e.g.,
\citealt{VossDiehl+:2009a}).  At later times energy input is dominated
by main sequence winds (e.g., \citealt{NaimanSoares-Furtado+:2013a}).
Mass input is also dominated by the massive stars for very young stellar
populations, but for most stellar ages the slow AGB winds of evolved
low-mass stars dominate the mass budget.

In Appendix \ref{app:windheat} we calculate the wind heating rate from
stellar winds, $v_{\rm w}^{\star}$, and the mass loss parameter,
$\eta$ (eq.~[\ref{eq:q}]), as a function of age, $\tau_{\star}$, of a
stellar population which is formed impulsively (Fig.~\ref{fig:vwImp}).
At the earliest times ($\tau_{\star} \lesssim 10^{7}$ yr), the wind
heating rate exceeds 1000 km s$^{-1}$, while much later ($\tau_{\star}
\sim t_{\rm h}$) stellar wind heating is dominated by main sequence winds
is much lower, $v^{\star}_{\rm w} \sim 50-100 $ km s$^{-1}$.  As will
be shown in $\S\ref{sec:combined}$, for the case of quasi-continuous
star formation representative of the average SFHs of low mass
galaxies, the heating rate from stellar winds can also be significant,
$v_{\rm w}^{\star} \sim 1000$ km s$^{-1}$.

Stellar winds thus contribute a potentially important source of both
energy and mass to the CNM.  However, two additional uncertainties are
(1) the efficiency with which massive stellar winds thermalize their
energy and (2) heating from core collapse supernovae, which is
potentially comparable to that provided by stellar winds.  We neglect
both effects, but expect they will act in different directions in
changing the total heating rate.

\subsection{Type Ia Supernovae} 

Type Ia SNe represent a source of heating, which unlike core collapse
SNe is present even in an evolved stellar population.  If each SN Ia
injects thermal energy $E_{\rm Ia}$ into the interstellar medium, and
the SN rate per stellar mass is given by $R_{\rm Ia}$, then the
resulting volumetric heating rate of $E_{\rm Ia}R_{\rm Ia}$ produces
an effective heating parameter (eq.~[\ref{eq:vw_eff}])
of \begin{align} v_{w}^{\rm Ia} =\sqrt{\frac{2 \th R_{\rm Ia} E_{\rm
        Ia}}{\eta}} \label{eq:vw_sne}.
\end{align} The thermal energy injected by each SN Ia, $E_{\rm Ia} \simeq
\epsilon_{\rm Ia} 10^{51}$ erg, depends on the efficiency $\epsilon_{\rm Ia}$
with which the initial blast wave energy is converted into bulk or turbulent
motion instead of being lost to radiation.  \cite{Thornton+98}
estimate a radiative efficiency $\epsilon_{\rm Ia} \sim 0.1$,
depending weakly on surrounding density, but \citet{Sharma+14} argues
that $\epsilon_{\rm Ia}$ can be considerably higher, $\sim 0.4$, if
the SNe occur in a hot dilute medium, as may characterize the CNM.  Hereafter we adopt $\epsilon_{\rm Ia} = 0.4$ as fiducial.

The SN Ia rate, $\RateIa$, depends on the age of the stellar
population, as it represent the convolution of the star formation rate
and the Ia delay time distribution (DTD) divided by the present
stellar mass.  In the limit of impulsive star formation, $\RateIa$ is
the DTD evaluated at the time since the star formation episode.  The
observationally-inferred DTD (Fig. 1 of \citealt{MaozMannucci+:2012a})
has the approximate functional form \begin{align} R_{\rm Ia}
  =1.7\times 10^{-14}\left(\tau_{\star}/t_{\rm h}\right)^{-1.12}
  M_{\odot}^{-1}\,{\rm yr^{-1}}
\label{eq:DTD}
  \end{align}
  where $\tau_{\star}$ is the time since star formation.

From equations (\ref{eq:vw_sne}), (\ref{eq:DTD}) we thus estimate that 
  \begin{eqnarray} 
    v_{w}^{\rm Ia} &\approx& 700(\epsilon_{\rm
      Ia}/0.4)^{0.5}(\tau_{\star}/t_{\rm h})^{-0.56}\eta_{0.02}^{-1/2}\,{\rm km
      \,s^{-1}} \nonumber \\
&\approx& 700(\epsilon_{\rm
      Ia}/0.4)^{0.5}(\tau_{\star}/t_{\rm h})^{0.09}\,{\rm km
      \,s^{-1}},
\label{eq:vIa}
  \end{eqnarray}
  where the second line assumes $\eta\simeq 0.02
  (\tau_{\star}/\th)^{-1.3}$ for a single burst of star formation
  (e.g. , \citealt{Ciotti+91}).

The high value of $v_{w}^{\rm Ia}$ implies that Ia SNe represent an
important source of CNM heating.  However, SNe can only be
approximated as supplying heating which is spatially and temporally
homogeneous if the rate of SNe is rapid compared to the characteristic
evolution time of the flow at the radius of interest
(\citealt{ShcherbakovWong+:2014a}).  We define the ``Ia radius"
  \begin{align}
    r_{\rm Ia} \sim \left(\frac{G}{R_{\rm Ia}\sigma_0}\right)^{1/2} \sim
    35 M_{\bullet,8}^{-0.1}(\tau_{\star}/t_{\rm h})^{0.56}\,{\rm pc}
    \label{eq:rIa}
  \end{align}
  as the location exterior to which the time interval between
  subsequent supernovae $\tau_{\rm Ia} \sim (M_{\rm enc}R_{\rm
    Ia})^{-1} \sim G/(r\sigma_0^{2}R_{\rm Ia})$ exceeds the local
  dynamical time-scale $t_{\rm dyn} \sim r/\sigma_0$, where we again
  adopt the Ia rate for an old stellar population and in the final
  equality we estimate $\sigma_0 \approx \sqrt{3}\sigma_{\bullet}$
  using $M_{\bullet}-\sigma$ (eq.~[\ref{eq:Msigma}]).

Assuming that $v_{w}^{\rm Ia} \gg \sigma_0$ then by substituting
$v_{w}^{\rm Ia}$ (eq.~\ref{eq:vIa}) into equation (\ref{eq:stag_simple})
for the stagnation radius, we find that
\begin{align}
  \left.\frac{r_{\rm Ia}}{r_{\rm s}}\right|_{\rm v_{w}^{\rm Ia}}&\approx
  \begin{cases}
    7\, \eta_{0.02}^{-1} M_{\bullet,8}^{-1.1}(\epsilon_{\rm
     Ia}/0.4) (\tau_{\star}/t_{\rm h})^{-0.56}  & {\rm core}\\
    14\, \eta_{0.02}^{-1} M_{\bullet,8}^{-1.1}(\epsilon_{\rm
     Ia}/0.4) (\tau_{\star}/t_{\rm h})^{-0.56} &{\rm cusp}\\
   \end{cases} \nonumber\\
 &\approx 
 \begin{cases}
    7\, M_{\bullet,8}^{-1.1}(\epsilon_{\rm
     Ia}/0.4) (\tau_{\star}/t_{\rm h})^{0.74}  & {\rm core}\\
    14\, M_{\bullet,8}^{-1.1}(\epsilon_{\rm
     Ia}/0.4) (\tau_{\star}/t_{\rm h})^{0.74} &{\rm cusp}\\
   \end{cases}
\label{eq:rIars}
\end{align}
SN Ia can only be approximated as a steady heating source near the stagnation radius for extremely
massive SMBHs with $M_{\bullet} \gtrsim 10^9 \Msun$ or for a very
young stellar population with $\tau_{\star} \ll t_{\rm h}$.

Even if SN Ia are rare near the stagnation radius, they may cap the
inflow rate (and thus the SMBH accretion) rate by periodically blowing
gas out of the nucleus of low mass galaxies.  Between successive SNe,
stars release a gaseous mass $M_{\rm g} \approx \eta
M_{\star}\tau_{\rm Ia}/t_{\rm h}$ interior to the Ia radius, which is
locally gravitationally bound to the SMBH by an energy $E_{\rm bind}
\gtrsim M_{\rm g}\sigma_0^{2}$.  From the above definitions it follows
that

\begin{equation}
 \frac{E_{\rm
    Ia}}{E_{\rm bind}} \lesssim \frac{(v_{w}^{\rm
    Ia})^{2}}{2\sigma_0^{2}}.
\label{eq:blowout}
\end{equation}

Hence, for low mass BHs with $\sigma_0 \ll v^{\rm Ia}_{w}$, SN Ia are
capable of dynamically clearing out gas from radii $\sim r_{\rm Ia}
\gtrsim r_{\rm s}$.  Thus even when heating is sufficiently weak that
the stagnation radius formally exceeds $r_{\rm Ia}$, the SMBH
accretion rate is still limited to a value
\begin{eqnarray}
\frac{\dot{M}_{\rm Ia}}{\dot{M}_{\rm edd}} &\approx& \frac{\eta M_{\bullet}}{\dot{M}_{\rm edd} t_{\rm h}}\left(\frac{r_{\rm Ia}}{r_{\rm inf}}\right)^{2-\Gamma} \approx \nonumber \\
 && \begin{cases}
    4.5 \times 10^{-4} M_{\bullet,8}^{-1.33}(\tau_{\star}/t_{\rm h})^{-0.2}
   & \text{core} \\
    2.2 \times 10^{-4} \Mbheight^{-0.84}(\tau_{\star}/t_{\rm h})^{-0.6}   & \text{cusp},
  \end{cases}
  \label{eq:eddr_Ia}
\end{eqnarray}
obtained substituting the Ia radius $r_{\rm Ia}$ (eq.~[\ref{eq:rIa}])
for $r_{\rm s}$ in the derivation leading to our estimate of $\dot{M}$
(eq.~[$\ref{eq:eddr_analytic}$]).  

The deep gravitational potential wells of high mass galaxies prevent
SN Ia from dynamically clearing out gas in these systems ($\sigma_0
\gg v^{\rm Ia}_{w}$).  Equation (\ref{eq:eddr_Ia}) nevertheless still
represents a cap on the accretion rate in practice because the Ia
heating rate (eq.~[\ref{eq:vIa}]) is usually high enough to prevent
the stagnation radius (calculated including the Ia heating) from
substantially exceeding $r_{\rm Ia}$. If the stagnation radius moves
inwards from $\rIa$, then decreased heating will force it outwards
again.  On the other hand, if the stagnation radius moves well outside
of $\rIa$, then the high level of Ia heating will force it inwards.

\subsection{Millisecond Pulsars}
Energy injection from the magnetic braking of millisecond pulsars
(MSPs) is a potentially important heating source.  If the number of
MSPs per unit stellar mass is $n_{\rm msp}$ and each contributes on
average a spin-down luminosity $\bar{L}_{\rm sd}$, then the resulting
heating per unit volume $\dot{e} \approx \bar{L}_{\rm sd}n_{\rm
  msp}\epsilon_{\rm msp}$ results in an effective heating rate (eq.~
[\ref{eq:vw_eff}]) of
\begin{eqnarray} v_{w}^{\rm MSP} \sim
30\left(\frac{\epsilon_{\rm msp}}{0.1}\right)^{1/2}\left(\frac{\bar{L}_{\rm
sd}}{10^{34}\,{\rm erg\,s^{-1}}}\right)^{1/2} \eta_{0.02}^{-1/2}\,{\rm
km\,s^{-1}},
 \label{eq:vmsp}
  \end{eqnarray} 
  where $\epsilon_{\rm msp}$ is the thermalization efficiency of the
  wind, normalized to a value $\lesssim 0.1$ based on that inferred by
  modeling the interstellar media of globular clusters
  \citep{NaimanSoares-Furtado+:2013a}.  Our numerical estimate assumes
  a pulsar density $n_{\rm msp} \sim 3 \times 10^{-40} $ MSPs
  g$^{-1}$, calculated from the estimated $\sim 30,000$ MSPs in the
  Milky Way (\citealt{Lorimer13}) of stellar mass $\approx 6\times
  10^{10}M_{\odot}$.

  Based on the ATNF radio pulsar catalog (\citealt{Manchester+05}), we
  estimate the average spin-down luminosity of millisecond pulsars in
  the field to be $\bar{L}_{\rm sd} \sim 10^{34}$ erg s$^{-1}$,
  resulting in $v_{w}^{\rm MSP} \lesssim 30$ km s$^{-1}$ for $\eta
  \gtrsim 0.02$.  For higher spin-down luminosities, $L_{\rm sd}\simeq
  10^{35}$ ergs s$^{-1}$ characteristic of some Fermi-detected
  pulsars, then the higher value of $v_{w}^{\rm MSP} \lesssim 300$ km
  s$^{-1}$ makes MSP heating in principle important under the most
  optimistic assumptions $\epsilon_{\rm msp} = 1$ and $\eta = 0.02$.

\subsection{SMBH Feedback}

Feedback from accretion onto the SMBH represents an important source
of heating which, however, is also the most difficult to quantify
(e.g., \citealt{Brighenti&Mathews03}, \citealt{DiMatteo+05};
\citealt{Kurosawa&Proga09}; \citealt{Fabian12} for a recent review).
A key difference between AGN heating and the other sources discussed
thus far is its dependence on the SMBH accretion rate
$\dot{M}_{\bullet}$, which is itself a function of the heating rate
(eq.~[\ref{eq:mdot_analytic}]).

\subsubsection{Compton Heating}

There are two types of SMBH feedback: kinetic and radiative.
Radiative feedback is potentially effective even in low luminosity AGN
via Compton heating (e.g., \citealt{Sazonov+04}, \citealt{Ciotti+10}),
which provides a volumetric heating rate (\citealt{Gan+14})
\begin{align}
\dot{e} = 4.1\times 10^{-35}n^{2}\xi T_{\rm C}\,{\rm erg\,cm^{-3}\,s^{-1}},
\end{align}
where $\xi = L/n r^{2}$ is the ionization parameter and $L$ is the SMBH luminosity with Compton temperature $T_{\rm C} \sim 10^{9}$ K $\gg T$ (e.g., \citealt{Ho99}, \citealt{Eracleous+10}).  

The importance of Compton heating can be estimated by assuming the
SMBH radiates with a luminosity $L = \epsilon \dot{M}c^{2}$, where
$\epsilon$ is the radiative efficiency and where $\dot{M}$ is
estimated from equation (\ref{eq:mdot_analytic}).  Then using
equations (\ref{eq:stag_simple}), (\ref{eq:mdot_analytic}),
(\ref{eq:rhors}), (\ref{eq:rhostarrs}) we calculate from equation
(\ref{eq:vw_eff}) that the effective heating rate at the stagnation
radius is given by
\begin{align} v_{w}^{\rm C} \simeq
  \begin{cases} 24 \eta_{0.02}^{0.5}T_{\rm
C,9}^{0.5}\epsilon_{-2}^{0.5} M_{\bullet,8}^{0.38}v_{500}^{-1.4}\,{\rm
km\,s^{-1}} &, \text{core}\\ 30 \eta_{0.02}^{0.5}T_{\rm
C,9}^{0.5}\epsilon_{-2}^{0.5} M_{\bullet,8}^{0.24}v_{500}^{-0.7}\,{\rm
km\,s^{-1}} &, \text{cusp},
  \end{cases}
  \label{eq:vC}
\end{align} where $T_{C,9} = T_{C}/10^{9}$ K and $\epsilon_{-2} =
\epsilon/0.01 \sim 1$.  We caveat that, unlike stellar wind heating,
Compton heating depends on radius, scaling as $v_{w}^{\rm C}(r)
\propto (n^{2}\xi/\rho_{\star})^{1/2} \propto
r^{(\Gamma-\densSlope-1)/2}$, i.e. $\propto r^{-0.6}$ and $\propto
r^{-0.75}$ for core ($\Gamma = 0.8$; $\densSlope \simeq 1$) and cusp
($\Gamma = 0.1$; $\densSlope \simeq 0.6$) galaxies, respectively.
Although our model's assumption that the heating parameter be radially
constant is not satisfied, this variation is sufficiently weak that it
should not significantly alter our conclusions.

If Compton heating acts alone, i.e. $\tilde{v}_{w} = v_{w}^{C}$, then
solving equation (\ref{eq:vC}) for $\tilde{v}_{\rm w}$ yields
\begin{align} v_{w}^{\rm C} \simeq
  \begin{cases} 140 \eta_{0.02}^{0.21}T_{\rm
C,9}^{0.21}\epsilon_{-2}^{0.21} M_{\bullet,8}^{0.16}\,{\rm km\,s^{-1}}
&, \text{core}\\ 95 \eta_{0.02}^{0.34}T_{\rm
C,9}^{0.29}\epsilon_{-2}^{0.29} M_{\bullet,8}^{0.14}\,{\rm km\,s^{-1}}
&, \text{cusp}.
  \end{cases}
  \label{eq:vC2}
\end{align} 
Compton heating is thus significant in young stellar populations
with relatively high mass loss rates, e.g. $v_{w}^{\rm C} \gtrsim 300$
km s$^{-1}$ for $\eta \gtrsim 1$.  

The inflow rate corresponding to a state in which Compton heating
self-regulates the accretion flow is given by substituting equation
(\ref{eq:vC2}) into equation (\ref{eq:eddr_analytic}):
\begin{align}
\frac{\dot{M}_{C}}{\dot{M}_{\rm edd}} \approx 
\begin{cases} 2\times 10^{-3} \eta_{0.02}^{0.2}T_{\rm
C,9}^{-0.8}\epsilon_{-2}^{-0.8} M_{\bullet,8}^{0.16}
&, \text{core}\\ 8\times 10^{-4} \eta_{0.02}^{0.3}T_{\rm
C,9}^{-0.7}\epsilon_{-2}^{-0.7} M_{\bullet,8}^{0.14}
&, \text{cusp}.
  \end{cases}
  \label{eq:MdotCa}
\end{align}

\citet{Sharma+2007} estimate the value of the radiative efficiency of
low luminosity AGN, $\epsilon_{\rm rad}$, based on MHD shearing box
simulations of collisionless plasmas.  For Eddington ratios of
relevance, their results (shown in their Fig.~6) are well approximated
by
\begin{align}
\epsilon_{\rm rad} \simeq 
\begin{cases}
  0.03 \left(\frac{\dot{M}_{\bullet}}{10^{-4}\dot{M}_{\rm edd}}\right)^{0.9} & \frac{\dot{M}_{\bullet}}{\dot{M}_{\rm edd}} \lsim 10^{-4} \\
 0.03 &  10^{-2} \gsim \frac{\dot{M}_{\bullet}}{\dot{M}_{\rm edd}} \gsim  10^{-4},
\end{cases}
\label{eq:efficiency}
\end{align}
where $\dot{M}_{\bullet}$ is the BH accretion rate. In general
$\dot{M}_{\bullet}$ will be smaller than the inflow rate $\dot{M}$
calculated thus far by a factor $\alpha < 1$ due to outflows from the
accretion disc on small scales.  Thus, the full efficiency relating
the mass inflow rate to the radiative output is $\epsilon = \alpha
\epsilon_{\rm rad}$, where we take $\alpha = 0.1$ following
\citet{Li+13}, who find that the fraction of the inflowing matter
lost to outflows equals the Shakura-Sunyaev viscosity parameter of the
disc.

Substituting $\epsilon$ into equation (\ref{eq:MdotCa}) and solving
for the inflow rate results in

\begin{align}
\frac{\dot{M}_{C}}{\dot{M}_{\rm edd}} \approx 
\begin{cases} 4\times 10^{-2} \eta_{0.02}^{0.2}T_{\rm
C,9}^{-0.8} M_{\bullet,8}^{0.16}
&, \text{core}\\ 9\times 10^{-3} \eta_{0.02}^{0.3}T_{\rm
C,9}^{-0.7} M_{\bullet,8}^{0.14}
&, \text{cusp}.
  \end{cases}
  \label{eq:MdotC}
\end{align}

Accretion may not be truly steady in cases where AGN heating
dominates, due to the inherent delay between black hole feedback and
the structure of the accretion flow on larger scales.  Equation
(\ref{eq:MdotC}) may nevertheless represent a characteristic average
value for the inflow rate if a quasi-equilibrium is achieved over
many cycles.

\subsubsection{Kinetic Feedback}

Kinetic feedback results from outflows of energy or momentum from
close to the black hole in the form of a disc wind or jet, which
deposits its energy as heat, e.g. via shocks or wave dissipation, over
much larger radial scales (e.g.~\citealt{McNamara&Nulsen07};
\citealt{Novak+11}; \citealt{Gaspari+12}).

Assume that the outflow power is proportional to the SMBH accretion
rate, $L_{\rm j} = \epsilon_{\rm j} \dot{M}_{\bullet}c^{2} =
0.1\epsilon_{\rm j} \dot{M}c^{2} $, where $\epsilon_{\rm j} < 1$ is an
outflow efficiency factor and we have again assumed a fraction $\alpha
= 0.1$ of the infall rate reaches the SMBH.  Further assume that this
energy is deposited as heat uniformly interior to a radius $r_{\rm
  heat}$ and volume $V_{\rm heat} \propto r_{\rm heat}^{3}$.  The
resulting volumetric heating rate $e = 0.1\epsilon_{\rm
  j}\dot{M}c^{2}/V_{\rm heat}$ near the stagnation radius results in a
heating parameter given by
\begin{eqnarray} 
  v_{w}^{\bullet} &\approx& \left(\frac{0.2 \th \epsilon_{\rm j}\dot{M}c^{2}}{V_{\rm heat}\eta
      \rho_{\star}|_{r_{s}}}\right)^{1/2} \\
  &\approx& \frac{700}{\sqrt{2-\Gamma}}\,{\rm km\,s^{-1}}\,\left(\frac{\epsilon_{\rm
        j}}{10^{-5}}\right)^{1/2}\left(\frac{r_{\rm s}}{r_{\rm
        heat}}\right)^{3/2} ,
\label{eq:vjet}
\end{eqnarray} 
where we have used the facts that $\dot{M} = \eta M_{\star}|_{r_{\rm
    s}}/t_{\rm h}$ and $\rho_{\star}(\rs) = M_{\star}(\rs)
(2-\Gamma)/(4 \pi \rs^3)$.  If the bulk of the energy from kinetic
feedback is released near the stagnation radius, then even a small
heating efficiency $\epsilon_{\rm j} \gtrsim 10^{-4}$ is sufficient
for $v_{w}^{\bullet}$ to exceed other sources of non-accretion powered
heating.  However, if this energy is instead deposited over much
larger physical scales comparable to the size of the galaxy,
i.e. $r_{\rm heat} \gtrsim $ 10 kpc $\sim 10^{4}r_{\rm s}$, then
kinetic feedback is unimportant, even for a powerful outflow with
$\epsilon_{\rm j} \sim 0.1$.

The time required for a jet of luminosity $L_{\rm j}$ and half opening
angle $\theta_{\rm j} = 0.1$ (characteristic of AGN jets) to propagate
through a gaseous mass $M_{\rm g}$ of radius $r$ is estimated from
\citet{Bromberg+11} to be
\begin{eqnarray}
  t_{\rm jet} \sim 4000\,{\rm yr}\left(\frac{{L}_{\rm j}}{10^{40}\,\rm erg\,s^{-1}}\right)^{-1/3}\left(\frac{r}{\rm pc}\right)^{2/3}\left(\frac{M_{\rm g}}{10^{8}M_{\odot}}\right)^{1/3} 
\end{eqnarray}
Approximating $M_{\rm g} \sim \dot{M}t_{\rm ff}$, the ratio of the jet
escape time-scale to the dynamical time-scale $t_{\rm dyn} \sim
r/\sigma$ is given by
\begin{equation}
  \frac{t_{\rm jet}}{t_{\rm dyn}} \sim 7\times 10^{-3}M_{\bullet,8}^{0.13}\left(\frac{\epsilon_{\rm j}}{10^{-6}}\right)^{-1/3},
\end{equation}
independent of $r$.  A jet with power sufficient to appreciably heat
the CNM on radial scales $\sim r_{\rm s}$ ($\epsilon_{j} \gtrsim
10^{-5}$) also necessarily has sufficient power to escape the nuclear
region and propagate to much larger radii.  Slower outflows from the
accretion disc, instead of a collimated relativistic jet, provide
a potentially more promising source of feedback in these systems.

As in the case of Compton heating, kinetic heating could in principle
`self-regulate' the accretion flow insofar as a lower heating rate
results in a higher accretion rate (eq.~[\ref{eq:vjet}]), which in
turn may create stronger kinetic feedback.  However, given the uncertainty
in the efficiency of kinetic heating, we hereafter neglect its effect and defer further discussion
to $\S\ref{sec:kinetic}$.

\begin{figure}
\includegraphics[width=\columnwidth]{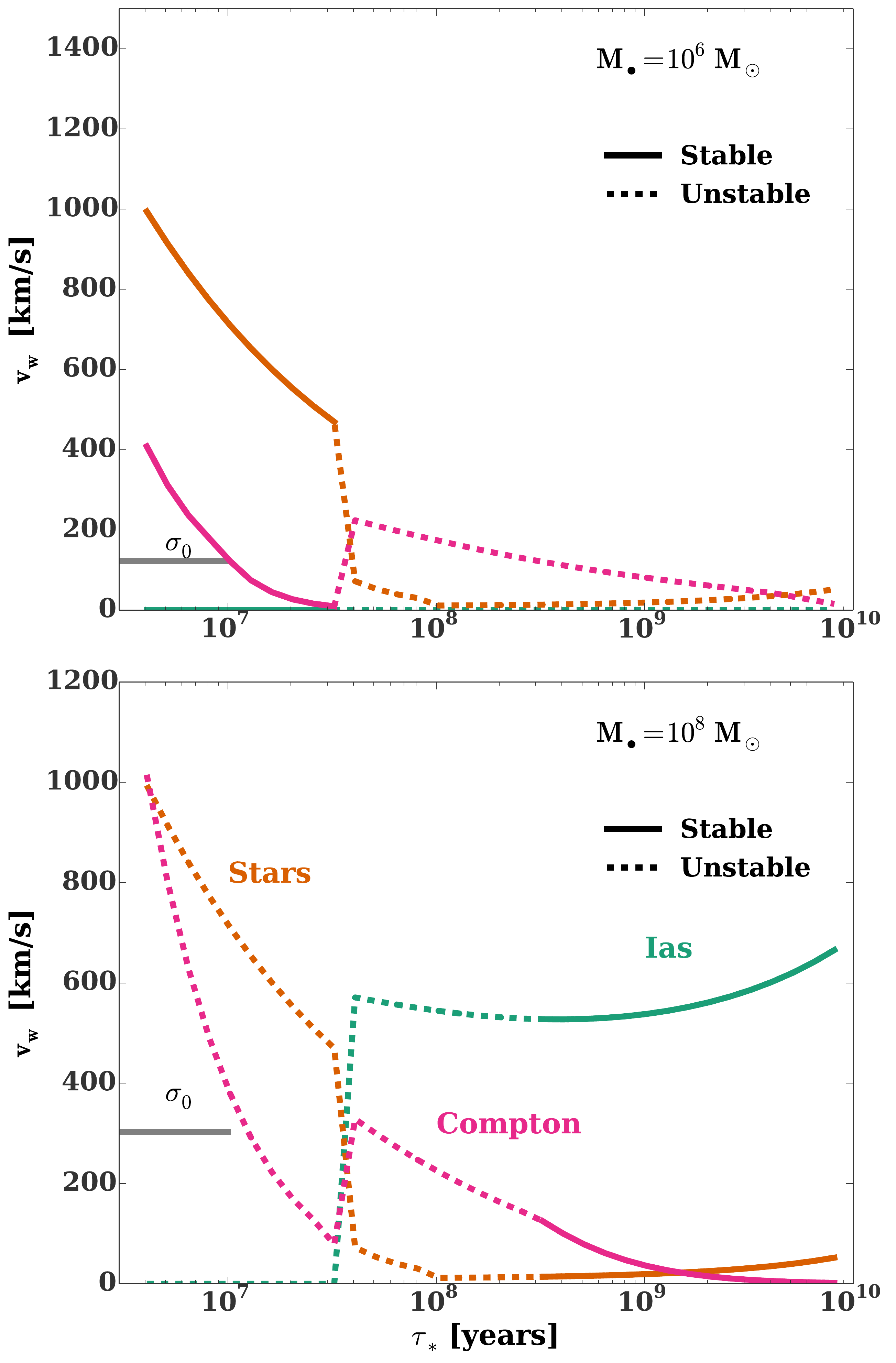}
\caption{\label{fig:vwSourcesImp} Sources contributing to
  the gas heating rate at time $\tau_{\star}$ after
  a single burst of star formation.  Top and bottom panels show black hole masses of $\Mbh=10^6 \Msun$ and $\Mbh= 10^8 \Msun$
  (both cusps with $\Gamma=0.8$).  Solid and dashed lines show the ranges of
  $\tau_{\star}$ for which the accretion flow is thermally stable and
  unstable, respectively, according to the ratio of $\dot{q}_{\rm
    heat}/|\dot{q}_{\rm rad}|$ near the stagnation radius
  (eq.~[\ref{eq:cooling2}]).  Shown with horizontal gray lines
  are the stellar velocity dispersion for each SMBH mass, estimated as $\sigma_0=\sqrt{3} \sigma_{\bullet}$ (eq.~[\ref{eq:Msigma}]).}
\end{figure}

\begin{figure}
\includegraphics[width=\columnwidth]{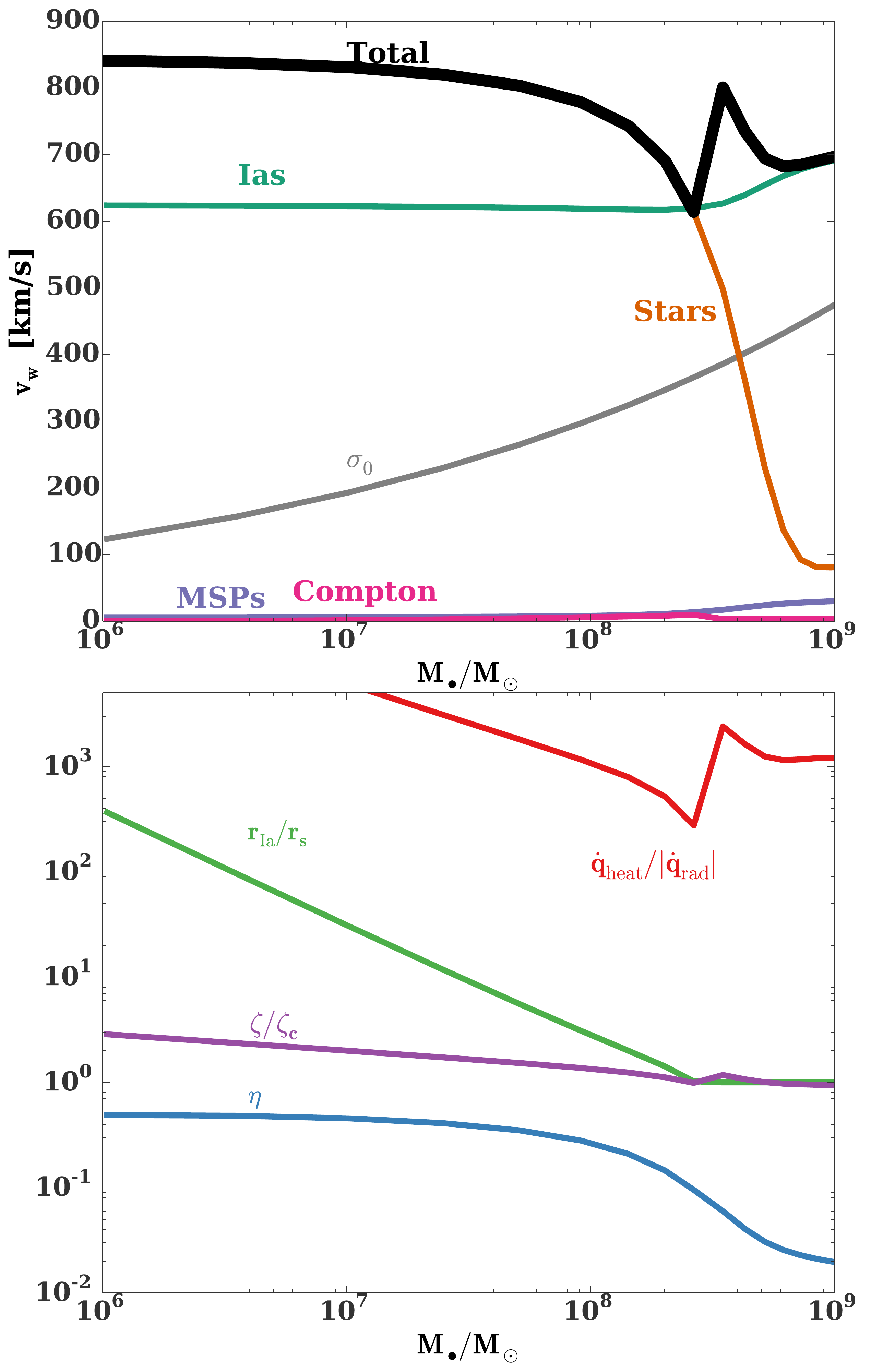}
\caption{\label{fig:vwSources} {\it Top Panel:} Sources contributing
  to the total heating rate of the CNM, $\vwO$ ({\it black}): stellar
  wind heating ({\it orange}), Ia supernovae ({\it green}),
  millisecond pulsars ({\it blue}), and compton heating ({\it pink}).
  Each heating source varies with black hole mass $\Mbh$ calculated
  for average SFHs from \citet{MosterNaab+:2013a}; see
  Appendix~\ref{app:windheat} for details.  {\it Bottom Panel:} The
  ratio of the total heating rate ($\dot{q}_{\rm heat} \propto
  v_{w}^{2}$) from the top panel to the radiative cooling rate
  ($|\dot{q}_{\rm rad}|$) at the stagnation radius (red), parameter
  $\eta$ characterizing stellar mass loss rate (eq.~[\ref{eq:q}]) as a
  function of black hole mass $\Mbh$, calculated for average star SFHs
  from \citealt{MosterNaab+:2013a} (blue), the ratio of Ia radius to
  stagnation radius as a function of $\Mbh$ (green), and the ratio of
  $\zeta/\zeta_{c}$ (purple), where $\zeta_c$
  (eq.~[\ref{eq:zetacrit}]) is the critical heating parameter $\zeta
  \equiv \sqrt{1+(v_w/\sigma_0)^2}$ below which outflows are
  impossible.  All quantities are calculated for core
  ($\Gamma=0.1$) galaxies. The results for cusp galaxies are
  qualitatively similar.}
\end{figure}

\subsection{Combined Heating Rate} 
\label{sec:combined}

The total external gas heating rate, 
\begin{equation}
\vwO = \sqrt{(v_{w}^{\star})^{2} + (v_{w}^{\rm MSP})^{2} + (v_{w}^{\rm Ia})^{2} + (v_{w}^{\bullet})^{2}},
\label{eq:vtot}
\end{equation}
includes contributions from stellar winds, supernovae, pulsars, and
radiative SMBH feedback. We implicitly assume the different sources of
energy injection mix efficiently. In reality, this may not be the case
as slower velocity sources cool and form high density structure due to
high pressure from the environment\citep{Cuadra+2005}.

The strength of each heating source depends
explicitly on the SMBH mass and the stellar population in the galactic
nuclear region.  The latter could best be
described by a single starburst episode in the past, or by a more
continuous SFH that itself varies systematically
with the galaxy mass and hence $\Mbh$.

\subsubsection{Single Starburst}

Figure~\ref{fig:vwSourcesImp} shows the contributions of heating
sources as a function of time $\tau_{\star}$ after a burst of star
formation for black holes of mass $\Mbh = 10^{6} \Msun$ (top) and
$\Mbh = 10^{8} \Msun$ (bottom). The heating and mass input parameters
due to stellar winds, $v_{w}^{\star}(\tau_{\star})$ and
$\eta(\tau_{\star})$, are calculated as described in
Appendix~\ref{app:windheat} (Fig.~\ref{fig:vwImp}).  The SN Ia and
Compton heating rates, $v_{w}^{\rm Ia}$ and $v_{\rm w}^{\rm C}$, are
calculated from equations~\eqref{eq:vIa} and~\eqref{eq:vC},
respectively, which also make use of $\eta(\tau_{\star})$ as set by
stellar winds.\footnote{Both $v_{\rm w}^{\rm C}$ and $v_{\rm w}^{\rm
    Ia}$ (through $r_{\rm s}/r_{\rm Ia}$) depend on the total heating
  rate (eq.~[\ref{eq:vtot}]), requiring us to simultaneously solve a
  series of implicit equations to determine each.}  SN Ia heating
depends also on a convolution of the DTD (eq. [\ref{eq:DTD}]) and the
SFH, which reduces to the DTD itself for a single star burst.  We
account for the expected suppression of SN Ia heating resulting from
its non-steady nature by setting $v_{w}^{\rm Ia} = 0$ when the
stagnation radius (calculated {\it excluding} Ia heating) exceeds the
Ia radius $r_{\rm Ia}$ (eq.~[\ref{eq:rIa}]). 

Figure \ref{fig:vwSourcesImp} shows that stellar winds are the most
important heating source at early times, $\tau_{\star} \lesssim
10^{7}$ years.  Compton heating becomes more important at later times
due to (1) the higher accretion rates that accompany the overall
decrease in all sources of heating, coupled with (2) the persistently
high mass loss rates and gas densities associated with the still
relatively young stellar population.  Interestingly, for $M_{\bullet}
= 10^{6}M_{\odot}$, there are two consistent solutions for times $10^8
{\rm yr} \lsim \tau_\star \lsim 10^9$ yr: one which has high Ia heating
and low compton heating and one with low Ia heating and high compton
heating. We choose the latter, but our model cannot truly distinguish
between these scenarios. Also, neither scenario would produce a
thermally stable flow. For high-mass black holes
(Fig.~\ref{fig:vwSourcesImp}, bottom panel), SN Ia dominate at all
times after $10^{7}$ yr.  Both Type Ia and Type II supernovae could be
an important heating source for times $\tau_\star \leq 40$ Myr, albeit
a non-steady one for $\Mbh \lsim 10^8 \Msun$. For simplicity we
neglect supernova heating for $\tau_\star \leq 40$, keeping in mind
that it could raise the overall heating by a factor of a few.

As the total heating rate declines with time, the flow inevitably
becomes thermally unstable according to the criterion $(\dot{q}_{\rm
  heat}/|\dot{q}_{\rm rad}|)_{r_{s}} \lesssim 10$
(eq.~[\ref{eq:cooling3}]), as shown by dashed lines in
Fig.~\ref{fig:vwSourcesImp}.  Compton heating is neglected in
calculating thermal stability because$-$unlike local stellar feedback
mechanisms$-$it is not clear that SMBH feedback is capable of
stabilizing the flow given its inability to respond instantaneously to
local changes in gas properties.  Thermally unstable flow is present
throughout the single starburst case except at very early times,
$\tau_{\star} \lesssim 10^7$ yr and at late times $\tau_{\star}
\gtrsim 10^{9}$ yr in the high $M_{\bullet}$ case.

Finally, MSP heating is negligibly small for fiducial
parameters and hence is not shown, while kinetic feedback is neglected
given its uncertain efficiency ($\S\ref{sec:kinetic}$).

\subsubsection{Continuous SFH}
A single burst of star formation does not describe the typical star
formation history of most galaxies.  Qualitatively, smaller galaxies
will on average experience more recent star formation, resulting in
energetic young stellar winds and supernovae which dominate the gas
heating budget.  Massive galaxies, on the other hand, will on average
possess older stellar populations, with their heating rates dominated
by SN Ia (on large radial scales) and SMBH feedback.  We estimate the
average value of $\vwO$ as a function of SMBH mass for each heating
source by calculating its value using the average cosmic star
formation histories of \citet{MosterNaab+:2013a}.  Appendix
\ref{app:windheat} describes how the average SFH is
used to determine the stellar wind heating $v_{\rm w}^{\star}$ and
mass return parameter $\eta$ as a function of $M_{\bullet}$
(Fig.~\ref{fig:vwSources}; top panel).  SFHs are also
convolved with the Ia DTD distribution to determine the SN Ia heating.

Figure~\ref{fig:vwSources} shows $\vwO(M_{\bullet})$ from each heating
source (stars, MSPs, SNe Ia, and black hole feedback) calculated for
the average SFH of galaxies containing a given black hole mass.  The
younger stellar populations characterizing low mass galaxies with $\Mbh\lsim
3\times 10^{8} \Msun$ are dominated by stellar winds, with
$v_{w}^{\star} \gtrsim 700$ km s$^{-1}$.  Only for $\Mbh\gsim 3\times
10^{8} \Msun$ does the lack of young stellar populations significantly
reduce the role of stellar wind heating.  For these massive
galaxies, however, the Ia radius is sufficiently small that SN Ia contribute
a comparable level of heating, $v_{w}^{\rm Ia} \gtrsim 500$ km
s$^{-1}$ (SN Ia do not contribute to the heating in low mass galaxies because $r_{\rm Ia} > \rs$; bottom panel).  

Strikingly, we find that galactic nuclei that experience the same
average SFH as their host galaxies possess thermally stable flows
across all black hole masses.  An important caveat, however, is that
$\vw$ is generally only a few times higher than the stellar velocity
dispersion, i.e. $\zeta \sim \zeta_c$ (eq.~[\ref{eq:zetacrit}]).  This
implies that the true stagnation radius and the inflow rate could be
larger than our analytic estimates, potentially resulting in thermal
instability in a significant fraction of galaxies.  The bottom panel
of Figure~\ref{fig:vwSources} also shows $\zeta/\zeta_c$, where
$\zeta_c$ is estimated using fits for $\rb$, $\rinf$, and $\Gamma$
derived from the \citet{LauerFaber+:2007a} sample.  In other words,
the CNM of a significant fraction of massive ellipticals may be
thermally unstable, not necessarily because they are receiving lower
stellar feedback than the average for their galaxy mass, but due to
the high heating required for stability given the structure of their
stellar potential.

Realistic variations (``burstiness") in the SFH can also
produce lower stellar wind heating rates (Appendix
\ref{app:windheat}).  Figure \ref{fig:NickPlot2} shows the heating
rate as a function of average black hole mass for non-fiducial cases
in which the current ($z = 0$) star formation is suppressed by a
factor $\iota$ from its average $z = 0$ value, for a characteristic
time-scale $\delta t_{\star}$.  For $\delta t_{\star} \lesssim 10^{7}$
yr, we see that $v_{\rm w}^{\star}$ is reduced by at most a factor of
$\approx 2$ from its average value, even for huge drops in the star
formation rate ($\iota \sim 10^{-3}$).  However, burstiness in the SFH
over longer time-scales can suppress heating more significantly.  When
$\delta t_{\star} \gtrsim 10^8~{\rm yr}$, $v_{\rm w}^\star$ becomes
very sensitive to $\iota$: at $\iota=0.1$, $v_{\rm w}^\star \approx
400~{\rm km~s}^{-1}$ in most galaxies, but for smaller $\iota$,
$v_{\rm w}^\star \approx 200~{\rm km~s}^{-1}$ (an exception to this is
in galactic nuclei with $M_\bullet \gtrsim 10^8~M_\odot$, where
heating is stabilized by SN Ia).  Our use of average cosmic SFHs is
appropriate provided local variations are either on short time-scales
($\delta t_{\star} \lesssim 10^7~{\rm yr}$), or limited in magnitude
($\iota \gtrsim 0.1$).  If both of these conditions are violated, then
the effective heating rates for all but the most massive galaxies
(where Ia explosions dominate) will fall by a factor $\gtrsim 4$ from
our fiducial volumetric averages, calculated using SFHs from
\citet{MosterNaab+:2013a}.

Another potential complication is the discreteness of local
sources. We assume heating and mass injection are smooth, but energy
and mass injection may be dominated by a handful of stars,
particularly for small $M_\bullet$. For example, for $\Mbh \lsim 10^8
\Msun$, O stars dominate the energy injection in the average star
formation histories of \citet{MosterNaab+:2013a}, but the expected
number of O stars inside the stagnation radius is 0.01 for $\Mbh=10^6
\Msun$ and only exceeds 1 for $\Mbh\gsim10^7 \Msun$, assuming circular
stellar orbits\footnote{If the stars providing the heating reside on
  elliptical orbits, a greater number will contribute at any time due
  to the portion of their orbital phase spent at small radii.}.

\section{Implications and Discussion}
\label{sec:discussion}

\subsection{Inflow and Black Hole Accretion Rates}
\label{sec:mdot}

\begin{figure}
\includegraphics[width=\columnwidth]{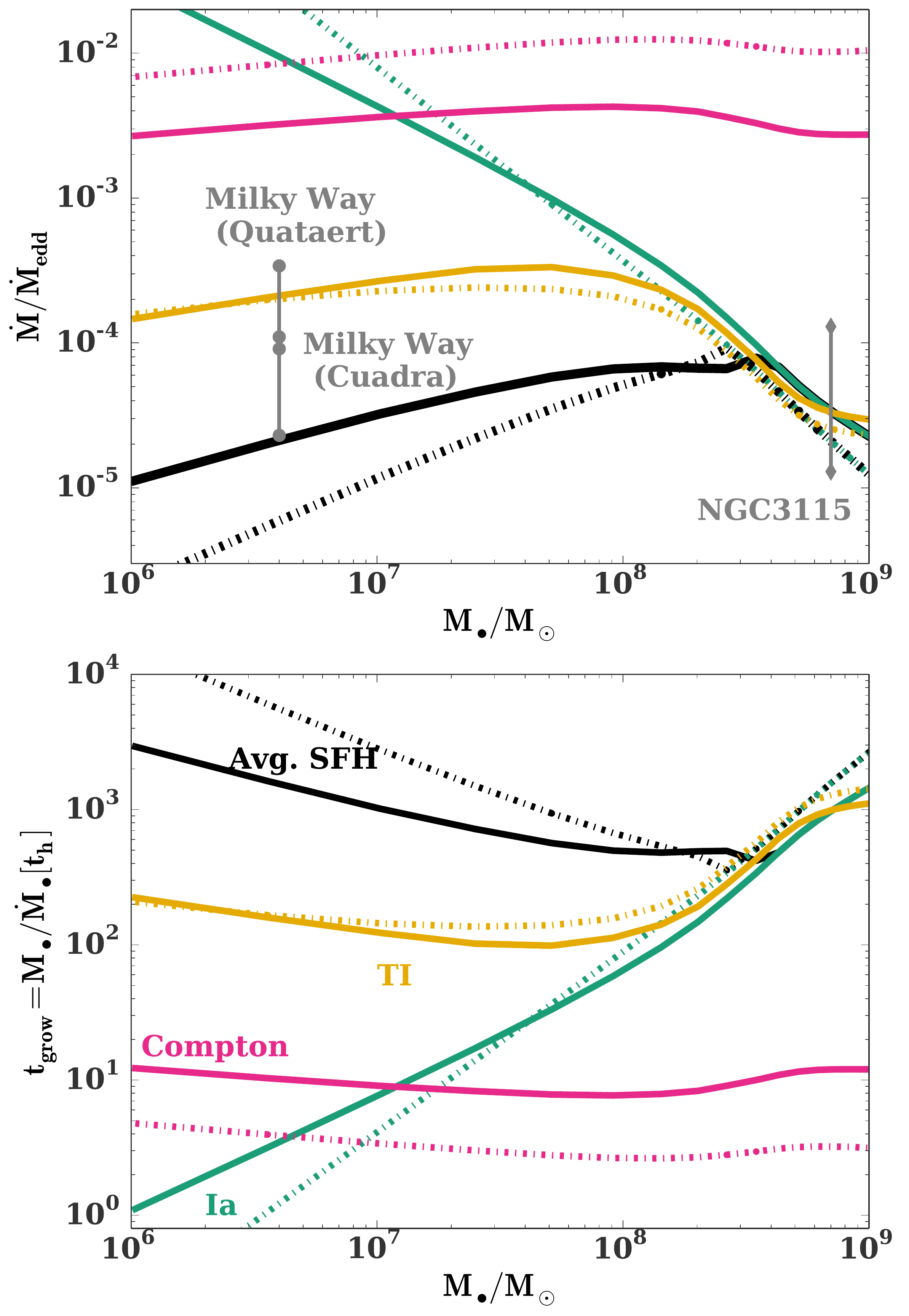}
\caption{\label{fig:bh_growth} {\it Top panel:} Gas inflow rate
  $\dot{M}/\dot{M}_{\rm edd}$ as a function of black hole mass
  $M_{\bullet}$, shown for core ($\Gamma=0.1$, dot-dashed) and cusp
  galaxies ($\Gamma=0.8$, solid).  Black lines show the inflow rate
  calculated from equation (\ref{eq:eddr_analytic}) using the heating
  rate provided at $z = 0$ by the average star formation histories for
  each galaxy mass (Fig.~\ref{fig:vwSources}).  Teal lines show the
  maximum accretion set by Ia supernovae blow-out or heating,
  $\dot{M}_{\rm Ia}$ (\ref{eq:eddr_Ia}).  Red lines show the inflow
  rate obtained if Compton heating acts alone, $\dot{M}_{\rm C}$
  (eq.~[\ref{eq:MdotC}]).  Yellow lines show the maximum inflow rate
  for a thermally stable flow near the stagnation radius,
  $\dot{M}_{\rm TI}$ (eq.~[\ref{eq:Mdotmax}]).  Also shown are the
  inferred inflow rates of SgrA* from \citet{Quataert:2004a} and
  \citet{Cuadra+2008} (gray circles) and for NGC3115 from
  \citet{ShcherbakovWong+:2014a} (gray diamonds).  {\it Bottom panel:}
  Growth times $t_{\rm grow} \equiv \Mbh/\dot{M}_{\bullet}$ in units
  of the Hubble time $\th$ for each of the accretion rates shown in
  the top panel, where $\dot{M}_{\bullet} = \alpha\dot{M}$, where
  $\alpha = 0.1$, to account for the fraction of inflowing mass lost
  to disc outflows on small scales.}
\end{figure}

Figure \ref{fig:vwSources} shows that the heating associated with the
average SFH of a galaxy is approximately constant
with SMBH mass, except for the largest black holes with $M_{\bullet}
\gtrsim 5\times 10^{8}M_{\odot}$.  For a fixed wind heating parameter,
the Eddington ratio $\dot{M}/\dot{M}_{\rm edd}$ increases $\propto
M_{\bullet}^{0.5(0.8)}$ for core(cusp) galaxies, respectively
(Fig.~\ref{fig:mdot_mass}; eq.~[\ref{eq:eddr_analytic}]).

Figure \ref{fig:bh_growth} (top panel) shows with solid lines the
inflow rate as a function of black hole mass for the average star
formation heating, calculated from equation (\ref{eq:eddr_analytic})
using our results for the total wind heating
(Fig.~\ref{fig:mdot_mass}).  This average inflow rate increases from
$\dot{M} \sim 10^{-6}-10^{-5}M_{\odot}$ yr$^{-1}$ for low-mass black
holes ($M_{\bullet} \sim 10^{6}M_{\odot}$) to $\dot{M} \sim
10^{-4}M_{\odot}$ yr$^{-1}$ for $M_{\bullet} \sim 10^{8}M_{\sun}$.
These fall below the maximum thermally stable inflow rate,
$\dot{M}_{\rm TI}$ (yellow lines). 

For low mass black holes, blow out from SN Ia caps the SMBH
accretion rate at a value $\dot{M}_{\rm Ia} \sim
10^{-6}-10^{-5}M_{\odot}$ yr$^{-1}$ (green lines;
eq.~[\ref{eq:eddr_Ia}]), which is however not low enough to prevent
otherwise thermally-unstable flow at smaller radii.  For higher
$M_{\bullet}$ the stagnation radius resides close to the Ia radius so Ia
heating contributes to the steady gas heating rate, even if the energy
released by a single Ia is insufficient to unbind gas from the stellar
bulge (eq.~[\ref{eq:blowout}]).  This explains why the black and teal
lines meet at high $M_{\bullet}$.

Shown also for comparison in Figure \ref{fig:bh_growth} (top panel) is
the inflow rate, $\dot{M}$, onto SgrA* calculated by
\citet{Quataert:2004a} and by \citet{Cuadra+2008}.  Also shown is the
range of $\dot{M}$ for the low-luminosity AGN NGC3115 derived through
detailed modeling by \citet{ShcherbakovWong+:2014a}, who find a range
of inflow rates depending on the assumed model for thermal
conductivity.  The inflow rate SgrA* estimated from
\citet{Quataert:2004a}\footnote{There is a typo in
  \citet{Quataert:2004a}: the inflow rate corresponds total stellar
  mass loss rate of $5 \times 10^{-4} \Msun$ yr$^{-1}$ not $10^{-3}
  \Msun$ yr$^{-1}$, as pointed out by \citet{Cuadra+2006}.} exceeds
our estimate of $\dot{M}$ due to the average SFH for the same
$M_{\bullet}$ by at least an order of magnitude. This is because the
mass source term in \citet{Quataert:2004a} exceeds ours by a factor of
$\sim 50$ (for cusp galaxies).  The star formation in the central
parsec of the Milky Way cannot be described as steady-state: feedback
is instead dominated by the stellar winds from the ring of young
massive stars of mass $\sim 10^{4}M_{\odot}$ and estimated age $\sim
10$ Myr (e.g., \citealt{Schodel+07}).  Such a star formation history
is better described by our impulsive (starburst) scenario, for which
the value of $\eta \sim 30$ at $\tau_{\star} = 10^{7}$ yr
(Fig.~\ref{fig:vwImp}) is a factor of $\sim 100$ times higher than the
value of $\eta \sim 0.4$ predicted in the average SFH case
(Fig.~\ref{fig:vwSources}, bottom panel). Note that the maximum
thermally stable accretion rate is dependent on the $\eta$ parameter
and would be higher in the case of an impulsive burst of star
formation.

However, simulations by \citet{Cuadra+2006} find that the inflow rate
for Sgr A* is reduced by up to an order of magnitude compared to the
result in \citet{Quataert:2004a} when the discreteness and disc
geometry of stellar sources is taken into account, and recent results
from \citet{Yusef-Zadeh+2015} suggest that the inflow rate could be
further reduced to $\eddr \sim 3\times 10^{-6}$ once clumping of
stellar winds is taken into account.

\subsection{Nuclear X-ray Luminosities}
\label{sec:Lx}

\begin{figure*}
\includegraphics[width=\textwidth]{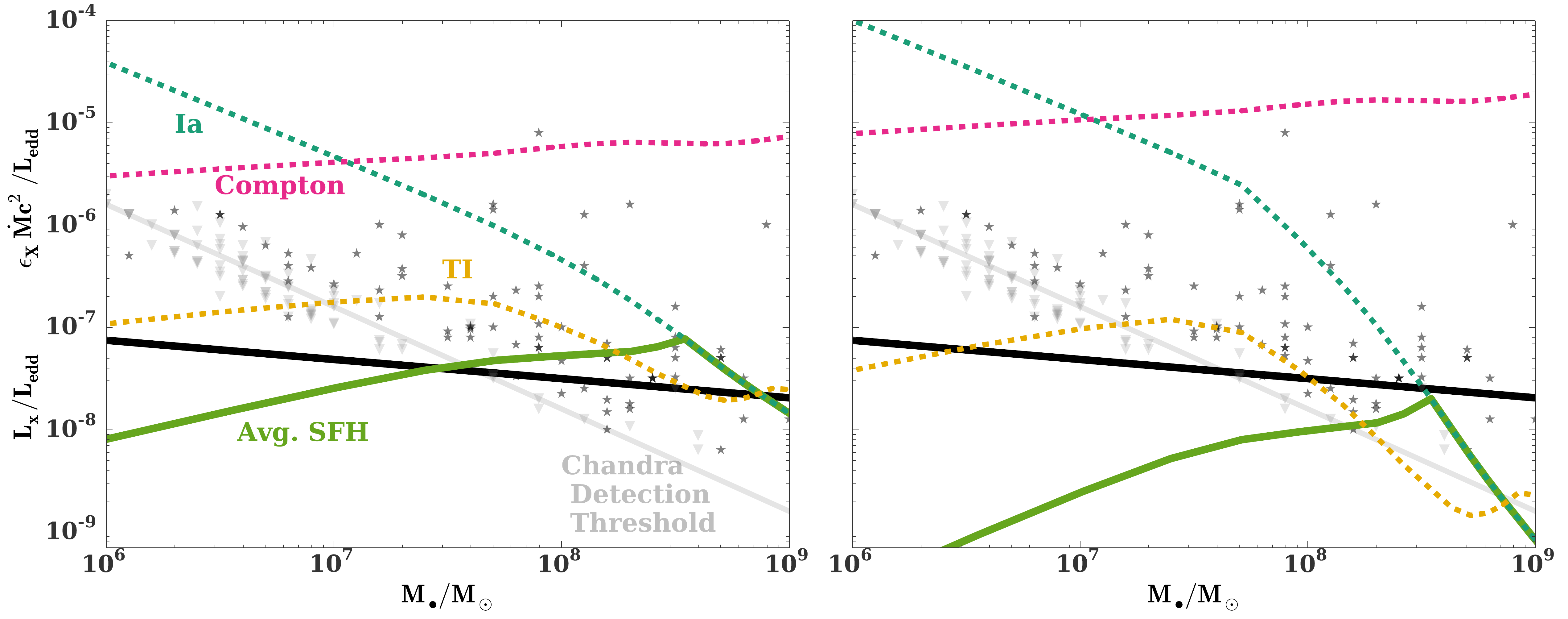}
\caption{\label{fig:miller} Average nuclear X-ray luminosity, $\langle
  L_X \rangle = \epsilon_X \Mdot c^2$, as a function of SMBH mass.
  Green lines show our prediction in the case of heating due to the
  average SFH for galaxies corresponding to each black hole mass
  (Fig.~\ref{fig:bh_growth}), calculated using $\Gamma = 0.7$ for
  $\Mbh < 4 \times 10^7 \Msun$, and $\Gamma=-0.3 \log_{10}
  \left(\Mbh/4\times 10^7 \Msun\right)$+0.7 for $\Mbh \geq 4 \times
  10^7 \Msun$, as derived from the \citet{LauerFaber+:2007a} sample.
  Shown for comparison are the measurements ({\it black stars}) and
  upper limits ({\it gray triangles}) for $L_X/L_{\rm edd}$ values,
  from the \citet{miller+2015} sample of early-type galaxies (a black
  line shows the best power-law fit $\langle L_X/L_{\rm edd}\rangle
  \propto M_{\bullet}^{\alpha}$, with $\alpha = -0.2$).  The left
  panel is calculated assuming a constant radiative efficiency
  $\epsilon_X=10^{-4} (\epsilon_{\rm bol}/0.1)$, while the right hand
  panel assumes $\epsilon_X= \alpha \epsilon_{\rm bol}\epsilon_{\rm
    rad}$, where $\epsilon_{\rm rad}$ is the radiative efficiency of
  low luminosity accretion discs calculated by \citet{Sharma+2007}
  (eq.~[\ref{eq:efficiency}]) and $\alpha = 0.1$ is the fraction of
  the mass inflowing on large scales that reaches the SMBH.  Shown for
  comparison are the X-ray luminosities calculated for the maximum
  thermally-stable accretion rate ({\it dashed orange line};
  eq.~[\ref{eq:Mdotmax}]); the SN Ia-regulated accretion rate ({\it
    dashed teal line}; eq.~[\ref{eq:eddr_Ia}]), and the Compton
  heating-regulated accretion rate ({\it dashed pink line};
  eq.~[\ref{eq:MdotCa}])\protect\footnotemark, again all calculated for the
  average SFH corresponding to each SMBH mass.}
\end{figure*}

The unresolved core X-ray luminosities of nearby galactic nuclei
provide a powerful diagnostic of the SMBH accretion rates and how they
vary with SMBH mass and other galaxy properties (e.g., \citealt{Ho08}
and references therein).  Figure~\ref{fig:miller} shows our
predictions for $\langle L_{X} \rangle =\epsilon_X \dot{M} c^2$ as a
function of black hole mass, where $\epsilon_{X} = \alpha\epsilon_{\rm
  rad}\epsilon_{\rm bol}$, $\alpha = 0.1$ accounts for the fraction of
the inflowing matter loss to outflows from the accretion disc,
$\epsilon_{\rm rad}$ is the radiative efficiency of the accreted
matter, and $\epsilon_{\rm bol} = 0.1$ is the assumed bolometric
correction into the measured X-ray band for low luminosity AGN
(\citealt{Ho08}).  In all cases the value of $\langle L_X \rangle$ is
calculated using the accretion rate from the top line of
equation~\eqref{eq:mdot_analytic} with $\Gamma = 0.7$ for $\Mbh < 4
\times 10^7 \Msun$, and $\Gamma=-0.3 \log_{10}
\left(\frac{\Mbh}{4\times 10^7 \Msun}\right)$+0.7 for $\Mbh \geq 4
\times 10^7 \Msun$. This functional form is designed to
approximately reproduce the behavior of $\Gamma(\Mbh)$ in the
\citet{LauerFaber+:2007a} sample.

The left panel of Figure \ref{fig:miller} is calculated assuming a
constant low value for the radiative efficiency of $\epsilon_X =
10^{-4}$, typical of those estimated for low luminosity AGN (e.g.,
\citealt{Ho:2009a}).  The right panel luminosities are calculated
instead assuming an efficiency of $\epsilon_X= \epsilon_{\rm
  bol}\epsilon_{\rm rad}$ that depends on the Eddington ratio as
predicted by MHD shearing box simulations by Sharma et al. (2007, see
eq.~[\ref{eq:efficiency}] and surrounding discussion).  Shown for
comparison are the X-ray measurements (black stars) and upper limits
(gray triangles) from the sample of early-type galaxies compiled by
Miller et al. (2015, cf.~\citealt{Gallo+10}).  A black line shows the
best power-law fit to the X-ray luminosity from \citet{miller+2015},
given by $\langle L_X/L_{\rm edd}\rangle \sim \Mbh^\alpha$ with
$\alpha = -0.2$ (see also \citealt{Zhang+09, Pellegrini10, Gallo+10}).
Also shown are the maximum accretion rates, respectively, for
thermally stable accretion (eq.~[\ref{eq:Mdotmax}]; yellow), as set by
SN Ia blow-out/heating (eq.~[\ref{eq:eddr_Ia}]; teal), and as allowed by
Compton heating feedback (eq.~[\ref{eq:MdotC}]; pink).

If the nuclei of elliptical galaxies are heated as expected for the
average SFH of galaxies with similar mass, then to first order we
predict that $\langle L_{X}/L_{\rm edd}\rangle$ should be an
increasing function of BH mass.  This result is in tension with the
observed down-sizing trend: the average Eddington ratio is seen to
decrease (albeit weakly) with $M_{\bullet}$, especially in the model
where $\epsilon_X$ increases with the Eddington ratio.  However, one
must keep in mind the enormous uncertainty in calculating the
luminosity of the accretion flow close to the black hole in a single
waveband based on the feeding rate on larger scales.  Also, if disc
outflows indeed carry away most of the infalling mass before it
reaches X-ray producing radii (e.g.~\citealt{Blandford&Begelman99};
\citealt{Li+13}), then the angular momentum of the infalling gas might
also influence the X-ray luminosity indirectly through the inflow
efficiency $\alpha$, in a way that could depend systematically on the
stellar population and hence $M_{\bullet}$.  However, in general the
efficiency with which inflowing matter reaches the SMBH (and hence
X-ray luminosity) would naively be expected to {\it decrease} with
decreasing $M_{\bullet}$ due to the increasing angular momentum of low
mass galaxies, exacerbating the tension between our findings and the
observed downsizing trend.

Perhaps a more readily addressable test is whether a steady-state
accretion picture developed in this work can account for the the large
scatter, typically of $2-3$ orders of magnitude, in $L_{X}/L_{\rm
  edd}$ at fixed $M_{\bullet}$.  Scatter could result from the strong
sensitivity of the accretion rate to the stellar wind velocity and
mass loss parameter, $\dot{M}/\dot{M}_{\rm edd} \propto
v_{w}^{-3.8(-2.4)}$ for core(cusp) galaxies, respectively
(Fig.~\ref{fig:mdot_mass}; eq.~[\ref{eq:eddr_analytic}]).  When
combined with the significant dependence of $v_w$ on stochastic
intermittency in the star formation history (Appendix
\ref{fig:NickPlot2}), this can lead to order of magnitude differences
in $\dot{M}$. However, we note that $v_w$ is not expected to vary by
more than a factor of a few for a thermally-stable flow, limiting the
allowed variation of $\dot{M}/\dot{M}_{\rm edd}$. The discrete nature
of stellar wind sources and their motions could result in an
additional order of magnitude variation $\dot{M}$
\citep{Cuadra+2008}.  

Variations in $L_{X}/L_{\rm edd}$ could also result from differences
in angular momentum of the infalling gas from galaxy to galaxy,
resulting in differences in the fraction of the gas lost to outflows.
Also potentially contributing is the order of magnitude difference, at
fixed $v_w$ and $M_{\bullet}$, between $\dot{M}$ for core and cusp
galaxies (Fig.~\ref{fig:bh_growth}).  Differences in $\dot{M}$ are
augmented by the theoretical expectation that $L_{\rm X} \propto
\dot{M}^{2} $ for radiatively inefficient flows.

\subsection{Steady Accretion versus outbursts}
\label{sec:cycle}

It is also possible, and indeed likely, that many of the
\citet{miller+2015} sample of galactic nuclei are not accreting in
steady state.  This is supported by the fact that many of the X-ray
luminosities in Figure \ref{fig:miller} lie above the predictions for
stable accretion, i.e., $L_X \gtrsim \epsilon_X \dot{M}_{\rm TI}c^{2}$
(yellow line), at least for our choice of radiative efficiency.

Within our steady state solutions, gas outside the stagnation radius
$\rs$ is blown out in wind. However, for very massive galaxies
($\Mbh\gsim 10^8 \Msun$), the energy injected into the gas is not
enough to truly eject it from the potential well of the galaxy, even
if the gas is blown out of the stellar bulge (e.g. even if the
$\zeta>\zeta_c$ criterion is satisfied--see
equation~\eqref{eq:zetacrit} and surrounding discussion). This gas is
likely to build up on the outskirts of the galaxy until a thermal
instability develops, causing the CNM to undergo cyclic oscillations
of rapid cooling and high accretion rates, followed by quiescent
periods once gas has been consumed and/or AGN feedback becomes
effective. Such cycles are seen in numerical simulations of elliptical
galaxies on large scales and long time-scales (\citealt{Ciotti+10}).
Evidence for such periodic outbursts includes observations showing
that a fraction of early-type galaxies in the local Universe have
undergone recent ($< 1-2$ Gyr old) star formation episodes
(\citealt{Donas+07}).

If the actual radiative efficiency is lower than our fiducial
assumption, then an even larger fraction of the X-ray detections would
lie above the predictions for stable accretion.  For example, we
assume that a fraction $\alpha = 0.1$ of the mass inflow rate reaches
the SMBH, but this fraction could be considerably smaller. In such a
case the X-ray detected galaxies will experience less heating than
we calculate for the average SFH, and may not be stably accreting at
all.  This is plausible given that the \citet{miller+2015} sample
includes only early-type galaxies with older stellar populations than
the average for their stellar mass.

Figure \ref{fig:bh_growth} (top panel) also shows that $\dot{M}_{\rm
  C} > \dot{M}_{\rm TI}$ across all black hole masses.  This indicates
that Compton heating is not sufficient to produce a thermally stable
accretion rate, at least for conditions corresponding to the average
SFH.

Similar cyclic AGN activity could occur on the smaller radial scale of
the sphere of influence (e.g.~\citealt{Yuan&Li11},
\citealt{cuadra+2015}).  In this case the large scatter in the X-ray
luminosities shown in Fig \ref{fig:miller} could result from the wide
range of inflow rates experienced over the course of a cyclic episode
between instabilities and periodic nuclear outbursts. 

If the galaxies with measured $L_{\rm X}$ at low $M_{\bullet}$ in
Figure \ref{fig:miller} are indeed in a state of thermally-unstable
outburst, then some of the galaxies with X-ray upper limits could be
contributing to a separate population of thermally-stable, steadily
accreting nuclei.  Such a population could include SgrA*, which has a
much lower X-ray luminosity for its SMBH mass than predicted by the
\citet{miller+2015} trend.  A potentially bimodal population of steady
and outbursting galactic nuclei calls into question the practice of
using simple extrapolations of power-law fits to the $L_{\rm
  X}(M_{\bullet}$) relationship to low $M_{\bullet}$ to constrain the
occupation fraction of SMBHs in the nuclei of low mass galaxies.

\subsection{SMBH Growth Times }
\label{sec:growth}

Low-mass AGN in the local Universe with $M_{\bullet} \lesssim 3\times
10^{7}M_{\odot}$ are observed to be growing on a time-scale comparable
to the age of the Universe, while the most massive SMBHs with
$M_{\bullet} \gtrsim 10^{9}M_{\odot}$ possess local growth times which
are more than 2 orders of magnitude longer (\citealt{Heckman+04};
\citealt{Kauffmann&Heckman09}).

The bottom panel of Figure~\ref{fig:bh_growth} shows the SMBH growth
time, $t_{\rm grow} \equiv M_{\bullet}/\dot{M}_{\bullet}$, for each accretion
rate shown in the top panel, recalling that $\dot{M}_{\bullet} = \alpha\dot{M}$, where $\alpha = 0.1$.  For SMBHs which accrete steadily at the
rate set by stellar wind heating due to the average star formation
history of their host galaxies, we see that $t_{\rm grow}$ exceeds the
Hubble time by 2$-$4 orders of magnitude across all $M_{\bullet}$.
Steady accretion therefore cannot explain the growth of low mass black
holes, a fact which is not surprising given that approximately half of
this growth occurs in AGN radiating within 10 per cent of their
Eddington rate (\citealt{Heckman+04}).  Such high accretion rates
likely instead require a source of gas external to the nuclear
region, triggered either by galaxy mergers associated with the
hierarchical growth of structure or thermal instabilities on larger,
galactic scales (e.g.~\citealt{Ciotti+10}, \citealt{Voit+15}).  

That the growth time associated with thermally-unstable accretion
(yellow line in Fig.~\ref{fig:bh_growth}) exceeds the Hubble time
across all SMBH masses highlights the fact that significant black hole
growth in the local Universe cannot result from thermally-stable
steady accretion of gas lost from the surrounding stellar population
studied in this paper.  Gas blow-out by SN Ia cannot alone prevent the
growth of low-mass black holes, as indicated by the low growth times
$\ll t_{\rm h}$ allowed by Ia heating (green lines), although Ia
heating could play in principle a role in capping the growth rate of
$\gtrsim 10^{8}M_{\odot}$ black holes, again depending on the
efficiency of Ia heating.

\subsection{TDE Jets}
\label{sec:TDE}

Our results also have implications for the environments encountered by
relativistic jets from TDEs.  For low-mass SMBHs with $M_{\bullet}
\lesssim 10^{7}M_{\odot}$, such as that responsible for powering the
transient {\it Swift} J1644+57 (\citealt{Bloom+11}), we predict gas
densities of a $n \sim $few cm$^{-3}$ on radial scales of 0.3 pc for
wind heating rates within the physically expected parameters $v_{\rm
  w} \sim 500-1000$ km s$^{-1}$ and $\eta=0.4$ (see
Fig.~\ref{fig:profiles}).  This is comparable to the density $\sim$
0.3-10 cm$^{-3}$ obtained by modeling the radio afterglow of {\it
  Swift} J1644+57(\citealt{BergerZauderer+:2012a},
\citealt{Metzger+2012}). However, there is considerable uncertainty in
the afterglow modeling. For example, \citet{Mimica+2015} find a much
higher density of 60 cm$^{-3}$ at 0.3 pc.

\citet{BergerZauderer+:2012a} infer a flattening of the gas density
profile of the host of J1644+57 on radial scales $r \gtrsim 0.4$ pc
(their Fig.~6) which looks qualitatively similar to the shape of the density
profile we predict for core galaxies (Fig.~\ref{fig:profiles}, solid
line).  To obtain such a flattening on the inferred radial scales
would require a black hole of $\Mbh\sim 10^7 \Msun$.  However, this is
inconsistent with the break in the gas density seen at $\sim 0.6$ pc
by \citet{BergerZauderer+:2012a}, which should correspond to the break
radius of the {\it stellar density} profile, thus implying an
unrealistically small SMBH mass of $\Mbh\simeq 100\Msun$ according to
the mean $\rb-M_{\bullet}$ relationship \citep{LauerFaber+:2007a}.
Alternatively, the observed density break in J1644+57 could result
from the outer edge of a sub-parsec nuclear star cluster (e.g.,
\citealt{Carson+2015}).  The high stellar densities of such a compact
star cluster could greatly enhance the TDE rate
(e.g.~\citealt{Stone&Metzger15}), possibly making this association
uncoincidental.  On the other hand, we note that some models for
J1644+57 (e.g., \citealt{Tchekhovskoy+2014}) favor a much smaller
black hole ($10^5-10^6 \Msun$) than we estimate would be required to
produce the observed inner break.

\subsection{Regulation by SMBH Kinetic Feedback}
\label{sec:kinetic}

\footnotetext{These equation correspond to the specific cases of
  $\Gamma=0.1$ and $\Gamma=0.8$, but it is straightforward to
  generalize these for arbitrary $\Gamma$ as we do for this figure.}
Our analysis of SMBH feedback has focused on Compton heating instead
of kinetic feedback, since the effects of radiative feedback are
relatively straightforward to calculate from the properties of the
accretion flow.  However, it is possible kinetic feedback from SMBH
outflows could play an equal or greater role in regulating accretion,
even in low luminosity AGN.

Our simple parameterization of kinetic feedback (eq.~[\ref{eq:vjet}])
assumes a uniform volumetric heating and predicts an effective wind
velocity which increases with SMBH mass as $v_{w}^{\bullet} \propto
M_{\bullet}^{0.38}$ for core and cusp galaxies, respectively.
Coupled with the dependence of the SMBH accretion rate on the wind
heating, $\dot{M}/\dot{M}_{\rm edd} \propto
M_{\bullet}^{0.76(0.48)}v_{w}^{-3.8(-2.4)}$ (Fig.~\ref{fig:mdot_mass};
eq.~[\ref{eq:eddr_analytic}]), a dominant source of kinetic feedback
of the form we have adopted leads to an Eddington ratio dependence on
SMBH mass of $\dot{M}/\dot{M}_{\rm edd} \propto
M_{\bullet}^{-0.7(-0.4)}$.

This prediction is more in line with the observed dependence of
$\langle L_{X}/L_{\rm edd} \rangle$ in elliptical galaxies
(Fig.~\ref{fig:miller}, green line), suggesting that accretion
regulation by kinetic feedback could play a role in determining the
X-ray luminosities of elliptical galaxies.  However, our assumption
that kinetic outflows heat the gas uniformly in volume is rather
arbitrary and would need to be more rigorously justified by numerical
studies of how jets or disc outflows couple energy to their gaseous
environment.  As already discussed, it is furthermore unclear whether
a steady-state accretion model is at all relevant in the case when
SMBH feedback dominates due to the time delay between the small scale
accretion flow and the thermodynamic response of the CNM on larger
scales.

\section{Conclusions}
\label{sec:conclusions}

We have calculated steady-state models for the hot gaseous
circumnuclear media of quiescent galaxies, under the assumption that
gas is supplied exclusively by stellar wind mass loss and heated by
shocked stellar winds, supernovae and black hole feedback.  We
numerically compute solutions for a range of different black hole
masses, heating rates, and observationally-motivated stellar density
profiles.  Then we use our numerical results (Table
\ref{table:models}) to verify and calibrate analytic
relationships (Appendix \ref{app:rs}).  We use the
latter to explore systematically how the SMBH accretion rate varies
with black hole mass and the galaxy's SFH.  Our results for
$\dot{M}(\Mbh)$ are compared with observed trends of the nuclear X-ray
luminosities of quiescent SMBHs and low luminosity AGN.  Our
conclusions are summarized as follows.

  \begin{enumerate}
  \item A stagnation radius, $\rs$, divides the nuclear gas between an
    accretion flow and an outgoing wind. In steady-state the gas
    inflow rate towards the black hole is proportional to the stellar
    mass enclosed inside of $\rs$.  In the limit of strong heating,
    the stagnation radius resides interior to the SMBH influence
    radius and coincides with the Bondi radius
    (eq.~[\ref{eq:rbondi}]).  In the limit of weak heating ($\zeta <
    \zeta_c$; eq.~[\ref{eq:zetacrit}]), the stagnation radius moves to
    large radii, near or exceeding the stellar break radius $\rb \gg
    r_{\rm inf}$, greatly increasing the density of gas on smaller
    radial scales.

  \item In the vicinity of the stagnation radius, including the effects
    of heat transport by electron conduction results in at most order
    unity changes to the key properties of the flow (e.g. stagnation
    radius, mass accretion rate, and thermal stability) for causal
    values of the conduction saturation parameter $\phi < 0.1$
    (eq.~[\ref{eq:conduction}]). However, in principle heat conduction
    of heat from the inner accretion flow can affect the solution
    properties on much larger scales \citep{Johnson+2007}. For
    example, \citet{Tanaka+2006} find that conductivity can contribute
    to driving bipolar outflows.

    Angular momentum of the gas will become important on small scales,
    where gas stopped by a centrifugal barrier and unable to cool
    could drive outflows near the equatorial and polar regions of the
    flow.

  \item Radiative cooling has a more pronounced influence on the flow
    structure when the radiative cooling rate exceeds the gas heating
    rate near the stagnation radius, where the gas is in nearly
    hydrostatic equilibrium (Fig.~\ref{fig:TI}).  This condition is
    approximately equivalent to the $t_{\rm cool} \lsim 10 t_{\rm ff}$
    criterion for thermal instability advocated by e.g. \citet{Sharma+2012},
    \citet{Gaspari+2012} and \citet{Li&Bryan14a}.  The transition in
    the flow properties that occurs as heating is reduced may
    represent a true ``thermal instability" (hot ISM condensing into a
    cooler clouds), or it may simply represent an abrupt transition
    from a steady inflow-outflow solution to a global cooling flow.
    We leave distinguishing between these possibilities to future
    work.

  \item The location of the stagnation radius, and hence the inflow
    rate, is a sensitive function of the gas heating rate $\propto
    v_w^{2}$.  Quantitatively, $\rs\propto v_w^{-2}$, implying that
    $\Mdot\propto v_w^{-3.8 (-2.4)}$ for fiducial core (cusp)
    galaxies.  However, $\Mdot$ can increase even more rapidly with
    decreasing $v_w$ for two reasons: (1) when $v_w$ becomes
    comparable to the stellar velocity dispersion ($\zeta < \zeta_c$;
    eq.~$[\ref{eq:zetacrit}]$), then gas remains bound to the stellar
    bulge and cannot produce an outflow interior to the stellar break
    radius (2) For low $v_w$, gas near the stagnation radius becomes
    thermally unstable, which based on the results of previous
    numerical studies (e.g.,\citealt{Ciotti+10}) is instead likely to
    result in a burst of high accretion ($\S\ref{sec:cycle}$).

  \item Stellar wind heating, supernovae, and AGN feedback depend
    explicitly on the SMBH mass as well as the SFH in the galactic
    nucleus.  A young starburst of age $\lesssim 10$ Myr produces mass
    and energy injection dominated by young stellar winds ($v_w \sim
    1000$ km s$^{-1}$) and supernovae, while for exclusively old
    stellar populations mass input is dominated by AGB winds and
    heating is dominated by AGN feedback and supernovae.  Ongoing,
    continuous star formation presents a hybrid situation, with energy
    input usually dominated by young stars and mass input dominated by
    old stars (Appendix \ref{app:windheat}).  Galactic nuclei with
    $\Mbh \lesssim 10^8\Msun$ that are heated according to their
    average SFHs (as derived from cosmological simulations) receive a
    stellar wind heating of $v_w \sim 700$ km s$^{-1}$.  This heating
    can be suppressed if the star formation is sufficiently
    intermittent (Fig.~\ref{fig:NickPlot2}).

\item Type Ia SNe only provide a continuous heating source exterior to
  the Ia radius $\rIa$ (eq.~[\ref{eq:rIa}]) where the time between
  subsequent SNe exceeds the dynamical time-scale.  Non-Ia heating due
  to the average SFH results in $\rs <\rIa$, except for the most
  massive galaxies (Fig.~\ref{fig:vwSources}).  However, if $\rs$ does
  approach $\rIa$, then SN Ia heating is usually
  large enough to prevent $\rs$ from greatly exceeding $\rIa$.
  SNe Ia thus regulate the SMBH accretion rate below the value
  $\dot{M}_{\rm Ia}$ (eq.~[\ref{eq:eddr_Ia}]), except possibly in high
  mass elliptical galaxies with large break radii and $\zeta(v_w^{\rm
    Ia}) \gtrsim \zeta_c$.

\item Unlike stellar heating, heating from SMBH feedback depends on
  the accretion rate, which itself depends on the heating.  Compton
  heating is generally unimportant compared to stellar wind heating
  for the average SFH, but it can be significant in the case of an impulsive
  starburst(Fig.~\ref{fig:vwSourcesImp}).  However, SMBH feedback may not
  be capable of truly stabilizing the flow given its inability
  to respond instantaneously to local changes in the properties of the
  gas.  Kinetic feedback could also be important in determining
  $\dot{M}(M_{\bullet}$) when stellar heating is weak
  ($\S\ref{sec:kinetic}$) but is more challenging to quantify.

\item Stellar wind heating from the average SFH may be sufficient to
  permit thermally-stable, steady-state accretion, depending on the
  accretion efficiency of the SMBH.  However, the resulting average
  Eddington-normalized accretion rate is predicted to increase with
  $M_{\bullet}$ (Fig.~\ref{fig:miller}), in tension with the (also
  weak) downsizing trend of measured X-ray luminosities in early-type
  galactic nuclei (e.g., \citealt{miller+2015}).  Thermally stable
  accretion models can reproduce the observed scatter in nuclear X-ray
  luminosities at fixed $\Mbh$ (two to four orders of magnitude) due to the
  combination of (1) differences in the stellar wind heating rate due
  to stochastic variation in SFH between galaxies, or within one
  galaxy due to burstiness in the SFH, as in
  ~Fig.~\ref{fig:NickPlot2}; (2) variations in the amount of inflowing
  mass which is lost to small scale outflows from the SMBH accretion
  disc, likely due to variations in the angular momentum of the
  accreting gas (cf.~\citealt{Pellegrini10}).

\item However, for lower X-ray radiative efficiencies, the accretion
  rates of early-type galaxies are above the maximum value for
  thermally stable flow, $\dot{M}_{\rm TI}$ (Fig.~\ref{fig:miller}).
  This implies that current X-ray detections could instead be
  comprised mostly of nuclei undergoing outburst due to thermal
  instabilities.  In such a case, there may exist a separate (usually
  undetected) branch of low-$L_X$ nuclei accreting stably.  The
  existence of such a bimodal population would call into question
  constraints on the SMBH occupation fraction in low mass galaxies
  (\citealt{miller+2015}) derived by extrapolating the $L_{\rm
    X}(M_{\bullet}$) relationship to small $M_{\bullet}$.

\item Low mass black holes grow in the low redshift Universe over
  time-scales comparable to the Hubble time (\citealt{Heckman+04}).
  The accretion rates so required are too high to be consistent with
  either the value predicted for the average SFH history or the
  maximum allowed for steady-state, thermally stable accretion
  (Fig.~\ref{fig:bh_growth}).  Perhaps unsurprisingly,
  low-$M_{\bullet}$ growth and AGN activity must instead be driven by
  a supply of gas external to the nucleus, such as galaxy mergers or
  thermal instabilities on larger, galactic scales
  (e.g.~\citealt{Voit+15}).

\end{enumerate}

We conclude by drawing attention to a few limitations of our model.
First, we have neglected any large scale inflow of gas to the nucleus
by assuming the only source of gas feeding the black hole is stellar
wind mass loss from the local stellar population.  Our estimates
represent a lower bound on the time averaged gas density of the CNM.
Although our model can, under certain assumptions, describe quiescent
galactic nuclei, it cannot account for AGN.

Predictions for the SMBH accretion rate are strongly tied to the age
and radial distribution of the stellar population within the sphere of
influence.  Our model therefore cannot make definitive predictions for
the accretion mode of individual galaxies without knowledge of their
specific SFH.  We adopt halo-averaged star formation rates measured
using multi-epoch abundance matching models by
\citet{MosterNaab+:2013a}, and neglect any spatial variation in the
star formation rate for any given galaxy. However, galaxies of mass
$\gsim 7\times 10^{10} \Msun$ assemble their stars from the inside out
\citep{PerezCidFernandes+:2013a}, resulting in galactic nuclei which
may be systematically older than assumed in our model and thus more
prone to thermal instability.

Finally, we do not account for the effects of non-spherical geometry
or discreteness of stellar wind sources \citep{Cuadra+2006,
  Cuadra+2008}. We assume all of the mass and energy sources are
efficiently mixed, and do not take into account the possibility of
slower wind material condensing into high density structures (as
described in \citealt{Cuadra+2005}) or some of the energy injection
leaking away in 3-D \citep{Harper-Clark+2009,Zubovas+2014}.

\section*{Acknowledgments}

We thank the referee, Sergei Nayakshin, for several insightful
comments. We acknowledge helpful conversations with Greg Bryan, Jenny
Greene, Jules Halpern, Kathryn Johnston, Jerry Ostriker, Eliot
Quataert, Massimo Gaspari and Takamitsu Tanaka.  This work relied on
Columbia University's Yeti computer cluster, and we acknowledge the
Columbia HPC support staff for assistance.  BDM gratefully
acknowledges support from NASA {\it Fermi} grant NNX14AQ68G, NSF grant
AST-1410950, and the Alfred P. Sloan Foundation.

  \clearpage
  \appendix
  \section{Derivation of stagnation radius}
  \label{app:rs}
  Here we derive an analytic expression for the time-independent
stagnation radius, $\rs$.  In steady-state, the requirement that the
net heating rate on the right hand side of the entropy equation
(eq.~[\ref{eq:dsdt}]) must equal zero at the stagnation radius (where $v = 0$) fixes the temperature at $\rs$:
\begin{align}
 \frac{\gamma}{\gamma-1} \left.\frac{p}{\rho}\right|_{\rs}=\frac{\vw^{2}|_{\rs}}{2} \Rightarrow \frac{\kb T|_{\rs}}{\mu \mp}=\gammafi \frac{\vw^{2}|_{\rs}}{2} 
\label{eq:appTanalytic}
\end{align}
Combining (eq.~[\ref{eq:dsdt}]) with the first law of thermodynamics,
\begin{align}
T\left.\dsdr\right|_{\rs} = \frac{1}{\gamma-1}\left.\ddr{(p/\rho)}\right|_{\rs}+\frac{\densSlope}{\rs}  \left.\frac{p}{\rho}\right|_{\rs}=\left.\underbrace{\frac{\Q}{\rho  v}}_{A}\right|_{\rs}, 
 \label{eq:first_law}
\end{align}
where $\densSlope\equiv -\left.d\ln\rho/d\ln r\right|_{\rs}$.  The right side can be evaluated using L'Hopital's rule, yielding

\begin{align}
  \lim_{r \rightarrow \rs} A &=& \frac{d}{dr} \left[\left(\ke -\gammaf \frac{p}{\rho}+\kew\right)\right]_{\rs} \nonumber \\
&\approx& -\gammaf
\left.\ddr{(p/\rho)}\right|_{\rs}-\frac{3}{2(2+\Gamma)} \frac{G \Mbh}{\rs^2},
\end{align}
where we have used the definition $\vw^2=\vwO^2+ \frac{3}{2+\Gamma} \frac{G \Mbh}{r} +
\sigma_0^2$ (eq.~[\ref{eq:sigmarel}]).  Substituting this expression back into equation (\ref{eq:first_law}) and using equation (\ref{eq:appTanalytic}) gives
\begin{align}
&\frac{\gamma+1}{\gamma-1}
\left.\ddr{(p/\rho)}\right|_{\rs}+ \frac{\gamma-1}{\gamma} \frac{\densSlope}{\rs} \frac{\vw|_{\rs}^{2}}{2} = -\frac{G
  \Mstar|_{\rs}}{2 \rs^2} -\frac{3}{2+\Gamma} \frac{G \Mbh}{2 \rs^2}.  \label{eq:rs1}
\end{align}
Combining this with condition of hydrostatic equilibrium (eq.~[\ref{eq:dvdt}]) at the stagnation point,
\begin{align}
\frac{1}{\rho}\left.\dpdr\right|_{\rs}=- \frac{G\Menc}{\rs^2} \Rightarrow
\left.\ddr{(p/\rho)}\right|_{\rs} +\frac{p}{\rho}
\underbrace{\frac{d\ln(\rho)}{dr}}_{-\densSlope/r} = -\frac{G \Menc}{\rs^2}, \label{eq:HSE}
\end{align}
results in the following expression for the stagnation radius:
\begin{align}
\rs=\frac{G \Mbh}{\densSlope \vw^{2}|_{\rs}}\left(4
  \frac{M_{\star}|_{\rs}}{\Mbh} +\frac{13+ 8\Gamma}{4+2\Gamma}\right) \nonumber \\
=\frac{G \Mbh}{\densSlope (v_w^{2}+\sigma_0^2)}\left[4
    \frac{\Mstar|_{\rs}}{\Mbh} +\frac{13+ 8\Gamma}{4+2\Gamma}-\frac{3\nu }{2+\Gamma}\right]
\label{eq:rs2main}
\end{align}
where we have assumed $\gamma=5/3$.  This relationship can be reparameterized as
\begin{align}
  \frac{\rs}{\rinf}=\frac{1}{3\nu \zeta^2} \left[4\frac{\Mstar|_{\rs}}{\Mbh} +\frac{13+8\Gamma}{4+2\Gamma}- \frac{3\nu }{2+\Gamma}\right],
  \label{eq:rsRinf}
\end{align}
where $\rinf=3 G \Mbh/\sigma_0^2$ and we have defined $\zeta \equiv \sqrt{1 + (v_w/\sigma_0)^2}$.  Because $M_{\star}|_{\rs} = M_{\bullet}(r_{\rm s}/r_{\rm inf})^{2-\Gamma}$, in general equation (\ref{eq:rsRinf}) must be solved implicitly for $r_{\rm s}/r_{\rm inf}$.  However, when $\Mstar|_{\rs} << \Mbh$ or equivalently $\rs \ll r_{\rm inf}$, equation (\ref{eq:rsRinf}) simplifies considerably:
\begin{align}
\rs=\frac{13+8\Gamma}{4+2\Gamma} \frac{G \Mbh}{\densSlope \vw^2}
\end{align}


  \section{Analytic model for dependence of wind heating $\vwO^{\star}$ on stellar age}
\label{app:windheat}

\subsection{Single Burst}

In the case of a single impulsive burst of star formation, the mass
and energy injection rate per unit stellar mass at time $t$ after the
burst are given, respectively, by
\begin{align} 
  \dot{\bar{m}}(t) &= f_{\rm TO} \dot{m}_{\rm TO} + f_{\rm OB}
  \dot{m}_{\rm OB}\\
\dot{\bar{e}}(t) &=f_{\rm OB} \, \dot{e}_{\rm OB}(t)+ f_{\rm MS}
\int_{m_0}^{m_{\rm T}(t)} \frac{\vw^2(\Mstar, t) \dot{m}(\Mstar, t)
  \mu|_{\Mstar} {\rm d}\Mstar}{\bar{m}_*},
  \label{eq:edotImp}
\end{align} 

where the $f$'s represent the efficiency with which each source of
mass/energy injection is thermalized (we take all of the f's to be 1).
In the top line, the first term corresponds to mass injection due
winds from post-main sequence stars. The second term corresponds to
mass injection due to stellar winds from massive
stars. For the latter we use population synthesis calculations by
\citet{VossDiehl+:2009a}. From the bottom panel of their Figure 7, the
stellar wind mass loss rate per massive star can be approximated to
within a factor of a few by

\begin{equation}
\dot{\mathcal{M}}=
\begin{cases}
10^{-6.5} \Msun {\rm yr^{-1}} & t < 4 {\, \rm Myr}\\
10^{-5.4} \Msun {\rm yr^{-1}} \left(\frac{t}{4\times 10^6 {\rm
       yr}}\right)^{-3} & 4 {\, \rm Myr} \leq t \leq 40 {\, \rm Myr}. \\
0 & t>40 {\rm Myr}.
\end{cases}
\end{equation}

$\dot{m}_{\rm OB}=f_{8} \dot{\mathcal{M}}/\bar{m}_{\star}$,
where $f_{8} =2.6 \times 10^{-3}$ is the fraction of the stellar mass
with $M_{\star} > 8M_{\odot}$ and $m_{\star}=0.35 \Msun$ is the mean
stellar mass for our assumed Salpeter IMF, $\mu\sim M_\star^{-2.35}$.

For the mass loss rate from post-main sequence winds, we take

\begin{equation}
  \dot{m}_{\rm TO}=\frac{\Delta M(t) |\dot{M}_{\rm TO}(t)|
    \mu|_{M_{\rm TO}(t)}}{\bar{m}_{\star}}  \,\,  t \geq {\rm 40 Myr},
\end{equation}

and 0 for earlier times. By truncating $\dot{m}_{\rm TO}$ at 40 Myr, we
are ignoring the non-steady mass injection by type II supernovae.

The quantity of mass lost in post-main sequence winds $\Delta M(t)$ is
estimated from the expression given by \citet{Ciotti&Ostriker07}
(their eq.~[10]),
\begin{align}
\Delta M=
\begin{cases}
0.945 M_{\rm TO}-0.503 & M_{\rm TO} < 9 \Msun\\
 M_{\rm TO}-1.4 \Msun &  M_{\rm TO} \ge 9 \Msun,
\end{cases}
\end{align}
where $M_{\rm TO}$ is the turn-off mass, which at time $t < t_{\rm h}$ is calculated as
\begin{align}
\log(M_{\rm TO})/M_{\odot} =0.24 + 0.068 x^2-0.34 x+4.76 e^{-4.58 x},
\end{align}
where $x=\log(t/10^9 {\rm yr})$.  This functional fit is designed
to reproduce the results of \citet{MaederMeynet:1987a} (their Table 9)
for massive stars while asymptoting to the formula provided by
\citet{Ciotti&Ostriker07} (their eq.~[9]) for intermediate and late
times ($t \gsim 10^8$ years).

The first term in equation~\eqref{eq:edotImp} corresponds energy
injection from massive stars, while the second term accounts
for energy injection from stars on the lower main sequence. Both terms
have a thermalization efficiency, $f$, which w e talk to be $1$. Note
we do not account for energy from Type II supernovae. 

The MS wind mass loss rate $\dot{m}(\Mstar, t)$ is calculated based on
the generalization of Reimer's law
\begin{align}
  \dot{m}=4 \times 10^{-13} \frac{L_* R_*}{\Mstar} \Msun \pyear,\
\end{align}
where  $R_*$, $L_*$, $T_{\rm eff}$ and $g_*$ are the stellar radius,
luminosity, effective temperature, and surface gravity, respectively, the latter normalized to its solar value $g_{\odot}$.  The stellar radius and luminosity are estimated as (\citet{Kippenhahn&Weigert90}; Figs.~22.2) 22.3)
\begin{align}
L_*=
\begin{cases}
L_{\odot} (\Mstar/\Msun)^{3.2} & \Mstar > \Msun \\
L_{\odot} (\Mstar/\Msun)^{2.5} & \Mstar \le \Msun
\end{cases}
\end{align}
\begin{align}
R_*=
\begin{cases}
R_{\odot} (\Mstar/\Msun)^{0.57} & \Mstar > \Msun \\
R_{\odot} (\Mstar/\Msun)^{0.8} & \Mstar \le \Msun
\end{cases}
\end{align}
The wind velocity of main sequence winds is assumed to equal $v_w
(\Mstar, t) =
v_{w,\odot}(M_{\star}/M_{\odot})^{1/2}(R_{\star}/R_{\odot})^{-1/2}$,
i.e.~scaling as the stellar escape velocity and normalized to the
velocity of the solar wind $v_{w,\odot} = 430$ km s$^{-1}$; this
produces an effective wind heating velocity for main sequence winds
alone of $\sim 100$ km s$^{-1}$ for $\tau_{\star} \sim t_{\rm h}$,
close to the value found by \citet{NaimanSoares-Furtado+:2013a} for
globular clusters based on a more sophisticated population synthesis
treatment.

Winds from lower main sequence stars only dominate the energy
injection a late times ~10 Gyr after an impulsive burst of star
formation. Even with 100\% thermalization efficiency these winds could
not thermally stabilize the CNM. 

The rate of energy injection due to winds from massive
stars, $\dot{e}_{\rm OB}(t) = f_{8}\dot{\mathcal{E}}/\bar{m}_*$, where
$f_{8} =2.6 \times 10^{-3}$ is the fraction of the stellar mass with
$M_{\star} > 8M_{\odot}$ for our assumed Salpeter IMF.  Here
$\dot{\mathcal{E}} (t)$ is the energy injection rate per massive star,
which we estimate as
\begin{align}
\dot{\mathcal{E}} (t)=  1.3 \times 10^{36} {\rm erg\,s^{-1}}
\begin{cases}
  1 & t<4 \times 10^6 {\rm yr}\\
  \left(\frac{t}{4 \times  10^6\,{\rm yr}}\right)^{-3.73} &  4\times
  10^{6} {\rm yr}  \le t \le 10^{8} {\rm yr},\\
  0 & t > 10^8 {\rm yr}
\end{cases}
\label{eq:voss}
\end{align}
based on the results of \citet{VossDiehl+:2009a} (their Fig.~7, top
panel), who use a population synthesis code to simulate the mass and
energy injection into the ISM from an OB association. Although
equation is valid only for $t \lsim$ 10 Myr, in practice the precise
functional form of $\dot{e}_{\rm OB}(t)$ is generally unimportant for
our purposes, so we adopt equation as being valid for all times.

\begin{figure}
\includegraphics[width=\columnwidth]{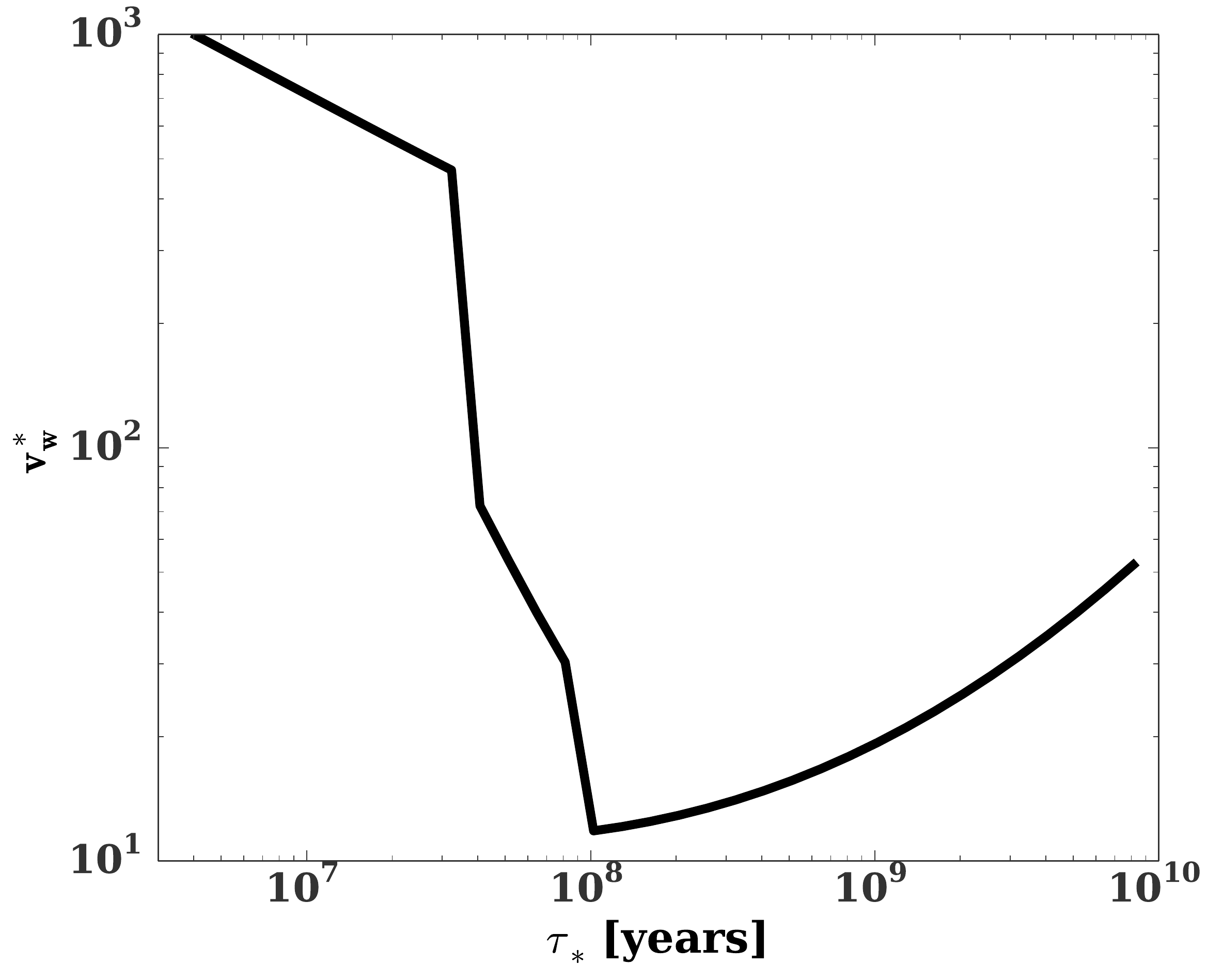}
\includegraphics[width=\columnwidth]{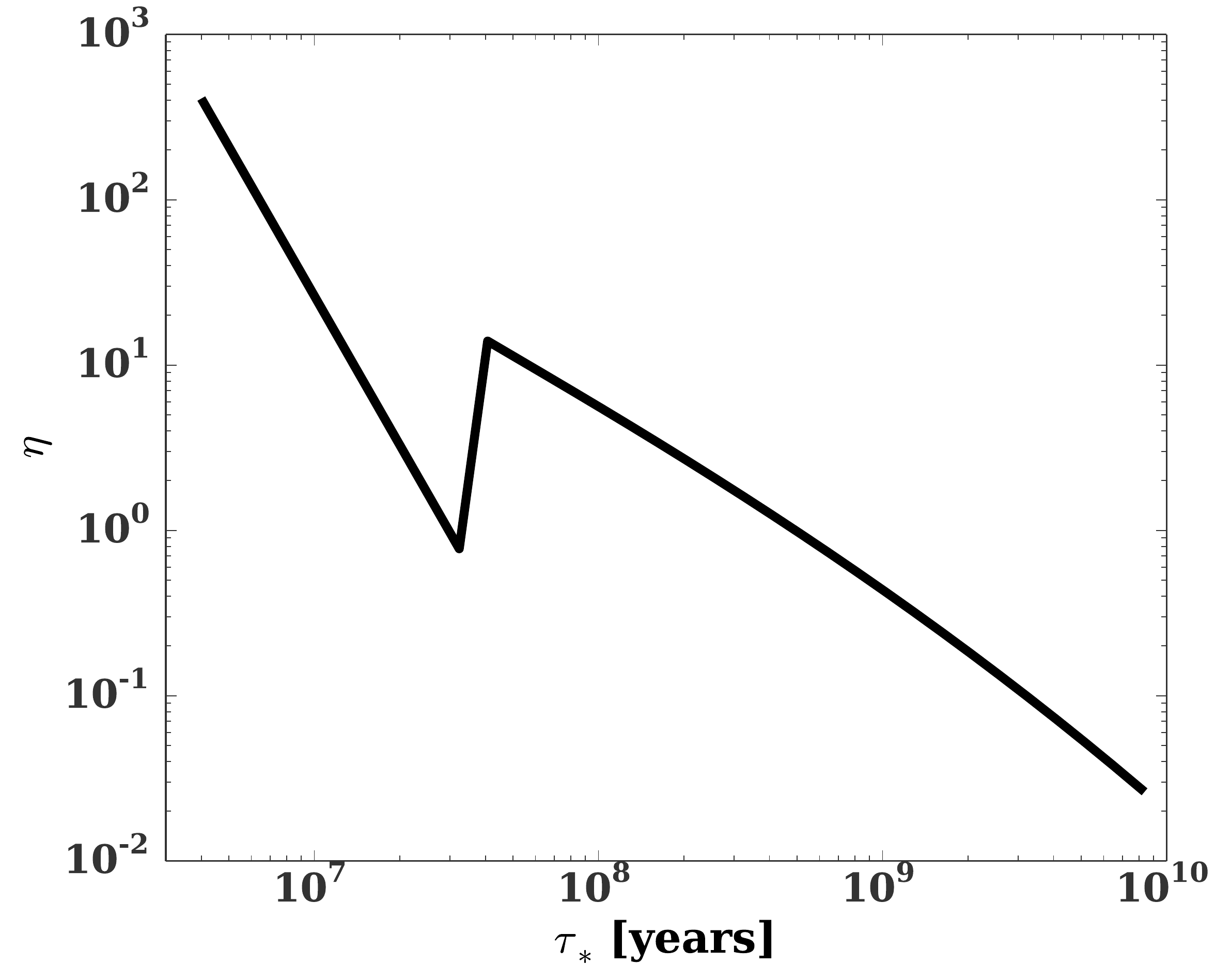}
\caption{\label{fig:vwImp} Effective heating rate, $v_w^{\star}$, and mass loss parameter, $\eta$ (eq.~[\ref{eq:q}]), resulting from stellar winds from a stellar population of age $\tau_{\star}$.}
\end{figure}

The effective wind velocity in the limit of an impulsive star
formation may then be written as
\begin{align}
\bar{v}_w(t)=2 \dot{\bar{e}}(t)/\dot{\bar{m}}(t)
\label{eq:vwImp}
\end{align}
while 
\begin{align}
\eta = \dot{\bar{m}}(t) \th
\label{eq:etaImp}
\end{align}
 Figure~\ref{fig:vwImp} shows the values of $\bar{v}_w(t)$ and $\eta(t)$ as a function of stellar age, $\tau_{\star}$.

\subsection{Over Star Formation History}
Generalizing to an arbitrary SFH $S(t)$, the total rate of mass and energy input can be written as
\begin{align} 
  \dot{M}(t) &= \int_0^t S(t_1) \dot{\bar{m}}(t-t_1){\rm
      d}t_1 \label{eq:MDotSFH}\\
  \dot{E}(t) &= \int_0^t S(t_1) \dot{\bar{e}}(t-t_1){\rm
      d}t_1, \label{eq:EDotSFH}
\end{align}
resulting in a wind heating parameter of 
\begin{align}
  v_w^2(t) &=2 \dot{E}(t)/\dot{M}(t).
\end{align}
The mass return parameter will be
\begin{align}
\eta = \frac{\dot{M}(t)}{\int_0^t S(t_1) {\rm d}t_1} \th
\end{align}

 \begin{figure}
  \includegraphics[width=\columnwidth]{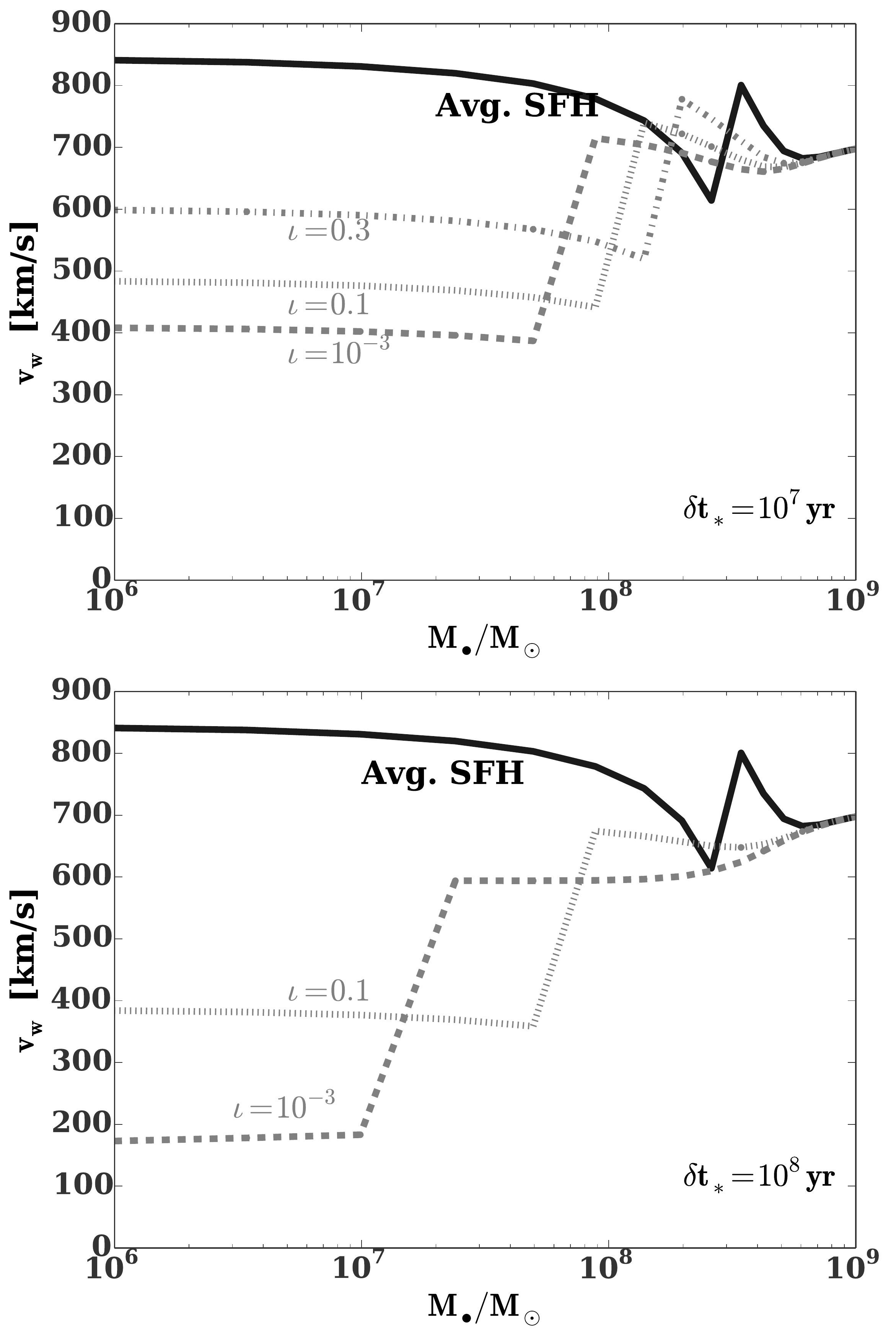}
  \caption{\label{fig:NickPlot2} Effective wind velocities for
    nonstandard star formation histories.  The black curves shows, for
    reference, $v_{\rm w}$ calculated using the halo-averaged $S(t)$,
    and gray curves show wind heating resulting from perturbed star
    formation histories given by equation~\eqref{eq:sfrPerturbed}. In
    the top panel the star formation rate declines for a time, $\delta
    t_*=10^{7}$ years to fractions $\iota= 0.001$ (dashed), $\iota
    =0.1$ (dotted), and $\iota = 0.3$ (dot-dashed) of the halo
    averaged value. The bottom panel shows results for $\delta
    t_*=10^{8}$ years and $\iota$=0.001 (dashed) and 0.1 (dotted).}
  \end{figure}
  We estimate the stellar wind heating provided by the {\it average}
  star formation history of galaxies of a given $\Mbh$ using the
  results of \citet[eqs.~17-20]{MosterNaab+:2013a}.  Note that the
  star formation histories in \citet{MosterNaab+:2013a} are in terms
  of halo mass. For a given $\Mbh$ we assign the halo mass whose star
  formation history would produce a bulge consistent with the
  $\Mbh-M_{\rm bulge}$ relationship from \citet{McConnellMa:2013a}.  A
  slight complication occurs for the largest mass halos, where much of
  the $z=0$ stellar mass has been acquired through accretion of
  satellite halos rather than {\it in situ} star formation.  To
  accommodate this, we incorporate analytic fits for mass accretion
  histories, taken from \citet[their eqs.~21-23]{MosterNaab+:2013a},
  assuming that the age distribution of the accreted stars is equal to
  the age distribution of those formed {\it in situ}.  This assumption
  may be conservative if the primary galaxy's accretion history is
  dominated by minor mergers with younger stellar populations.  On the
  other hand, the dynamical friction inspiral time for small satellite
  galaxies is quite long, generally much greater than the $\sim
  10^7~{\rm yr}$ for which young stars can dominate the heating
  budget.  The mass of stars accreted for halo masses, $M_{\rm halo}< 3
  \times 10^{12} \Msun$, and redshifts, $z>4$, is small and is neglected.

  To find the total mass ($\Mdot'(t)$) and energy ($\dot{E}'(t)$)
  injection rates, including the contribution of accreted stars, we
  add a convolution of the specific mass and injection rates with the
  accretion history $A(t)$ to the mass and energy injection rates from
  star formed in-situ.  Thus,

\begin{align}
\Mdot '(t)&=\Mdot(t)+\int_{0}^{t} A(t_{\rm acc}) \frac{\Mdot(t_{\rm acc})}
{\int_{0}^{t_{\rm acc}} S(t_1) dt_1} {\rm dt_{ acc}}\\
\dot{E}'(t)&=\dot{E}(t)+\int_{0}^{t} A(t_{\rm acc}) \frac{\dot{E}(t_{\rm acc})}
{\int_{0}^{t_{\rm acc}} S(t') dt_1} {\rm dt_{acc}}
\end{align}

The corresponding wind heating parameter $v_w'$ and mass return
parameter, $\eta'$, will be given by

\begin{align}
\eta'&=\frac{\Mdot'(t)}{\int_0^t S(t_1) {\rm d}t_1+\int_0^t A(t_1) {\rm
    d}t_1} \th\\
v_w'^2&=2 \frac{\dot{E}'(t)}{\Mdot'(t)}
\end{align}

Figure \ref{fig:NickPlot2} shows how the wind heating varies as star formation histories deviate from their halo-averaged values.  In particular, we show
\begin{align} 
  \dot{\mathcal{M}}(t) &= \int_0^t \mathcal{S}(t_1) \dot{\bar{m}}(t-t_1){\rm
      d}t_1\\
  \dot{\mathcal{E}}(t) &= \int_0^t \mathcal{S}(t_1) \dot{\bar{e}}(t-t_1){\rm
      d}t_1,\\
  \mathcal{V}_w^2(t) &=2 \dot{\mathcal{E}}(t)/\dot{\mathcal{M}}(t)
\end{align}
In these equations, 
\begin{equation}
\mathcal{S}(t) = S(t) \times
\left(\frac{2}{\pi}(1-\iota)\arctan(t/\delta t_{\star}) + \iota
\right)
\label{eq:sfrPerturbed}
\end{equation}
This function convolves the recent ($z \approx 0$) halo-averaged star
formation history with local variation to give a more pessimistic
estimate for the value of $\tilde{v}_{\rm w}$.  In particular,
replacing $S(t)$ with $\mathcal{S}(t)$ reduces the recent star
formation rate to a fraction $\epsilon$ of its halo-averaged value,
and does so for a characteristic time $\delta t_{\star}$ into the
past.  As we can see in Fig. \ref{fig:NickPlot2}, this dramatically
lowers the effective wind speed when both $\delta t_{\star} \gtrsim
10^7 ~{\rm yr}$ and $\epsilon \lesssim 0.1$, but otherwise has too
modest of an effect to change the thermal stability properties of the
flow (although the location of $r_{\rm s}$ and the value of $\dot{M}$
may change significantly).


  \footnotesize{
    \bibliographystyle{mnras}
    \bibliography{master}
  }
\end{document}